\RequirePackage{snapshot} 

\RequirePackage{sty/bootstrap}

\WithOptions
\documentclass[runningheads]{llncs}

\ProvideIfOptions {final}
\ProvideIf[true]{draft}
\iffinal\draftfalse\fi

\usepackage{hyperref}

\usepackage{comment}
\usepackage{amsmath,mathtools}
\usepackage{multido}
\usepackage{enumitem,xcolor}
\usepackage{tikz}
\usepackage{graphicx}
\usepackage{caption}
\usepackage{subfigure}
\usepackage[utf8]{inputenc}
\usepackage{pgfplots} 
\usepackage{pgfgantt}
\usetikzlibrary{spy}
\usetikzlibrary{shapes.multipart}
\usepackage{xfrac}
\usepackage{colortbl}
\usepackage{amssymb}
\usepackage{cancel}
\usepackage{mleftright} 
\usetikzlibrary{datavisualization.formats.functions}
\usetikzlibrary{arrows, decorations.markings}
\usetikzlibrary{arrows.meta}

\newcommand{\h}{\mathsf{H}}

\usepackage{booktabs}
\usepackage{tabularray}

\usepackage{etoolbox}
\usepackage{hyperref}
\apptocmd{\UrlBreaks}{\do\f\do\m}{}{}

\newcommand{\setx}{\ensuremath{\mathcal{X}}}
\newcommand{\sety}{\ensuremath{\mathcal{Y}}}
\newcommand{\setz}{\ensuremath{\mathcal{Z}}}
\newcommand{\setd}{\ensuremath{\mathcal{D}}}
\newcommand{\setn}{\ensuremath{\mathcal{N}}}

\newcommand{\upp}{u^{\prime\prime}}

\newcommand{\Mpp}{M^{\prime\prime}}

\DeclarePairedDelimiterX\set[1]\lbrace\rbrace{#1}

\ExplSyntaxOn
\ifdraft
  \newcommand \newauthormark [2] {
    \cs_new:cpn {#1} ##1 {\textcolor{#2}{[#1:~##1]}}
  }
\else
  \newcommand \newauthormark [2] {
    \cs_new:cpn {#1} ##1 {}
  }
\fi
\ExplSyntaxOff

\ClistMap{
  , {Wafa}      {blue}
  , {Nilesh}    {purple}
  , {Johannes}  {green}
  , {Juan}      {teal}
  , {Holger}    {yellow}
  , {Christian} {brown}
  , {Frank}     {brown}
  , {Marc}      {red}
  }
  { \newauthormark #1 }

\title{Identification Codes and Post-Shannon Communication:\
Theory, Architectures, and Emerging Applications}

\titlerunning{Identification Codes and Post-Shannon Communication}

\author{
Wafa Labidi\inst{1} \and
Kumar Nilesh\inst{1} \and
Johannes Rosenberger\inst{1} \and
Juan Cabrera\inst{2,6} \and
Holger Boche\inst{1,5,6,7} \and
Christian Deppe\inst{3} \and
Frank H.P. Fitzek\inst{2,6} \and
Marc Geitz\inst{4}
}

\institute{
Technical University of Munich, Germany \and
Technische Universität Dresden \and
Technische Universität Braunschweig, Germany \and
T-Labs, Deutsche Telekom \and
Munich Center for Quantum Science and Technology \and
Centre for Tactile Internet with Human-in-the-Loop (CeTI) of Technische
Universität Dresden 
\and Munich Quantum Valley (MQV)}

\authorrunning{W. Labidi, K. Nilesh, J. Rosenberger, J. Cabrera, H. Boche, C. Deppe, F.H.P. Fitzek and M. Geitz}
\begin{document}
\maketitle

\begin{abstract}
Identification (ID) coding, introduced by Ahlswede and Dueck, extends Shannon’s classical communication paradigm by replacing message reconstruction with hypothesis testing. Instead of decoding the transmitted message, the receiver only decides whether a particular message was sent. A fundamental result of ID theory is the double-exponential growth in the number of identifiable messages with respect to (w.r.t.) the blocklength. This scaling behavior enables fundamentally new communication architectures for large-scale distributed systems and forms a key building block of post-Shannon communication.

While ID cannot replace classical communication in general, it is particularly
well-suited for scenarios in which full message reconstruction is unnecessary,
such as monitoring, alarming, and control systems.

In this survey, we review the theoretical foundations of ID coding and discuss
emerging communication architectures and application domains based on this
paradigm. Particular emphasis is placed on practical use cases, including
monitoring systems, special-purpose data storage, joint identification and
sensing (JIDAS), semantic communications, mobile-network control systems
and networked consensus testing systems. We
further highlight recent system concepts, industrial perspectives, and
implementation examples that illustrate how ID-based principles can be realized
in practical communication systems.
\end{abstract}

\begingroup 
  \setcounter{tocdepth}{2}
  \makeatletter
    \def\l@title#1#2{}
    \def\l@author#1#2{}
    \def\authcount#1{}
  \makeatother
  %
  \tableofcontents
\endgroup
\newpage

\section{Introduction}\label{sec:intro}


Trustworthiness is a cornerstone of next-generation communication networks, as these systems will serve as an essential infrastructure and support emerging mission-critical services, including applications in healthcare, medical technologies, robotics, and autonomous driving. Ensuring trustworthiness requires communication systems to satisfy key properties such as robustness, resilience, security, scalability, reliability, and privacy \cite{6Gandtrustworthiness,6Gcomm}. 
Achieving these objectives calls for new communication paradigms that extend beyond classical Shannon theory, with post-Shannon theory emerging as a promising framework in this context. Within this framework, message identification (ID) constitutes a fundamental building block of post-Shannon theory.


Shannon’s classical communication theory focuses on the reliable transmission
of messages over noisy channels, where the receiver aims to reconstruct the
transmitted information with vanishing error probability. This paradigm has
formed the foundation of modern communication systems and remains central to
their design.
\begin{figure}
    \centering
    \includegraphics[width=0.8\linewidth]{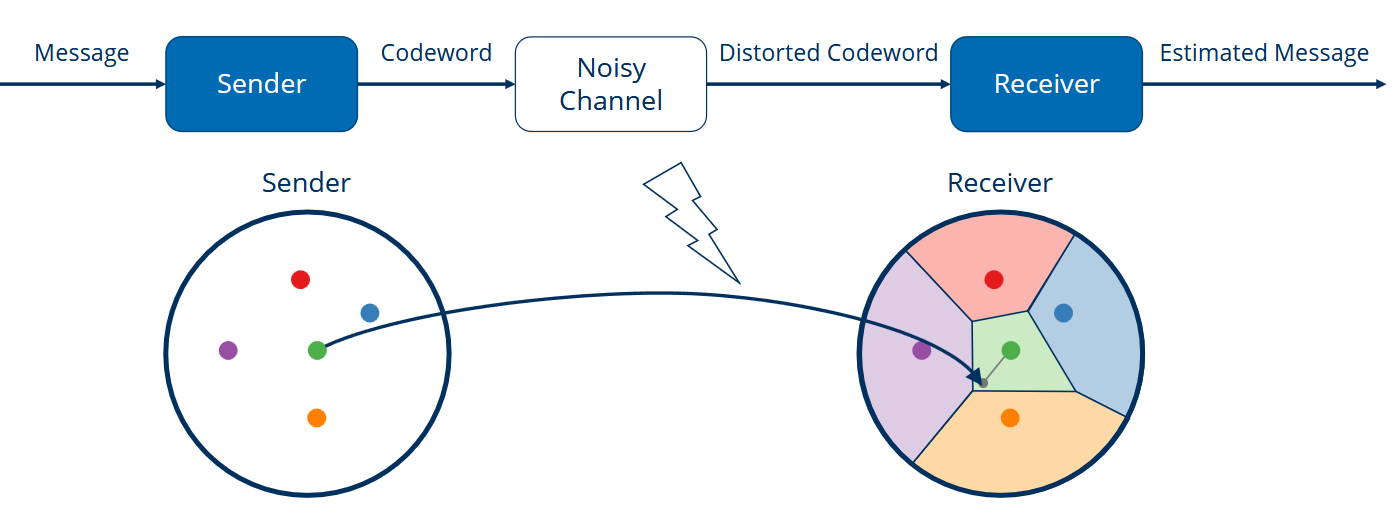}
    \caption{Shannon transmission: each message is represented by a unique (\emph{deterministic}) codeword at the sender. The receiver employs a decoding rule that partitions the observation space into disjoint, convex regions, with each region corresponding to exactly one message. Adapted from \cite{Schaefer2022PostShannonTutorial}}
    \label{TransmissionCodes}
\end{figure}
However, many emerging applications challenge the necessity of full message
reconstruction. In large-scale distributed systems, the primary objective is
often not to recover detailed data, but to determine whether a certain message,
event, or condition is relevant. This shift is particularly pronounced in
scenarios involving massive numbers of devices, such as Internet-of-Things (IoT)
networks, monitoring and control systems, and distributed sensing architectures.
In such settings, classical communication approaches face fundamental
scalability limitations, as signaling overhead, addressing complexity, and
resource consumption grow with the number of devices.

\begin{figure}
    \centering
    \includegraphics[width=0.8\linewidth]{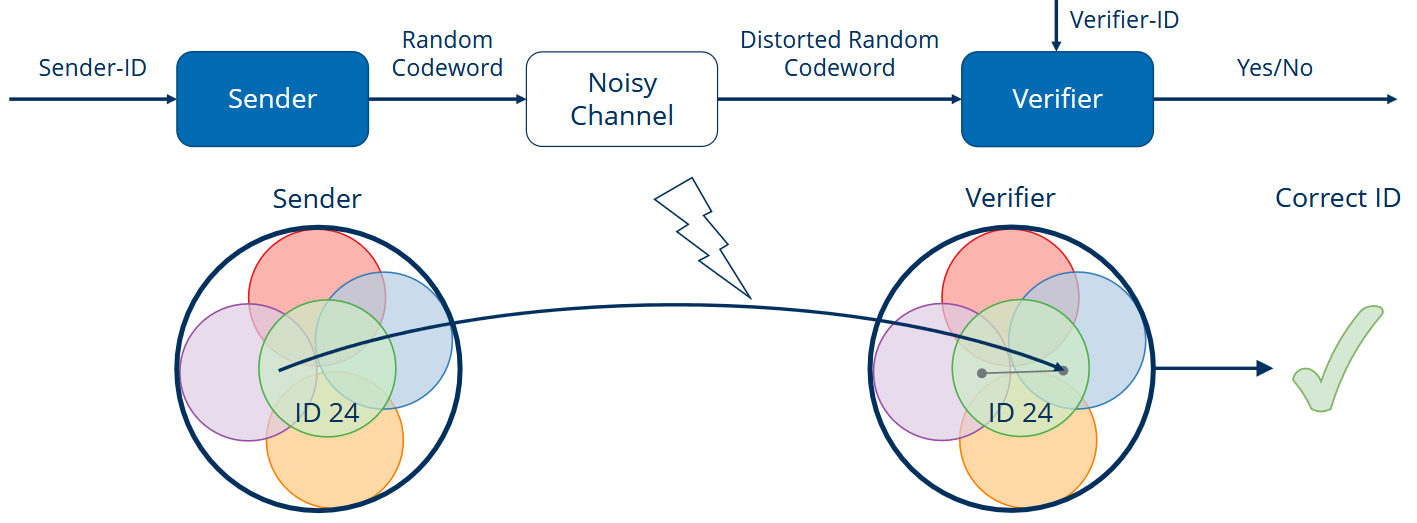}
    \caption{In the ID setting, each identity is associated with a set of possible codewords, from which the sender selects one at random for transmission. The corresponding acceptance regions at the verifier may overlap and are not required to be convex. Adapted from \cite{Schaefer2022PostShannonTutorial}}
    \label{figIDcodes}
\end{figure}

The ID scheme, introduced by Ahlswede and Dueck, provides an
alternative communication paradigm tailored to these requirements. Instead of
reconstructing the transmitted message, the receiver evaluates whether a given
message applies to it. A fundamental result of ID theory is the
double-exponential growth in the number of identifiable messages with respect
to the blocklength, i.e.,
\(
N \approx 2^{2^{nR}},
\)
which stands in sharp contrast to the exponential scaling in classical
transmission. Unlike transmission codes (see Fig.~\ref{TransmissionCodes}), ID codes introduced in \cite{ahlswede1989identification} allow decoding sets to overlap. This feature enables a double-exponential growth in the number of messages, as noted earlier, but also gives rise to another type of error, which will be defined subsequently (see Fig.~\ref{figIDcodes}).

Building on these insights, recent work has introduced the concept of
\emph{post-Shannon communication}, in which ID replaces data
transmission as the primary communication primitive. In such systems,
communication signals are not interpreted as data streams, but as queries or
triggers. Receivers locally decide whether they are addressed and only react
when relevant. This perspective leads naturally to broadcast-based architectures
that can support very large populations of devices with minimal signaling
overhead. Practical realizations of these ideas have already been proposed in
several system architectures and patents~\cite{BocheDeppeEP3937117B1,BaurBocheDeppeUS20210117126,BocheDeppeGeitzRosenbergerDisaster2024}.

From a system's perspective, message ID is particularly
attractive in regimes characterized by sparse relevance: although the overall
address space is extremely large, only a small subset of messages, devices, or
events is relevant at any given time. In such scenarios, replacing payload
transmission by relevance testing can significantly reduce communication
overhead and improve scalability. At the same time, ID does not aim
to replace classical transmission in general, but rather complements it by
providing an efficient mechanism for control, signaling, and decision-making.

This observation motivates a range of emerging application domains such as machine-to-machine and human-to-machine systems \cite{CompoundChannel}, digital watermarking \cite{AhlswedeWatermarking,MOULINwatermarking,SteinbergWatermarking}, Industry 4.0 \cite{industry4.0}, 6G communication systems \cite{6Gandtrustworthiness,6Gcomm}, as well as molecular communication systems \cite{macromolecular2,bienau2026molecular}. The ID scheme naturally aligns with monitoring and alarming systems, where the goal is
to detect whether an event has occurred. It also enables new architectures for
special-purpose data storage, where queries are answered in a binary manner
instead of reconstructing full records. Furthermore, ID can be
combined with sensing to directly map observed system states to distributed
decisions, and it is one of the fundamental problems in the field of semantic and goal-oriented communication, in
which the transmitted signal represents an alarm, control, or command signal rather than a
bit string. In addition, ID-based signaling can enhance control-plane
communication in mobile networks by enabling scalable broadcast coordination of
large device populations.

In this survey, we review the theoretical foundations of ID coding
and discuss its role as a key enabler of post-Shannon communication. We provide
a structured overview of emerging application domains, including monitoring
systems, special-purpose data storage, joint identification and sensing (JIDAS),
semantic communication, mobile network control, and networked consensus testing.
We also highlight recent
system implementations and patent-based architectures that demonstrate the
practical relevance of message ID~\cite{BocheDeppeEP3937117B1,BaurBocheDeppeUS20210117126,BocheDeppeGeitzRosenbergerDisaster2024}.

The goal of this survey is to bridge the gap between the information-theoretic
foundations of ID and its growing set of applications, and to
outline promising directions for future research. Beyond its
information-theoretic foundations, this survey is intended for
a broad audience, including both theorists and practitioners
from communication theory, wireless systems, distributed control,
sensing, embedded systems, and future mobile-network architectures.
The objective is not only to summarize the theoretical development
of identification coding, but also to provide intuition for how
identification-based principles can influence practical
communication-system design. In particular, the survey aims to
highlight the role of identification as a key building block for
post-Shannon, task-oriented, semantic, and large-scale distributed
communication systems.

{\bf Notation} Calligraphic letters $\setx, \sety,\setz, \ldots$ are used for finite or infinite sets. Lowercase letters $x,y,z,\ldots$ stand for constants and values of random variables. Uppercase letters $X,Y,Z,\ldots$ stand for random variables. $H(\cdot)$, $I(\cdot ;\cdot)$ are the entropy and mutual information, respectively. $H_2$ denotes the binary entropy. The Kullback-Leibler divergence between $P_X$ and $P_Y$ is denoted by $D\!\big(P_X\,\big\|\,P_Y\big)\;$. The total variational distance between $P_X$ and $P_Y$ is denoted by $d(P_X,P_Y)$. We use the notation $X\to Y\to Z$ to indicate a Markov chain.
all logarithms and information quantities are taken to the base $2$; the set of probability distributions on the set $\mathcal{A}$ is denoted by $\mathcal{P}(\mathcal{A})$; $|\boldsymbol{a}|$ denotes the $L_1$ norm of a vector $\boldsymbol{a}$; 

\begin{table}[h]
\caption{List of abbreviations}
\centering
\begin{tabular}{p{3cm} p{10cm}}
\toprule
\textbf{Abbreviation} &  \textbf{Definition} \\
\midrule
AWGN  & Additive white Gaussian noise \\
DMC   & Discrete memoryless channel \\
BC    & Broadcast channel \\
ID    & Identification \\
i.i.d. & Independent and identically distributed \\
JIDAS & Joint identification and sensing \\
MAC   & Multiple access channel \\
MIMO  & Multiple-input multiple-output \\
w.r.t. & With respect to \\
\bottomrule
\end{tabular}
\end{table}

\section{Message identification}\label{sec:ident}
In this section, we review the concept of ID via channels and its fundamental differences from classical transmission. The goal is to introduce the key definitions and results underlying the ID paradigm, while maintaining a clear system-oriented interpretation. To set the stage, we revisit the key definitions and auxiliary results for ID over discrete memoryless channels (DMC)s.

\subsection{Fundamentals of identification}
\label{subsec:funda}
The concept of ID via was introduced by Ahlswede and Dueck \cite{ahlswede1989identification} and represents a fundamental departure from Shannon’s classical communication paradigm \cite{shannon1948mathematical}. The ID scheme is conceptually different from the classical Shannon transmission. 

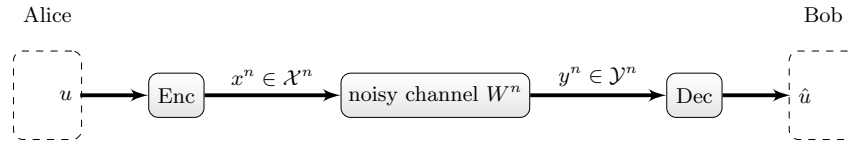
\begin{figure}[!hbt]
\centering
\tikzstyle{block} = [draw, fill=none, rectangle, rounded corners,
minimum height=4em, minimum width=1
cm]
\tikzstyle{block1} = [draw, fill=none, rectangle, rounded corners,
minimum height=4em, minimum width=4
cm]
\tikzstyle{farbverlauf} = [ top color=white, bottom color=white!80!gray] 
\tikzstyle{blocked} = [draw, rectangle, rounded corners,
minimum height=2em, minimum width=2.5em,farbverlauf]
\tikzstyle{input} = [coordinate]
\usetikzlibrary{arrows}
\scalebox{.9}{\begin{tikzpicture}[auto, node distance=2cm,>=latex']
\node[](m){$u$};
   \node[blocked, right=1cm of m] (encoder) {Enc};
    \node[blocked,right=2cm of encoder] (channel) {noisy channel $W^n$ };
    \node[blocked, right=2cm of channel](decoder) {Dec};
    \node[ right=1cm of decoder] (mhat){$\hat{u}$};
    \node[block,dashed, left=3.8cm of channel] (alice) {};
    \node[block, dashed, right=3.8cm of channel] (bob) {};
    \node[above=.3cm of alice](a){Alice};
    \node[above=.3cm of bob](b) {Bob};
\draw[->,ultra thick] (encoder) -- node[above]{$x^n \in \setx^n$} (channel);
\draw[->,ultra thick] (channel) -- node[above]{$y^n \in \sety^n$} (decoder);
\draw[->,ultra thick] (m) --(encoder);
\draw[->,ultra thick] (decoder) -- (mhat);
\end{tikzpicture}}
\caption{Shannon transmission. The encoder transmits a message $u \in \mathcal{M}$ over a noisy channel. The decoder outputs $\hat{u}$, representing the reconstruction of the message $u$ from the channel output.}
\label{fig:Tra_model}
\end{figure}

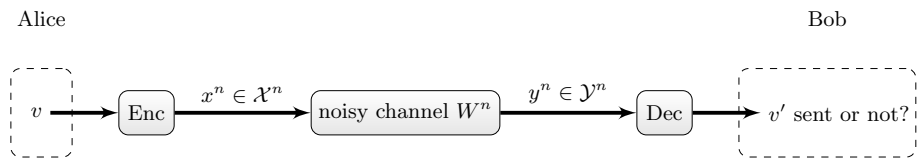
\begin{figure}[!hbt]
\centering
\tikzstyle{farbverlauf} = [ top color=white, bottom color=white!80!gray]
\tikzstyle{block} = [draw, fill=none, rectangle, rounded corners,
minimum height=4em, minimum width=0.9
cm]
\tikzstyle{block1} = [draw, fill=none, rectangle, rounded corners,
minimum height=4em, minimum width=2.6
cm]
\tikzstyle{blocked} = [draw, rectangle, rounded corners,
minimum height=2em, minimum width=2.5em,farbverlauf]
\tikzstyle{input} = [coordinate]
\usetikzlibrary{arrows}
\scalebox{.9}{\begin{tikzpicture}[auto, node distance=2cm,>=latex']
\node[](m){$v$};
   \node[blocked, right=1cm of m] (encoder) {Enc};
    \node[blocked,right=2cm of encoder] (channel) {noisy channel $W^n$ };
    \node[blocked, right=2cm of channel](decoder) {Dec};
    \node[ right=1cm of decoder] (mhat){$v^\prime$ sent or not?};
    \node[block,dashed, left=3.5cm of channel] (alice) {};
    \node[block1, dashed, right=3.5cm of channel] (bob) {};
    \node[above=.5cm of alice](a){Alice};
    \node[above=.5cm of bob](b) {Bob};
\draw[->,ultra thick] (encoder) -- node[above]{$x^n \in \setx^n$} (channel);
\draw[->,ultra thick] (channel) -- node[above]{$y^n \in \sety^n$} (decoder);
\draw[->,ultra thick] (m) --(encoder);
\draw[->,ultra thick] (decoder) -- (mhat);
\end{tikzpicture}}
\caption{Post-Shannon ID. The encoder transmits an identity $v \in \mathcal{N}$ over a noisy channel, while the decoder evaluates a query $v'$ and decides whether $v = v'$. In contrast to classical transmission, the goal is not message reconstruction but a binary decision.}
\label{fig:ID_model}
\end{figure}

In the classical transmission setting, a sender encodes a message $u \in \mathcal{M}$ into a channel input sequence, and the receiver aims to reliably reconstruct $u$ from the channel output. The objective is therefore to accurately reconstruct the transmitted message, denoted by $\hat{u}$ (see Fig.~\ref{fig:Tra_model}).
In contrast, ID considers a different task. The encoder selects an identity $v \in \mathcal{N}$, while the decoder is not interested in reconstructing $v$ itself. Instead, for a given query $v' \in \mathcal{N}$, the decoder only decides whether $v = v'$ or not (see Fig.~\ref{fig:ID_model}). Thus, ID can be interpreted as a family of binary hypothesis tests, one for each possible identity. A crucial aspect is that the encoder does not know the query $v'$ at the receiver. Therefore, the encoding must enable reliable decisions for all possible queries based on a single channel output.

Table~\ref{table:Comparison} provides a summary of the key differences between classical Shannon transmission and post-Shannon identification coding, emphasizing contrasts in communication goals, scaling behavior, error types, and the roles of feedback and common randomness. Further details will be discussed in the following sections.

%

\begin{table}[h]
\caption{%
  \label{table:Comparison}%
  Summary of key differences between Shannon transmission and post-Shannon identification.}
  \vskip\belowcaptionskip
  \centering
  \begin{tblr}{
    colspec = {X X},
    hline{1,Z} = {1pt},
    hlines,
    cell{1}{-} = {font=\bfseries},
    column{1} = {leftsep  = 0pt},
    column{Z} = {rightsep = 0pt},
  }
  Shannon Transmission & Post-Shannon / Identification \\
  Receiver's goal: Which message $u$ was sent?
    & Receiver's goal: Was the message $v'$ sent? \\
  Exponential rate: $2^{nR}$
    & Various scalings: $2^{2^{nR}},\, 2^{(n \log n)R}$,\ $2^{2^{nR}}$ \\ 
  The decoding regions of a transmission code must be disjoint. & The decoding regions of an ID code may overlap. \\
  Error only due to channel noise.
    & Type I error (channel noise) and type II error (ID code construction). \\
  Feedback generally does not increase the transmission capacity (e.g., DMC).
    & Feedback can increase the ID capacity (e.g., DMC).\\
  Common randomness generally does not increase the transmission capacity (see correlation-assisted capacity of a DMC).
    & Common randomness can increase the ID capacity (see correlation assisted ID capacity of a DMC).
  \end{tblr}
\end{table}



To better highlight the key differences between classical message transmission and message ID, we begin by reviewing the definition of transmission codes. Let $ W=\{W^n=P_{Y^n|X^n} \colon \setx^n \rightarrow \mathcal{P}(Y^n)\}_{n=1}^\infty$ be an arbitrary channel.
\begin{definition}
An $(n, M, \lambda)$ \emph{deterministic} transmission code for a channel $W$ consists of codewords $u_i \in \mathcal{X}^n$, decoding sets $D_i \subset \mathcal{Y}^n$, and a decoding error that satisfy
\begin{align}
W^n(D_i^c | u_i) &\leq \lambda, \\
D_i \cap D_j &= \emptyset, 
\end{align}
for all $i,j=1,\ldots,M$ with $i\neq j$ and some $\lambda \in (0,1)$. 
\end{definition} 
\emph{Randomized} transmission codes (Shannon) replace each codeword by a distribution over $\mathcal{X}^n$. However, randomization does not increase the achievable transmission rate.\\
The classical channel coding theorem \cite{shannon1948mathematical} states that
\begin{equation}
C(W) = \lim_{n \to \infty} \frac{1}{n} \log M(n,\delta) = \sup_{P_X} I(X;Y), \quad \text{ for }\delta \in (0,1).
\end{equation}

Ahlswede and Dueck introduced the \emph{randomized} ID problem motivated by JaJa’s earlier work on \emph{deterministic} ID \cite{JaJa85}, which considers codewords as \emph{deterministic} functions of messages in a communication complexity setting.
Unlike transmission codes, ID codes introduced in \cite{ahlswede1989identification} permit overlapping decoding sets. This overlap results in a double exponential growth in the number of messages, as will be shortly outlined in Theorem \ref{theorem:IDCoding}, and introduces a different type of error, which we will define subsequently. 

\begin{figure}[!hbt]
    \centering
        \includegraphics[width=0.8\textwidth]{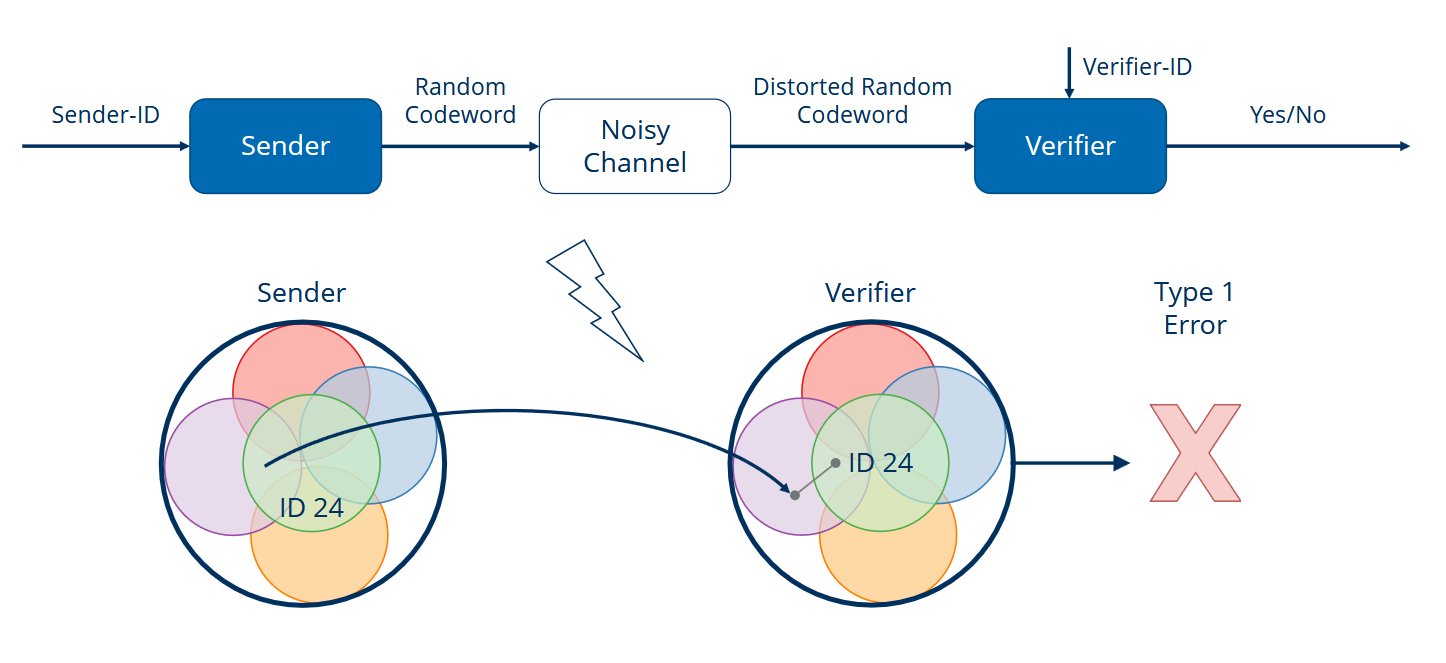}
        \caption{Illustration of the error of the first kind induced by channel noise; the probability that the decoder fails to detect the identity “ID 24” given that “ID 24” was sent. Adapted from \cite{Schaefer2022PostShannonTutorial}}
        \label{fig:error1}
        \end{figure}

\begin{figure}[!hbt]
    \centering
        \includegraphics[width=0.8\textwidth]{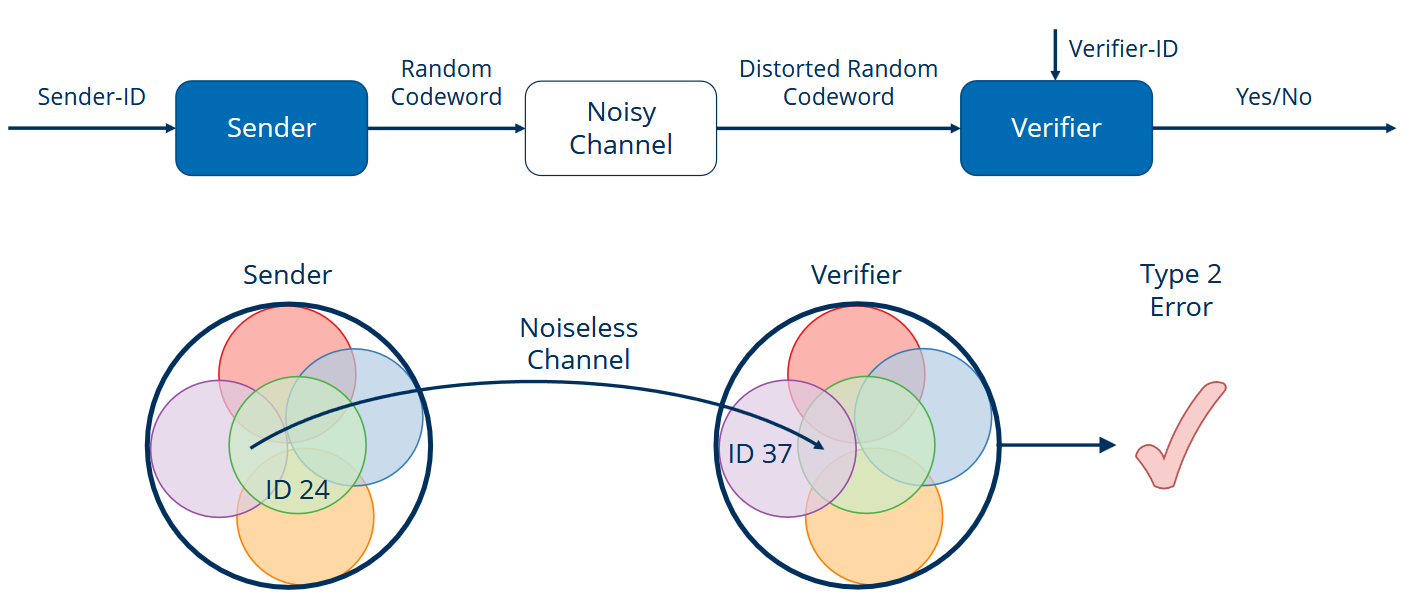}
        \caption{Illustration of the error of the second kind. Assuming perfect error correction, the only remaining error is due to the ID code construction; the probability that the decoder selects the identity “ID 37” when a different identity, “ID 24,” was actually sent. Adapted from \cite{Schaefer2022PostShannonTutorial}}
        \label{fig:error2}
        \end{figure}

In the following, we present a definition of a (\emph{randomized}) ID code. Let $ W=\{W^n=P_{Y^n|X} \colon \setx^n \rightarrow \mathcal{P}(Y^n)\}_{n=1}^\infty$ be an arbitrary channel.
\begin{definition}
  An $(n, N, \lambda_1, \lambda_2)$ ID code for an arbitrary channel $W$ is a family of pairs \\$\{(Q_i,\setd_i),\quad i=1,\ldots,N\}$ with probability distributions $Q_i$ over $\mathcal{X}^n$ and decoding sets $D_i \subset \mathcal{Y}^n$, and with errors of the first and second kinds that satisfy 
 \begin{align}
\mathbb{E}_{Q_i}\left[ W^n(\mathcal D_i^c|X^n) \right] &\leq \lambda_1,  \label{eq:error1}\\
\mathbb{E}_{Q_i}\left[W^n(\mathcal D_j|X^n) \right] &\leq \lambda_2, \label{eq:error2}
\end{align}
for all $i,j=1,\ldots, N$ with $i \neq j$ and some $\lambda_1+\lambda_2 < 1$. The expectation $\mathbb{E}_{Q_i}$ is with respect to (w.r.t.) the distribution $Q_i$.
\end{definition}
The error described in \eqref{eq:error1} is known as \emph{the error of the first kind}. It is produced by channel noise and matches the same error definition as that used for a transmission code. The error described in \eqref{eq:error2} is known as \emph{the error of the second kind}, resulting from the construction of ID codes, i.e., the overlap between the decoding sets. The errors of the first and second kinds are depicted in Fig.~\ref{fig:error1} and Fig.~\ref{fig:error2}, respectively.

\begin{remark}
A \emph{deterministic} ID code is a special case of a \emph{randomized} ID code in which the stochastic encoding distribution $Q_i$ is \emph{deterministic}, i.e., it assigns probability one to a single codeword. \end{remark}
\begin{remark}
In contrast to transmission codes, \emph{local randomness} at the encoder is essential to achieve the double exponential growth of ID codes over a DMC when considering maximal error definition. By \emph{local randomness}, we refer to randomness that is exclusive to the encoder. The decoder does not need any knowledge of this randomness. 
\end{remark}
As already highlighted in Table~\ref{table:Comparison}, the ID capacity exhibits fundamentally different behavior compared to the classical Shannon transmission. In contrast to the classical transmission, where the codebook size scales exponentially with the blocklength, ID schemes can exhibit a much wider range of growth regimes, including exponential, super-exponential, and double-exponential scaling, depending on the underlying system model, the encoding strategy, etc. This diversity of scaling behaviors motivates the introduction of a rate-function framework, which will be used in the sequel to characterize the growth of rates in the context of the ID capacity. Let $\{\zeta_i\}_{i=1}^3$ be a family of functions
$$\zeta_i : \mathbb{N} \times \mathbb{R} \to \mathbb{R}, \qquad i \in \{1,2,3\},$$

defined by
\begin{equation}
\begin{aligned}
\zeta_1(n,R) &:= 2^{nR}, \\
\zeta_2(n,R) &:= 2^{ (n\log n) R}, \\
\zeta_3(n,R) &:= 2^{2^{nR}}.
\end{aligned} \label{ratefct}
\end{equation}
 We refer to such functions as \emph{rate functions}, where $n$ and $R$ denote the blocklength and the rate, respectively.

The \emph{rate function} used in the traditional definition of message transmission capacity is $\zeta_1(n,R)=2^{nR}$. The double-exponential increase in blocklength of the number of identities achievable in ID for DMCs with \emph{randomized} encoding is reflected in the use of the \emph{rate function} $\zeta_3(n,R)=2^{2^{nR}}$ in the corresponding capacity definition. More details will be discussed below (see Table~\ref{summary1} and Table~\ref{summary2}).

In the following, we revisit the definition of an achievable ID rate and the ID capacity of an arbitrary channel $W$. 
\begin{definition} \label{def:AchievableRatedeterministic}
Let $W=\{W^n=P_{Y^n|X^n} \colon \setx^n \rightarrow \mathcal{P}(\mathcal Y^n)\}_{n=1}^\infty$ be an arbitrary channel. Let $\zeta_i$ be a \emph{rate function} as defined in \eqref{ratefct}.
	\begin{enumerate}
		\item The ID rate $R$ for $W$ is said to be achievable w.r.t. the \emph{rate function} $\zeta_i$ if for $\lambda \in (0,\frac{1}{2})$ there exists an $n_0(\lambda)$, such that for all $n\geq n_0(\lambda)$ there exists an $(n,\zeta_i(n,R),\lambda,\lambda)$ ID code for $W$.
		\item The ID capacity $C_{ID}\left(W,\zeta_i \right)$ of the channel $W$ w.r.t. the \emph{rate function} $\zeta_i$ is the supremum of all achievable rates.
	\end{enumerate}
\end{definition}

The \emph{deterministic} ID capacity of a DMC under power constraints is established in \cite{deterministicDMC}.   
Let $W_{\text{D}}$ be a DMC. Let $\phi(x)$ be a given non-negative bounded cost function, and let $\phi^n(x^n)=\sum_{t=1}^n \phi(x_t)$. Given an input constraint $A>0$, the channel input $x^n$ must satisfy $\phi^n(x^n)\leq A$. The following theorem characterizes the \emph{deterministic} ID capacity of the DMC $W_{\text{D}}$ under input constraint  $\phi^n(x^n)\leq A$. Let ${C}^{(d)}_{ID}(W_{\text{D}},\zeta_1) $ denote the \emph{deterministic} ID capacity of the DMC $W_{\text{D}}$ and w.r.t. the \emph{rate function} $\zeta_1$. 
\begin{theorem}[\cite{deterministicDMC}]
     The \emph{deterministic} ID capacity of a DMC $W_{\text{D}}$ under the input constraint described above is given by 
     \begin{equation}{C}^{(d)}_{ID}(W_{\text{D}},\zeta_1) = \mathop {\max }\limits_{{p_X}:\mathbb{E}\{ \phi (X)\} \leq A} H(X).\end{equation}
     \label{theorem:deterministic}
\end{theorem}
In the \emph{deterministic} setup, the number of messages that we can identify over a DMC scales only exponentially with the blocklength ($\zeta_1(n,R)=2^{nR}$). However, the rate is still larger than the transmission rate in the exponential scale. Ahlswede's and Dueck's goal was to show that incorporating randomness, similar to ideas of Yao \cite{yao1979some} in communication complexity, can substantially outperform \emph{deterministic} schemes, leading to an exponential increase in the achievable codebook size. 
The following ID coding theorem shows that the number of ID messages over a DMC grows double-exponentially with the blocklength.
\begin{theorem}[\cite{ahlswede1989identification,han2003information}]
The \emph{randomized} ID capacity of the DMC $W_{\text{D}}$ is given by
	\begin{equation*}
	C_{ID}(W_{\text{D}},\zeta_3)=\sup_{P_X} I(X;Y).
	\end{equation*} 
    \label{theorem:IDCoding}
\end{theorem}

\begingroup
  \def\MessageSet{\mathcal M}
  \def\SharedSeedSet{\mathcal S}
  \def\SharedSeedRV{S}
  \def\HashSet{\mathcal H}
\paragraph{Sketch of the direct proof in \cite{Feedbackidentification,dissertationWafa}}
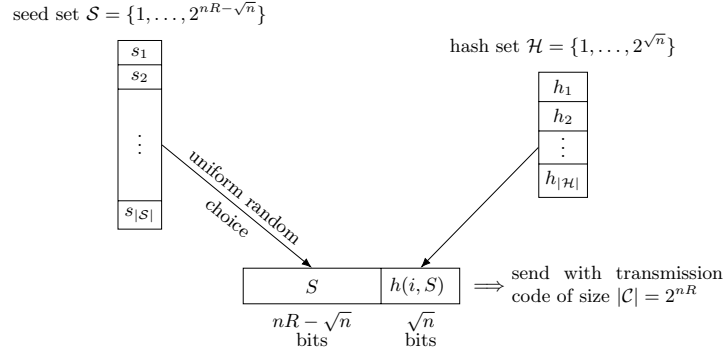
\begin{figure}[hbt!]
    \centering
    \makeatletter
\DeclareRobustCommand\vdotsNoKern{%
  \vbox{\baselineskip4\p@ \lineskiplimit\z@%
    \hbox{.}\hbox{.}\hbox{.}}
}
\makeatother
\scalebox{.8}{
\def\upp{h}
\def\Mpp{|\mathcal H|}
\begin{tikzpicture}[
  font=\footnotesize,
  pblock/.style = {
    draw,
    inner sep=1ex,
    rectangle split, rectangle split horizontal,
    rectangle split parts=2, align=center,
    text centered,
  },
  ->/.style = {-Latex},
]
 
\node[pblock] (C) at (2.5,-4) {
  \nodepart[text width=2cm]{one}$S$
  \nodepart{two}$h(i,S)$
};
\node[below] at (C.one south) {$\substack{\textstyle nR -\sqrt n \\ \textstyle \text{bits}}$};
\node[below] at (C.two south) {$\substack{\textstyle \sqrt n \\ \textstyle \text{bits}}$};

\

\node[rectangle split, rectangle split parts=4,draw] (A) at (-1,-1.5)
{$s_1$\nodepart{two}$s_2$\nodepart{three}$\vbox{\kern2em\vdotsNoKern\kern2em}$\nodepart{four}$s_{|\mathcal S|}$ }
;
\node[above=1ex of A] (A1) {seed set $\mathcal S = \{1,\dots,2^{nR - \sqrt{n}}\}$};

\node[rectangle split, rectangle split parts=4,draw] (B) at (6,-1.5)
{$\upp_1$\nodepart{two}$\upp_2$ \nodepart{three}$\small \vdotsNoKern$\nodepart{four}$\upp_{\Mpp}$} ;
\node[above=1ex of B] (B1) {hash set $\mathcal H = \{1,\dots,2^{\sqrt n}\}$};
\draw[->] (A.three east)
  -- node[above, sloped] {uniform random} 
     node[below, sloped] {choice}
  (C.one north)
  ;
  
\draw[->] (B.three west) -- (C.two north);

\draw
(C.two east) +(3pt,0)
node[anchor=west] {$\Longrightarrow$~\parbox{3.5cm}{send with transmission code of size $|\mathcal C| = 2^{nR}$}};

\end{tikzpicture}}
    \caption{\label{fig:CodeConcatenation}%
    Construction of an ID code via $\epsilon$-almost universal hash functions
    which hash the message and a uniform random number called a seed.
    Seed and the hash value are then transmitted via a transmission code.}
\end{figure}
In \cite{Feedbackidentification}, Ahlswede and Dueck constructed an ID code with
rate $R \geq C$, based on hash functions%
\footnote{These have also been called coloring functions \cite{CompoundChannel},
but we use name “hash functions” which is more wide-spread across multiple fields.},
with error probability bounded by $\epsilon = 2^{-\sqrt{n}}$.
Their code was created at random, but explicit constructions have
been discovered subsequently (see Section~\ref{sec:realizationRanIDCodes}).
Formally, the hash functions are defined as follows:
\begin{definition}
  \label{def:auhf}
  For finite sets $\MessageSet, \SharedSeedSet$ and $\HashSet$,
  a function $h : \MessageSet \times \SharedSeedSet \to \HashSet$ is called
  an \emph{$\epsilon$-almost universal hash function} if
  for all $m, m' \in \MessageSet$, $m \ne m'$ and
  a $S$ that is uniformly distributed on $\SharedSeedSet$,
  \begin{equation*}
    P_S \mleft[ h(m, S) = h(m', S) \mright] \le \epsilon .
  \end{equation*}
  This probability is called the \emph{hash collision probability}.
\end{definition}
This means that for any two messages, the probability that they have
the same hash is bounded by $\epsilon$.

The construction of an ID code using hash functions is described in the following and depicted in Fig.~\ref{fig:CodeConcatenation}.
\begin{enumerate}
  \item When the transmission capacity $C > 0$, then
    for any rate $R < C$
    and any error probability $\lambda > 0$,
    there exists a reliable transmission code
    of size $2^{nR}$.
  \item 
    To send an identity $i \in \MessageSet = \{1,\ldots,N\}$,
    we can prepare \cite{rosenbergerBoche2026kid_itw}
    an $\epsilon$-almost universal hash function as defined above,
    where $|\HashSet| = 6\epsilon^{-1} = 6 \cdot 2^{\sqrt n}$,
    $|\SharedSeedSet| \cdot |\HashSet| = 2^{nR}$,
    and
    \begin{equation}
      \log N < \epsilon |\SharedSeedSet| = \epsilon^2 2^{nR} / 6
      = 2^{n R - 2 \sqrt n - \log 6}.
    \end{equation}
  \item The sender, Alice, chooses a random number $S$ (“seed”)
    uniformly at random and computes the hash value $h(i,S)$.
  \item Alice transmits the pair $(S, h(i,S))$ over the
    DMC $W$ with the aforementioned transmission code.
    This is possible since 
    $|\SharedSeedSet| \cdot |\HashSet| = 2^{nR}$.
  \item The receiver, Bob, decodes the received signal. His decoder output
    is an estimate $(\hat S, \hat H)$.
  \item Suppose Bob is interested in some ID message $i' \in \MessageSet$.
    He computes the hash $h(i', \hat S)$ under the decoded seed $\hat S$.
  \item Then, Bob compares the received and the computed hash.
    He declares that $i = i'$ if $\hat H = h(i', \hat S)$.
    Otherwise he declares that $i \ne i'$.
\end{enumerate}
This scheme is reliable, since the transmission code has an arbitrarily
small error probability $\lambda > 0$ and the ID code does not add much error
to that: An error can occur when the transmission is erroneous,
i.e., if $(\hat S, \hat H) \ne (S, h(i, S))$.
The probability of this event is at most $\lambda$.
Otherwise, an error can occur if $i' \ne i$ and
$h(i', \hat S) = h(i', S) = \hat H = h(i, S)$.
By the $\epsilon$-almost universality, this has probability
$\le \epsilon = 2^{-\sqrt n}$.
Therefore, $\frac 1 n \log\log N \to R$, for $n \to \infty$,
and hence $R$ is achievable.
\endgroup

\begin{remark}
  A crucial observation from the code construction in the proof is that
  most channel resources are used only to establish common randomness,
  i.e., make a random number (the “seed”) known to both sender and receiver.
  The amount of common randomness that can be established
  determines the achievable ID rates (in the double-exponential scaling $\zeta_3$).
  When the seed is exchanged, the sender and receiver hash their respective
  messages, each. The sender transmits her hash value to the
  receiver, who compares the hashes.
  Therefore, when common randomness is established before the
  communication event, it can be used to enhance the rate or even
  replace completely the transmission of the seed. Then, almost no channel
  resources are used (only $\sqrt{n}$ logical symbols per $n$ physical channel
  symbols).
  In applications, this effect may be used, for example, to
  generate common randomness in times of low network usage. This
  can then enable highly efficient burst ID even in when network
  usage is high (see Section~\ref{sec:AlarmingSystems}).
  Further, when a feedback channel from the receiver to the sender is available,
  the noise of the feed forward channel can be used as a source of randomness,
  which is fed back to the sender via the feedback channel \cite{Feedbackidentification}.
  This feedback also occurs in the sensing problem in Section~\ref{sec:JIDAS}.
  Whereas more channel noise decreases the transmission capacity,
  even when an ideal feedback link is available,
  more noise, "viewed as a source of randomness", in the feed forward channel increases the ID capacity in this case.
\end{remark}

\begin{remark}
    The \emph{deterministic} ID capacity of the additive white noise Gaussian (AWGN) channel is infinite w.r.t. the exponential scaling $\zeta_1(n,R)=2^{nR}$ and zero w.r.t. the double-exponential scaling $\zeta_3(n,R)=2^{2^{nR}}$. It was shown in \cite{SPBD21f} that the ID capacity w.r.t.\ the rather unusual intermediate rate scaling $\zeta_2(n,R)=2^{(n \log n) R}$ is positive and finite. Hence in this problem setup, the maximal number of identities grows super-exponentially in the blocklength, but slower than doubly-exponentially. Generally, one can observe that ID is more sensitive w.r.t. different problem setups than message transmission (see Table~\ref{summary2}). 
\end{remark}
\subsection{Capacity results}
\label{subsec:capa}
As discussed in the aforementioned section, in the classical Shannon transmission, the goal is to reliably convey one message out of an exponentially growing set of size $2^{nR}$. This exponential scaling is the natural benchmark because it yields a sharp threshold between achievable and non-achievable rates.
However, the ID capacity has a completely different behavior. ID problems exhibit a variety of distinct scaling behaviors for the number of ID messages, depending on the channel model, coding constraints, etc. 
As highlighted in Theorem \ref{theorem:IDCoding}, for a DMC, the maximal number of ID messages grows doubly exponentially with the blocklength ($\zeta_3(n,R)=2^{2^{nR}}$) if \emph{randomized} encoding is used. If one restricts to \emph{deterministic} ID over a DMC (see Theorem \ref{theorem:deterministic}), the scaling collapses back to the classical exponential form $\zeta_1(n,R)=2^{nR}$. 
The situation becomes even more nuanced in infinite alphabet settings such as the AWGN channel. It was shown in \cite{SPBD21f} that the \emph{deterministic} ID capacity of the AWGN channel grows super-exponentially in the blocklength but slower than doubly-exponentially. This illustrates that the ID capacity is not tied to a single universal growth law, but instead depends delicately on both the channel model and the allowed coding strategies (e.g., \emph{deterministic} vs. \emph{randomized} encoders). 
 In the following, we summarize the main capacity results for \emph{deterministic} and \emph{randomized} ID over different channel models.

 The channel models listed in Table~\ref{summary1} and Table~\ref{summary2} represent several canonical communication scenarios encountered in information theory and wireless communications. 
 \begin{itemize}
     \item A DMC is a channel model with finite input and output alphabets, where the channel law satisfies memorylessness across successive uses. It serves as a fundamental abstraction of digital communication systems operating over noisy channels. 
\item The AWGN channel models point-to-point communication corrupted by additive thermal noise and serves as a fundamental baseline model in communication theory, especially for radio and wireless communication systems. 
\item The discrete-time memoryless Poisson channel is commonly used in optical and molecular communication systems, where the received signal is governed by counting statistics. 
\item The single-user multiple-input multiple-output (MIMO) channel models communication systems employing multiple transmit and receive antennas, as used in cellular networks, and is relevant in high-capacity wired, wireless, satellite, and deep-space communication links to enhance reliability and spectral efficiency.
\item The compound channel is a family of possible channel laws where the actual channel is unknown but remains fixed during transmission. It captures practical communication scenarios with channel uncertainty, where coding schemes are designed to guarantee reliable performance for all admissible channel conditions. 
\item The broadcast channel (BC) models downlink communication in which a single transmitter sends information simultaneously to multiple receivers. It captures practical multi-user scenarios such as cellular downlink transmission, where a base station communicates with several users over a shared wireless medium. 
\item The multiple access channel (MAC) models uplink communication scenarios in which multiple users simultaneously transmit information to a common receiver over a shared medium. They provide a fundamental abstraction for multi-user systems such as the uplink of cellular networks, where signals from different devices interfere at the base station and must be jointly decoded.
\end{itemize}
 Table~\ref{summary1} summarizes the \emph{randomized} ID capacity results for several communication channel models, including DMC, AWGN, discrete-time Poisson and compound channels, where the ID capacity expressions are characterized through mutual information optimization under channel constraints. It also presents \emph{randomized} ID capacity regions for MIMO, BC and MAC channels. 
Since the channels considered in the following are memoryless, we use the notation $W=P_{Y|X}$.
\begin{table*}[hbt!]
\caption{\label{summary1}%
  Capacity results for \emph{randomized} ID and various channel models. For all
these models, the scaling is $\zeta_3(n, R) = 2^{2^{nR}}$.}
  \vskip\belowcaptionskip
  \centering
\begin{tblr}{
  colspec = {X[2] X[7] l},
  hline{1,Z} = {1pt},
  hline{2},
  row{2-Y} = {rowsep+ = 0.2\baselineskip},
  row{Z} = {abovesep+ = 0.2\baselineskip},
  columns  = {colsep = 4pt},
  column{1} = {leftsep  = 0pt},
  column{Z} = {rightsep = 0pt},
  cell{1}{-} = {font=\bfseries},
  cell{2-Z}{2,3} = {mode=dmath},
}
Channel Model $W$ 
               & Capacity $C_{ID}(W, \zeta_3)$ & Ref.
\\
DMC
  & \max_{P_X} I(X;Y)
  & \cite{ahlswede1989identification,Feedbackidentification,HanVerdu}
\\
AWGN
  & \frac{1}{2} \log\mleft(1+ \frac{P}{\sigma^2}\mright),
    \,\quad \mathbb{E}[X^2]\leq  P,
    \, Z \sim  \mathcal{N}(0,\sigma^2)
  & \cite{han2003information,Burnashev}
\\
discrete-time memoryless Poisson
  & \max_{\substack{P_X  \\ X \leq P_{\text{max}} \\ \mathbb{E}[X] \leq P_{\text{avg}} }} I(X;Y)
 &\cite{EW2024}
\\
single-user MIMO
 & \max_{\mathbf{Q}:\ \substack{tr(\mathbf{Q})=P\\ Q \succeq 0}} \log \det
 \left[\mathbf{I}_{N_R}+\frac{1}{\sigma^2}\mathbf{H} \mathbf{Q} \mathbf{H}^\h \right]
 & \cite{labidi2021identification} \\
discrete memoryless compound
  & \max_{P} \min _{P_{Y|X} \in \mathcal W } I(X;Y), \mathcal{W}=\{W_t\}_{t \in \theta},\quad \text{$\theta$ finite}
  & \cite{han2003information,CompoundChannel}
\\
discrete-time BC $P_{Y_1,Y_2|X}$
  & \bigcup_{P_{X} } \left\{  \begin{aligned} (R_{1}, R_{2}): & R_{1} \leq I(X; Y_{1}), \\ & R_{2} \leq I(X; Y_{2}) \end{aligned} \right\}
  & \cite{ID_BC} \\
discrete memoryless MAC
$P_{Y|X_1,X_2}$
  & 
    \bigcup_{P_U P_{X_1|U} P_{X_2|U} P_{Y|X_1X_2}}
    \kern-3pt
     \left\{
             \begin{aligned}
     (R_1,R_2):
     \quad
     R_1 &\le I(X_1;Y|X_2,U),\\
     R_2 &\le I(X_2;Y|X_1,U),\\
     R_1+R_2 &\le I(X_1,X_2;Y|U)
     \end{aligned}
     \right\}
  & \cite{ahlswede2008general}
\\
\end{tblr}
\end{table*}

Table~\ref{summary2} summarizes the \emph{deterministic} ID capacity results for various channel models. For the DMC, the \emph{deterministic} ID capacity is achieved by maximizing the input entropy subject to a bounded cost constraint, w.r.t. the \emph{rate function} $\zeta_1(n,R)=2^{nR}$. For the discrete-time memoryless Poisson and binomial channels, the \emph{deterministic} ID capacity is bounded between $\frac{1}{4}$ and $\frac{2}{3}$, w.r.t. $\zeta_2(n,R)=2^{(n\log n)R}$. For the AWGN channel, under the same scaling $\zeta_2(n,R)$, the corresponding \emph{deterministic} ID capacity is equal to $\frac{1}{2}$. For the MAC, the \emph{deterministic} ID capacity region, w.r.t. the exponential scaling $\zeta_1(n,R)=2^{nR}$ is characterized by inner and outer bounds derived through optimization over entropy and mutual information terms. These results further emphasize that the ID capacity has a completely different behavior, compared to the classical Shannon transmission. 

\begin{table}[hbt!]
\caption{\label{summary2}%
  Capacity results for \emph{deterministic} ID and various channel models.}
  \vskip\belowcaptionskip
  \centering
  \scalebox{0.95}{
  \begin{tblr}{
    width   = 1.05\linewidth,
    colspec = {X[2] l X[9] l},
    hline{1,Z} = {1pt},
    hline{2},
    row{2-Y} = {rowsep+ = 0.5\baselineskip},
    row{Z} = {abovesep+ = 0.5\baselineskip},
    columns  = {colsep = 2pt},
    column{1} = {leftsep  = 0pt},
    column{Z} = {rightsep = 0pt},
    cell{1}{-} = {font=\bfseries},
    cell{2-Z}{2,3} = {mode=dmath},
  }
  Channel Model $W$ & Scaling & Main Result $C_{ID}^{(d)}(W, \zeta_i)$ & Ref.
  \\
  DMC
    & \zeta_1
    &\max_{P_X:\mathbb{E}\{ \phi (X)\} \leq A } H(X),\,\,   \substack{\phi(x) \text{ non-negative bounded} \\\text{ cost function}}
    & \cite{IDwithoutRandom,deterministicDMC}
  \\
  discrete-time memoryless Poisson
    & \zeta_2
    & \frac{1}{2}
    & \cite{NewResultsBernoulli}
  \\
  discrete-time binomial
    & \zeta_2
    & \frac{1}{2}
    & \cite{NewResultsBernoulli}
  \\
  AWGN
    &\zeta_2
    & \frac{1}{2}
    & \cite{determinsticGaussianPau}
  \\
  discrete-memoryless MAC $P_{Y|\boldsymbol{X}}$
    &\zeta_1
    &\bigcup\limits_{\substack{ p{\mathbf{X}} \in \mathcal{P}(\mathcal{X}): \\ \mathbb{E}[\phi ({\mathbf{X}})]\preceq \Phi }} {{\mathcal{R}^l}} ({\mathbf{X}},Y) \subseteq {\mathcal{C}_{ID}^{(d)}(W_{\text{MAC}},\zeta_3)} \subseteq \bigcup\limits_{\substack{ {p_{\mathbf{X}}} \in \mathcal{P}(\mathcal{X}): \\ \mathbb{E}[\phi ({\mathbf{X}})]\preceq \Phi }} {{\mathcal{R}^u}} ({\mathbf{X}}).
    \newline
    \begin{aligned}
            {\mathcal{R}^u}({\mathbf{X}}) &= \left\{ {{\mathbf{R}} \in {\mathbb{R}^K}:0 \leq {R_k} \leq H\left({{X_k}}\right),\forall k \in [K]} \right\}
    \\[1em]
            {\mathcal{R}^l}(X,Y) &= \left\{ \begin{array}{c} {\mathbf{R}} \in {\mathbb{R}^K}:\forall k \in [K], \\ 0 \leq {R_k} \leq \mathop {\min }\limits_{\mathop {{p_{{X_k}X_k^\prime {S_k}Y}}}\limits_{ \in \mathcal{Q}\left({{p_{{X_k}}},\left\{ {{p_{{S_k}}}} \right\},W}\right)} } I\left({X_k^\prime ;XY}\right) \end{array} \right\}
    \end{aligned}
   & \cite{deterministicMAC}
  \\
 \end{tblr}
  }
\end{table}
For a more comprehensive overview of the capacity results, we refer the reader to the corresponding references listed in Table~\ref{summary1} and Table~\ref{summary2}.

\subsection{K-Identification and opportunistic communication}
\label{subsec:kide}
ID is a very specific communication task – there
are not many other tasks which can be achieved with the help
of only one equality test. While applications of ID
exist, e.g., in certain authentication, alarming and integrity checking problems, they are somewhat scarce. On the other hand,
transmission is inefficient when the actual task is an ID task,
but also whenever
the receiver is only interested in a relatively small subset of all messages
which is unknown to the sender.
This can be useful, for example, in alarming systems
(see Section~\ref{sec:AlarmingSystems})
or some distributed storage
systems like Venti \cite{QuinlanDorward2002Venti}.
This setting has been modeled by $K$-identification (K-ID) $K$-decoding
\cite{ahlswede2008general,rosenbergerBoche2026kid_itw}. $K$-decoding
is also called opportunistic communication since the sender
sends opportunistically whenever they have a message,
oblivious of the receiver's interest, while the receiver
only wants to know understand the message if it is interesting to them.
$K$-identification is defined just like ID, but
the receiver tests for $K$ many messages at the same time.

\begin{definition}
An $(n, N, K, \lambda_1, \lambda_2)$ \emph{randomized} K-ID code is a pair
$(Q, D)$ where the function $Q$ assigns
to every message $i=1,\dots,N$ an encoding distribution $Q_i = Q(\cdot|i)$
over $\mathcal{X}^n$ and $D$ assigns a decoding set $D_{\mathcal T} \subseteq \mathcal{Y}^n$
to every subset $\mathcal T \subseteq \{1,\dots,N\}$
of cardinality $|\mathcal T| = K$.
We have again two error types and require that
\begin{align}
  \mathbb E_{Q_i} W^n(D_{\mathcal T}^c|X^n) &\le \lambda_1, \\
  \mathbb E_{Q_i} W^n(D_{\mathcal T'} |X^n) &\le \lambda_2,
\end{align}
for all $\mathcal T, \mathcal T' \subseteq \{1,\dots,N\}$, $|\mathcal T| = K$,
and $i \in \mathcal T \setminus \mathcal T'$.
\end{definition} 
The error types have the following interpretation, similar to 1-ID.
\begin{enumerate}
    \item While the message is interesting, i.e., $i \in \mathcal T$
    of interest, the receiver declares that it is boring.
    \item The message is boring, i.e., $i \notin \mathcal T$,
    but the receiver declares it is interesting.
\end{enumerate}

Opportunistic communication is defined similarly:
The encoder is defined as for $K$-ID, but the
decoder now outputs either that the sent message is not the test set,
or otherwise it decodes the message. The behavior of sender and receiver
can be considered opportunistic: The sender, Alice, sends
whenever she has a message, even if she does not know if it is interesting
to the receiver, Bob. Bob, on the other hand, has a set of interest of
size $K$, and he does not care about the message if it is not interesting.
However, if it is, he wants to know it exactly, taking every opportunity
to use interesting information, without knowing in advance that
Alice's message will be interesting to him.

\begin{definition}
An $(n, N, K, \lambda_1, \lambda_2, \lambda_3)$-opportunistic code is a pair
$(Q, D)$ where the function $Q$ assigns
to every message $i=1,\dots,N$ an encoding distribution $Q_i = Q(\cdot|i)$
over $\mathcal{X}^n$ $D$ assigns a decoding set $D_{\mathcal T, j} \subseteq \mathcal{Y}^n$
to every subset $\mathcal T \subseteq \{1,\dots,N\}$
of cardinality $|\mathcal T| = K$
and every interesting message $j \in \mathcal T$.
We have now three error types:
First, it is possible that $i \in \mathcal T$, but
the decoder declares that the message is not interesting,
i.e., the channel output $Y^n \notin \bigcup_{i \in \mathcal T} D_{\mathcal T, i}$.
We call this error number zero, to keep the other error labels
similar to ID.
Second, the message is boring ($i \notin \mathcal T$),
but the receiver finds it interesting, i.e.,
$Y^n \in \bigcup_{j \in \mathcal T} D_{\mathcal T, j}$.
The third error event has the sender sending a message
that is interesting to the receiver ($i \in \mathcal T$),
and the receiver finds it interesting
but decodes it wrongly, i.e.,
$Y^n \in \bigcup_{i' \ne i} D_{\mathcal T, i'}$.
The corresponding error bounds are
\begin{align}
  \mathbb E_{Q_i} W^n\mleft(\textstyle \bigcap_{j \in \mathcal T} D_{\mathcal T, j}^c|X^n\mright) &\le \lambda_1,
  \\
  \mathbb E_{Q_i} W^n\mleft(\textstyle \bigcap_{j \in \mathcal T'} D_{\mathcal T', j}|X^n\mright) &\le \lambda_2,
  \\
  \mathbb E_{Q_i} W^n\mleft(\textstyle \bigcap_{j \in \mathcal T \setminus \{i\}} D_{\mathcal T, j}|X^n\mright) &\le \lambda_3,
\end{align}
for all $\mathcal T, \mathcal T' \subseteq \{1,\dots,N\}$, $|\mathcal T| = K$,
and $i \in \mathcal T \setminus \mathcal T'$.
\end{definition}
For opportunistic communication, the first two error types correspond to
the respective $K$-ID errors.
The third type is the event that a message is interesting,
and the receiver correctly acknowledges this, but the decoder outputs
the wrong message $j \ne i.$

Opportunistic communication has transmission and 1-ID as special cases:
When we choose $K = N$, the set of interest is the whole
message set, and every message must be decoded. The first
two types of error are then trivial.
When $K=1$, the receiver only tests if the message matches one
message of his choice, and the third type of error becomes trivial.

In the information theoretic study of codes for $K$-ID and
opportunistic communication, were the total number of messages is not the only
quantity of interest. Rather, one studies the tradeoff between the number of messages
and the size of the test set of interest.
Hence, we define achievable rate pairs and capacity regions as follows.
\begin{definition}
\begin{enumerate}
  \item
    A rate pair $(R_N, R_K)$ is called achievable if,
    for all $0 < \lambda_1$, $\lambda_2$, $\lambda_3 > 0$
    there exists $n_0 > 0$ such that for every $n > n_0$,
    there exists an $(n, \zeta_3(n, R_N), \zeta_1(n, R_K), \lambda_1, \lambda_2)$-ID code
    resp. an \[(n, \zeta_3(n, R_N), \zeta_1(n, R_K), \lambda_1, \lambda_2, \lambda_3)\]-opportunistic code.
  \item The \emph{capacity region} for $K$-ID, $\mathcal C_{ID}(\zeta_3, \zeta_1, W)$,
    resp. opportunistic communication, $\mathcal C_{OC}(W, \zeta_3, \zeta_1)$, is the
    closure of the set of all achievable rate pairs.
\end{enumerate}
\end{definition}

In general, only upper and lower bounds are known for the capacities.
Only for the special case where $\lambda_1 = 0$ and the
channel is a noiseless identity channel, the capacity is known:
for any DMC $P_{Y|X}$ let
\begin{equation}
  \mathcal R_a(P_{Y|X}) = \big\{
    (R_N, R_K) :
    0 \le R_N + a \cdot R_K \le \max_{P_X \in \mathcal P(\mathcal X)} I(X; Y)
  \big\}
  \,.
\end{equation}
\begin{theorem}[{\cite{rosenbergerBoche2026kid_itw}}]
  Consider rate functions $\zeta_3(n, R) = 2^{2^{nR}}$
  and $\zeta_1(n, R) = 2^{nR}$.
  Under the additional constraint that $\lambda_1 = 0$,
  and the channel $P_{Y|X}$ is noiseless, i.e., $Y = X$,
  the capacities $\mathcal C_{ID}^{\lambda_1 = 0}, \mathcal C_{OC}^{\lambda_1 = 0}$
  for $K$-ID and opportunistic communication satisfy
  \begin{equation}
  C_{ID}^{\lambda_1 = 0}(\zeta_3, \zeta_1, P_{Y|X}) = 
  C_{OC}^{\lambda_1 = 0}(\zeta_3, \zeta_1, P_{Y|X}) = \mathcal R_2(P_{Y|X})
  \,,
  \end{equation}
  i.e., $R_N + 2 R_K \le \log |\mathcal X|$.
\end{theorem}
In general, we have the following.
\begin{theorem}[{\cite{ahlswede2008general}}]
  For the rate functions $\zeta_3(n, R) = 2^{2^{nR}}$,
  $\zeta_1(n, R) = 2^{nR}$ and
  any DMC $P_{Y|X}$, the capacities
  for $K$-ID and opportunistic communication satisfy
  \begin{equation}
  \mathcal R_2(P_{Y|X}) \subseteq
  C_{OC}(P_{Y|X}, \zeta_3, \zeta_1)
  \subseteq
  C_{ID}(P_{Y|X}, \zeta_3, \zeta_1)
  \subseteq \mathcal R_1(P_{Y|X})
  \,.
  \end{equation}
\end{theorem}

\subsection{Common randomness generation and message identification}
\label{sec:CR_and_ID}

There is a strong connection between the problem of common randomness generation and the ID problem.
It has been shown in \cite{ahlswede2008general} that the value of the ID capacity in the double exponential scaling of the simplest channels, namely DMCs, coincides with the common randomness capacity.
For instance, in the case of ID with a noiseless feedback link over an AWGN channel, incorporating feedback enables the establishment of a correlated random experiment between the sender and the receiver. The size of this random experiment can be used to determine the growth of the ID rate. The problem of ID with noiseless feedback over AWGN channels can, to some extent, be reinterpreted as the problem of generating common randomness from Gaussian sources \cite{dissertationWafa}.

\begin{figure}[!hbt]
    \centering
   \tikzstyle{block} = [draw, top color=white, bottom color=white!80!gray, rectangle, rounded corners,
minimum height=2em, minimum width=3cm]
\tikzstyle{blockchannel} = [draw, top color=white, bottom color=white!80!gray, rectangle, rounded corners,
minimum height=2cm, minimum width=.1cm]
\tikzstyle{input} = [coordinate]
\usetikzlibrary{arrows}
\scalebox{.8}{
\begin{tikzpicture}[auto, node distance=2cm,>=latex']
\node[] (m) {\small $M$};
\node[block,right=1cm of m] (enc) {\small Encoder};
\node[blockchannel, right=1.5cm of enc](channel) {\small Channel
$W^n$};
\node[block, above=.7cm of channel] (source) {$P_{XY}$};
\node[block,right= 1.5 cm of channel.360] (bob) {\small Decoder};
\node[right=1cm of bob, align=center] (what) {\small Is $\hat{M}$ sent? \\ Yes,No?};

\draw[->] (m) -- (enc);
\draw[->] (enc) -- node[above]{$T^{n}$}  (channel);
\draw[->] (channel.360) -- node[above]{$Z^{n}$} (bob);
\draw[->] (bob) -- (what);
\draw[->] (source) -- node[above]{$X^n$} (enc);
\draw[->] (source) -- node[above]{$Y^n$} (bob);
\end{tikzpicture}}
   \caption{Correlation-assisted ID scheme.}
    \label{fig:IdwithCR}
\end{figure}
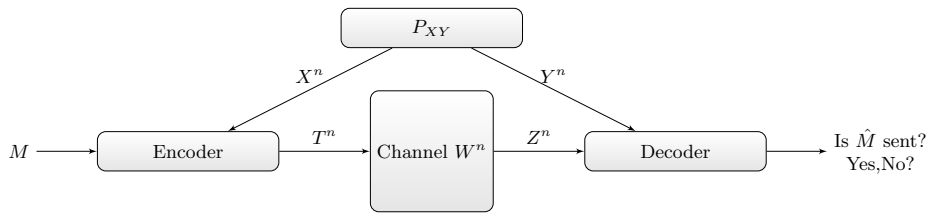

Since the common randomness capacity is typically easier to compute, it can offer useful insight into the ID capacity. For a long time, it was believed that the two capacities coincide, meaning that understanding common randomness generation would directly resolve the ID problem. However, this assumption has been disproven by counterexamples demonstrating that the ID capacity can be strictly larger than the common randomness capacity (see \cite{ahlswede2008general} and \cite{Kleinewächter2006}). Moreover, \cite{Kleinewächter2006} also provides an example of a channel for which the common randomness capacity exceeds the ID capacity.
It has been proved in \cite{correlationAssisted} that common randomness, as an additional resource, can enhance the ID capacity. In the correlation-assisted ID problem, the sender and the receiver have additionally access to a correlated source
$P_{XY}\in \mathcal{P} \left(X \times Y\right)$ as visualized in Fig.  \ref{fig:IdwithCR}.

A different approach to distribute randomness within a network is to generate it locally and distribute it during times of reduced network load, for example during night time. To do so, a distribution broker of randomness can be deployed centrally and network clients reading the published random numbers \cite{YinSquire}.

\subsection{Key properties and system implications}

The ID paradigm exhibits several distinctive properties particularly relevant for large-scale systems.

\textbf{Double-Exponential Scaling:}  
ID allows selecting among extremely large sets of possible actions or device subsets using very limited communication resources. This behavior extends beyond DMCs to Poisson, Gaussian and MIMO channels \cite{labidi2021identification,EW2024}.

\textbf{Role of Common Randomness:}  
Common randomness between sender and receiver increases the ID capacity additively \cite{ahlswede2008general}. Such randomness can be generated via feedback or sensing mechanisms \cite{ezzine2020common,CR_TSPwithRami}.

\textbf{Feedback and Interaction:}  
Feedback can serve not only to improve reliability but also as a resource for generating shared randomness, thereby enhancing ID performance \cite{wiese2022identification,isit21,labidi2025joint,globecom_with_yaning,itw_with_yaning}.

\textbf{Control-Oriented Interpretation:}  
In the JIDAS framework, the sender simultaneously transmits an identification (ID) message while sensing the channel state \cite{labidi2025joint} (see Section \ref{sec:JIDAS}), ID provides a mechanism for the selection of subsets. Each device performs a local binary decision (“Is this message intended for me?”), which directly maps to a control action. This interpretation is essential for scalable control under communication constraints \cite{DistributedPlatoon}.

\bigskip

\medskip

Overall, these properties indicate that message ID is particularly well suited for scenarios involving massive numbers of devices and stringent communication constraints, where classical transmission-based approaches become inefficient or infeasible.

\section{Use cases}\label{sec:use}

The following sections illustrate how identification (ID) coding can be integrated into practical communication and distributed-system architectures. While the underlying communication principle remains the same, namely relevance testing instead of full message reconstruction, the resulting system architectures differ significantly depending on the application domain and operational requirements. The considered use cases can be grouped according to the role ID plays within the overall system.
\begin{enumerate}
\item  In \textbf{monitoring and alarming} systems , ID enables scalable event detection and sparse alarm dissemination.
\item In \textbf{special-purpose data storage} ID transforms data retrieval into efficient membership testing.
\item \textbf{Joint identification and sensing} combines sensing, communication, and distributed decision making,
\item while \textbf{semantic communications} interprets ID as a relevance-oriented communication primitive.
\item In \textbf{control plane} ID acts as a scalable pre-control layer for activating and coordinating large device populations.
\item Finally \textbf{consensus testing} extends ID principles to distributed agreement and verification problems.
\end{enumerate}
 Although these applications differ substantially in their system objectives, they share a common architectural principle: communication resources are primarily used to determine which devices, events, or actions are relevant before conventional payload communication is initiated. This enables scalable and low-overhead communication architectures particularly suited for future large-scale distributed systems and mobile networks.

\subsection{Monitoring and alarming systems}
\label{sec:AlarmingSystems}


Large-scale monitoring and alarming systems represent a natural application domain.
The reason is that many monitoring tasks are already formulated in a binary way. A receiver typically does not need to reconstruct a detailed sensor report, but only needs to decide whether a relevant event has occurred. In environmental monitoring, infrastructure supervision, or disaster warning, the essential question is often whether a threshold has been exceeded, whether an abnormal condition has been detected, or whether an alarm applies to a particular geographic region. In such settings, ID coding aligns particularly well with the communication objective.

Consider a network of environmental sensors distributed over a large area. In a conventional system, each sensor would transmit measurement values or alarm packets, which would then have to be aggregated and interpreted. In an ID-based architecture, communication can shift from value transmission to event signaling. Each sensor or aggregation node emits a short ID signal representing a particular alarm condition, region, or event class. The network infrastructure and end devices then only test whether the corresponding event is relevant to them.

\begin{figure}[t]
\centering
\includegraphics[width=0.88\textwidth]{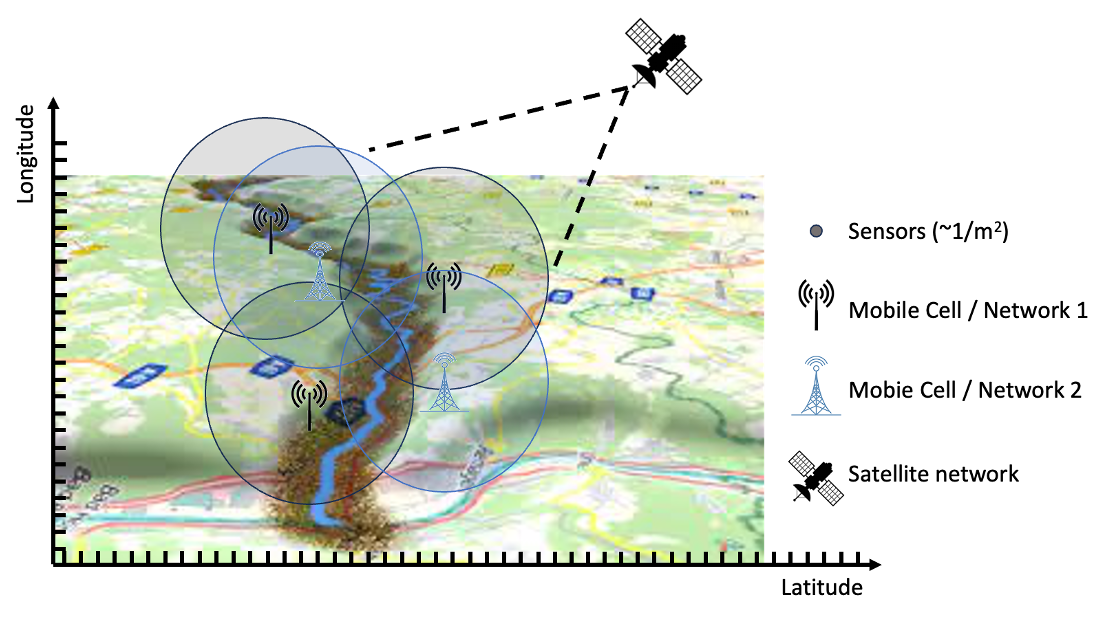}
\caption{Illustration of an ID-based disaster warning architecture. A dense population of geographically distributed sensors reports short alarm-oriented ID messages via mobile access infrastructure, while the network aggregates event indications over a region and escalates to a classical warning message only once the configured alarm threshold is exceeded.}
\label{fig:monitoring-disaster-warning}
\end{figure}

This change in perspective is important for scalability. A very large number of sensors can participate in the system, while only a fraction of end devices or control nodes need to react to a given alarm. In this sense, ID coding provides a natural mechanism for disseminating sparse alerts over a common channel. It is especially attractive when the same physical observation must be checked by many receivers simultaneously, for example, users in an endangered area, machines in a production environment, or network nodes supervising critical infrastructure. 

A related disaster communication system is proposed in \cite{BocheDeppeGeitzRosenbergerDisaster2024} and illustrated in Fig.~\ref{fig:monitoring-disaster-warning}. The figure shows a flood-warning scenario with geographically distributed sensors, mobile radio access, and a subsequent warning dissemination step. In this setting, ID serves as a first-stage trigger: short alarm-oriented ID messages can be transmitted with minimal overhead and very low latency, while richer follow-up information after ID decoding is delivered later through conventional channels. This two-stage design is especially useful when the immediate priority is to activate devices, initiate local safety procedures, or decide whether detailed warning information needs to be retrieved. It also integrates naturally with established cell-broadcast warning systems, in which ID messages indicate that a hazardous situation is emerging in a region, and the corresponding public warning text is then disseminated through the existing broadcast infrastructure.

An important practical advantage is that such short ID messages can be transmitted efficiently over a random-access channel of a mobile network. Because the messages are extremely short, the occupation time of radio resources is small, and the collision probability remains correspondingly low even when many sensors contend for access within a short time interval. This makes the approach particularly attractive for bursty alarm scenarios, in which many devices become active nearly simultaneously after the same external event.

The monitoring use case is also naturally compatible with threshold-based decision logic. In many warning systems, the receiver is not interested in every single sensor report in isolation, but in the aggregated alarm intensity over a geographic region or infrastructure segment. The network can therefore count or combine incoming alarm IDs and convert them into a classical warning message only when a predefined threshold has been reached. This aggregation step substantially improves robustness. If an individual sensor report is dropped from time to time, the overall decision may still remain unchanged because many other sensors observing the same phenomenon contribute to the same regional alarm state. Conversely, occasional false alarms from individual sensors can be absorbed by the thresholding mechanism, because a single spurious report need not trigger a system-wide warning.

Compared with classical approaches, the resulting architecture offers several system-level advantages. The messages are shorter, the first-stage decisions can be taken automatically and with very low latency, and the warning process can be coupled directly to network-side aggregation and broadcast triggering. In disaster scenarios this can save critical seconds and thereby contribute directly to the protection of human life and infrastructure.

\paragraph{K-identification} may play a role for the ID of sensor clusters. A cluster may consist of different sensor groups, i.e., sensors that are grouped according to their measured physical quantity, sensing principle, or application context. For example, an environmental monitoring cluster may include atmospheric sensors such as temperature, humidity, air-pressure, and air-quality sensors, while a structural-health or seismic-monitoring cluster may combine inertial sensors, vibration sensors, acoustic sensors, and ground-motion sensors. Similarly, a mobility-related cluster may comprise positioning sensors such as Global Navigation Satellite Systems (GNSS), Ultra Wide Band (UWB), Radio Detection and Ranging (RADAR), Light Detection and Ranging (LiDAR), and inertial measurement units, whereas a medical or human-centric cluster may include biosensors such as ElectroCardioGram (ECG), pulse-oximetry, glucose, temperature, and pressure sensors. Since individual sensors can belong to multiple categories, for instance an earthquake sensor may be viewed as a seismic sensor, a vibration sensor, and often as an accelerometer-based inertial sensor, K-ID can support the discovery of meaningful higher-level groupings and relationships between heterogeneous sensor types.

\subsection{Special-purpose data storage}
ID coding can also be brought to bear on the problem of efficient database queries. The idea is to step away from full retrieval and instead let the system answer questions of the form
\[
\text{``Is record } x \text{ stored?''}
\]
in a purely binary way, i.e., yes or no. An architecture that implements exactly this concept is described in \cite{BaurBocheDeppeUS20210117126}.

The central insight is that many storage systems simply do not need to support arbitrary reconstruction of the data they hold. In fact, several of the most important questions a modern system asks of stored data are not reconstruction questions at all. A border guard does not need to retrieve every photograph of every traveler; she needs only to verify whether the person standing in front of her matches the watchlist entry under inspection. A web browser does not need a copy of every malicious URL on Earth; it needs only the ability to ask, one URL at a time, ``is this in the bad set?'' In cases like these, the storage problem becomes closely related to identification. Rather than storing and retrieving complete records in the classical sense, the system can be optimized for membership testing over a very large logical record space.

To put this into accessible perspective, consider what the switch from transmission to identification buys us. If a classical storage system using a given block of memory can store $N \approx 2^{nR}$ distinct messages, an ID-based system using the very same physical memory can identify $N \approx 2^{2^{nR}}$ messages. To make the gap concrete: if the classical system could store a few thousand distinct identities, the ID system could in principle differentiate between a number of identities exceeding the count of atoms in the observable universe.


\subsubsection{Presence Verification}
\label{subsec:presence-verification}

The use of mobile-network data to answer whether a specific device was present in a specific cell at a specific time is called \emph{presence verification}. For example, an authorised query in an investigation might ask: ``Was device $X$ connected to cell $Y$ around time $Z$?''

Mobile networks generate this kind of evidence as a matter of course. Each device is attached to a radio cell, and the serving cell can be logged over time. Stored classically, this produces a movement history for every device, which is both privacy-sensitive and expensive: reconstructing individual movements is legally restricted, and continuous spatio-temporal logs for millions of devices quickly become enormous. ID-based storage changes the very purpose of the database. Instead of storing trajectories that can later be reconstructed, the system stores compact encoded records that support only membership queries: ``Was this device in this cell at this time?'' The answer is yes or no. The database is therefore designed for verification, not reconstruction. This both reduces the storage burden and limits what an adversary can learn by reading the stored data.

\paragraph{Data model.}
Time is divided into fixed intervals, for example minutes, and the network area is divided into mobile cells. For each device and time interval, the system observes the cell to which the device was attached, or registers no cell if the device was absent or unavailable. The ID-based store encodes this observation so that an authorised party can later test a single triple:
\[
(\text{device}, \text{cell}, \text{time}).
\]
The database should answer this membership question correctly, while avoiding any ability to reconstruct a complete location history from the stored records.

\begin{figure}[h]
\centering
\includegraphics[width=0.88\textwidth]{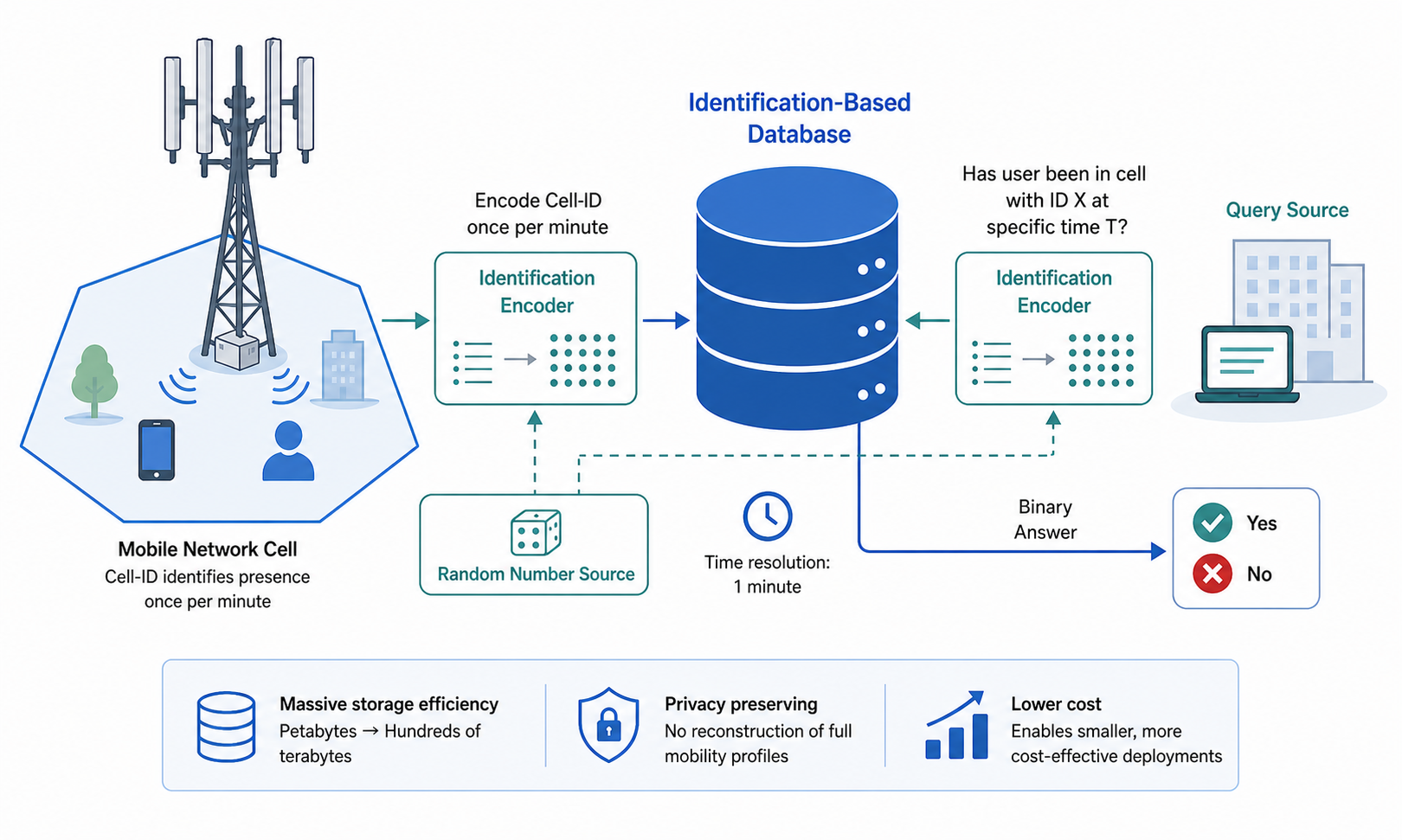}
\caption{ID-based database architecture for mobile-network presence verification. A mobile cell stores encoded cell-ID presence information once per minute, while external queries are independently encoded and checked against the database to return only a binary yes/no answer. Shared randomness supports the ID encoders, enabling privacy-preserving and storage-efficient verification without reconstructing full mobility profiles.}
\label{fig:storageID}
\end{figure}

\paragraph{Mobile-cell case study.}
For Germany, we work with round figures of about $90{,}000$ mobile cells \cite{deutschetelekom_5jahre5g_2024} and $70$ million \cite{smartweb_mobilfunk_2026} active SIM profiles. If each SIM is registered once every ten minutes, this gives $144$ records per SIM per day, or about $9\times 10^{14}$ records per day nationally. A conventional representation using roughly $100$ bytes per record would require about $90$ PB per day. With ID-based storage, the same verification functionality can be represented using only a few bits per entry, reducing the German daily volume to roughly $500$ TB. The figures are summarised in Table~\ref{tab:presence-verification-storage}, and a concrete realisation is shown in Fig.~\ref{fig:storageID}.

\begin{table}[!hbt]
\caption{Illustrative storage comparison for mobile-cell presence verification in Germany.}
\centering
\begin{tabular}{l r}
\hline
Quantity & Approximate value \\
\hline
Mobile cells in Germany & $90{,}000$ \\
Active SIM profiles in Germany & $70$ million \\
Records per SIM per day & $144$ \\
Total records per day & $900\times 10^{12}$ \\
Classical record size & $100$ bytes \\
Classical daily storage & $90$ PB \\
ID-based storage per entry & $4$--$5$ bits \\
ID-based daily storage & $500$ TB \\
Reduction factor & $150$--$200\times$ \\
\hline
\end{tabular}

\label{tab:presence-verification-storage}
\end{table}

\subsubsection{Privacy-preserving storage}
\label{subsec:privacy-storage}

Special-purpose storage of this kind is naturally privacy-friendly: only yes/no answers are ever revealed, and reconstructing full profiles would require many such queries together with careful aggregation, which can be controlled by access policies such as rate limits and per-query authorisation. The strongest version of the idea enforces privacy \emph{at the bit level}, by combining ID coding with a physical source of secret randomness.

The construction works as follows. A secret data message $D$ is never placed in the public database in directly decryptable form. Instead, the storage mechanism generates a uniform random key $S$ and uses it as a one-time pad, $M=S\oplus D$, so that the codeword $M$ is what goes into the public, untrusted database. The key $S$ must be reproducible by the legitimate user later without being stored alongside the data: keeping it on a server invites compromise, and deriving it from a password invites guessing. This is precisely what motivates hardware-intrinsic anchors such as the Physical Unclonable Function (PUF). The stored value can then be identified only by someone in possession of the same PUF device; even if the database is breached, the raw ID tags remain protected \cite{information}.

A PUF is a piece of hardware that, on a challenge input, produces a response output determined by uncontrollable manufacturing variations specific to that device. Two chips from the same wafer give different responses to the same challenge, while a single chip's response is reproducible over time, up to a small amount of noise \cite{baur2018robust,quantum}. PUFs are, in this sense, silicon fingerprints. The classical secrecy-dichotomy theorem for storage tells us that the secure ID storage capacity equals the non-secure one whenever the secrecy capacity is positive, and vanishes otherwise \cite{csecure}. In other words, a PUF is useful for secure ID-based storage \emph{if and only if} it delivers at least one bit of secrecy per mobile cell (in the running example) that is unknown to the eavesdropper; and as soon as it does, the secure rate matches the non-secure rate. There is no middle ground. These results were later extended to Quantum PUFs (QPUFs), where quantum randomness, together with the no-cloning theorem and the indistinguishability of non-orthogonal quantum states, is used to guarantee security \cite{quantum6,quantum5,quantum3,quantum1}.

While single-source PUF-based ID already represents a major leap forward in secure storage, the physical limitations of the hardware can occasionally bottleneck the rate at which private quantum randomness is generated. To circumvent this limitation and push the capacity even higher, the special-purpose storage framework introduces a more advanced mechanism: public second-source-aided ID \cite{quantum2}. Here a freely available stream of randomness, shared by encoder and decoder, raises the ID rate without raising the privacy-leakage rate.

\subsubsection{Query repetition and robustness}
\label{subsec:query-repetition}

All of the capacity statements above are \emph{asymptotic}; any real deployment at finite $n$ and $k$ will admit a positive error of each kind. Real systems use finite-length codes on noisy hardware, with users who occasionally err, so the right question is not ``is the error zero?'' but ``can it be driven arbitrarily low at acceptable cost per query?'' The standard answer is \emph{query repetition with majority voting}. To check ``is record $x$ stored?'', a user encodes and sends the query three independent times; the database replies $a_1,a_2,a_3\in\{0,1\}$, and the user takes $\hat a=\mathrm{maj}(a_1,a_2,a_3)$, correcting isolated errors at the cost of a few extra queries. For $n$ independent repetitions, each with per-query error $p<\tfrac{1}{2}$, the error probability is
\[
P_{\mathrm{err}}(n)=\sum_{k=\lceil n/2\rceil}^{n}\binom{n}{k}p^k(1-p)^{n-k},
\]
which decreases exponentially in $n$. In the mobile-network worked example, a per-query error of $10^{-3}$ becomes a triple-redundant error on the order of $10^{-9}$, at the cost of a threefold increase in queries, typically negligible compared with full reconstruction. Repetition is especially useful in wireless and IoT settings, where interference can corrupt a short message. Thresholding and aggregation extend the same idea: a fraud-detection system may raise an alert only when several suspicious presence verifications occur within a window, absorbing isolated false alarms; the same logic underpins warning-system architectures in monitoring systems.

\subsubsection{Geofencing and compliance use cases}
\label{subsec:geofencing-compliance}

Modern data-protection regimes (GDPR in the EU, CCPA/CPRA in California, China's PIPL, and comparable instruments worldwide) are organised around principles such as data minimisation and purpose limitation: collect and retain only what is necessary for a stated purpose, and do not silently repurpose it. ID-based storage architectures enforce both principles at the architectural level. A database built for one binary question literally cannot answer another, because the information needed to do so was never stored in the first place. There is no GPS field to query, only the answer to ``was the user in this cell at this time?''; an internal team or partner wanting a daily movement profile cannot extract one, even in principle. The architecture is intrinsically purpose-limited, a particularly clean realisation of what privacy lawyers call \emph{data protection by design}.

A \emph{geofence} is a virtual perimeter around a real-world location; services raise events when a device enters or leaves it. Commercial geofences serve advertising (push a coupon when the user enters a shopping district), fleet management (alert when a vehicle leaves its assigned zone), compliance (block a transaction originating from outside the user's home country), and digital-rights management (a film licensed only for Germany must refuse to play in France). Corporate policies that grant access to sensitive data only when the device is on campus can likewise rely on a membership database that checks ``is user $U$ presently, or recently, in building $A$?'' Each of these is, at heart, a presence-verification query of the kind formalised in Section~\ref{subsec:presence-verification}. The ID-based replacement is direct: for each user $u$, candidate region $r$, and time bin $t$, the system precomputes the binary answer ``was $u$ in $r$ during $t$?'' and encodes it into a public storage codeword via a PUF or QPUF; queries from permitted services then use the PUF to derive the decoding key. An eavesdropper who copies the database obtains essentially no information, not even whether any particular user was in any particular region. The same template covers a range of adjacent scenarios: \emph{payment-fraud screening} (the payment provider checks the consistency of a claimed transaction location with cardholder network presence, without acquiring any history); \emph{regional content licensing and geoblocking} (a streaming service confirms that the viewer is in a permitted region at access time, not where they have been); \emph{border and travel-related compliance} (a regulator confirms presence in a jurisdiction at a given time without acquiring a full travel history); and \emph{public safety} (law enforcement, under judicial oversight, queries whether a suspect already identified by independent means was registered in a particular cell at a particular time, with the architecture granting exactly that access and no more).

Accordingly, special-purpose data storage based on ID should not be read as a replacement for conventional databases. It is, rather, a highly specialised architecture for massive record spaces with sparse binary queries.

From a survey perspective, this application class is noteworthy because it extends identification beyond communication channels in the narrow sense. The system is not designed to reproduce information in full, but to determine whether a specific hypothesis is true. This makes special-purpose storage one of the clearest examples of how ID theory can inspire non-traditional communication and information-processing architectures.

\subsection{Joint identification and sensing}
\label{sec:JIDAS}

Joint communication and radar/radio sensing (JCAS) means that communication and sensing are jointly designed by sharing the same bandwidth. Sensing and communication systems are usually designed separately such that resources are dedicated to either sensing or data communications. Joint sensing and communication approach is a solution to overcome the limitations of a separation-based approach. Recent works \cite{sensing1,sensing2,sensing3,sensing4} explored JCAS and showed that this approach can improve spectrum efficiency and minimize hardware costs.
For instance, fundamental limits of joint sensing and communication for a point-to-point channel have been studied in \cite{MariPaper}. The problem of JIDAS and sensing over a state-dependent DMC was first investigated in \cite{isit23} . A complete characterization of the ID capacity-distortion function was established in \cite{isit23,labidi2025joint}. 
This section presents a theoretical analysis that characterizes fundamental limits and trade-offs of the JIDAS problem, followed by a discussion of practical applications that illustrate how these concepts can be leveraged in real-world scenarios.
 
\subsubsection{Theoretical analysis}
\begin{figure}[hbt!]
\centering
\tikzstyle{farbverlauf} = [ top color=white, bottom color=white!80!gray]
\tikzstyle{block} = [draw,top color=white, bottom color=white!80!gray, rectangle, rounded corners,
minimum height=2em, minimum width=3em]
\tikzstyle{blockD} = [draw,top color=white, bottom color=white!80!gray, rectangle, rounded corners,
minimum height=1.5em, minimum width=1.5em]
\tikzstyle{block1} = [draw, fill=none, rectangle, rounded corners,
minimum height=8em, minimum width=2
cm]
\tikzstyle{input} = [coordinate]
\tikzstyle{sum} = [draw, circle,inner sep=0pt, minimum size=2mm,  thick]

\scalebox{1}{
\tikzstyle{arrow}=[draw,->] 
\begin{tikzpicture}[auto, node distance=2cm,>=latex']
\node[] (M) {$i \in \mathcal{N}$};

\node[block,right=.5cm of M] (enc) {Encoder};
\node[block, above=.7cm of enc] (est) {Estimator};
\node[left=.4cm of est] (S) {$\hat{S}^n$};
\node[block, right=3.2cm of enc] (channel) {DMC};
\node[block1, dashed] at (1.75,.75) (tr) {};
\node[above=.3cm of tr] (tran) {sender};
\node[block,below=.7cm of channel](state){$P_S$};
\node[block, right=1cm of channel] (dec) {Decoder};
\node[right=.4cm of dec] (Mhat) {\begin{tabular}{c} {\small Is ${i}^\prime$ sent?} \\ {\small Yes or No?} \end{tabular}};
\node[input,right=.7cm of channel] (t1) {};

\node[blockD, above=2cm of t1] (delay) {\small $D$};

\node[above=2cm of channel] (t4) {};
\node[below=1.5cm of t1] (t5) {};
\node[input,above=2.3cm of t1] (t2) {};
\node[input,left=4cm of t2] (point) {};
\node[input,below=1.65cm of point] (ttpoint) {};
\node[input,left=4.3cm of t2] (tt) {};
\node[input,below=1.65cm of tt] (ttt) {};
\node[input,left=0.4cm of ttt] (ttn) {};
\node[input, right=.5cm of enc] (t3) {};
\node[input, below=0.35 cm of est] (tte) {};
\draw[-{Latex[length=1.5mm, width=1.5mm]},thick] (M) -- (enc);
\draw[-{Latex[length=1.5mm, width=1.5mm]},thick] (enc) --node[above]{ $X_t=f_i^t(Y^{t-1})$} (channel);
\draw[-{Latex[length=1.5mm, width=1.5mm]},thick] (channel) --node[below]{$Y_t$} (dec);
\draw[-{Latex[length=1.5mm, width=1.5mm]},thick] (dec) -- (Mhat);
\draw[-] (t1) -- (delay);
\draw[-{Latex[length=1.5mm, width=1.5mm]},thick] (delay) -|  (est);
\draw[-{Latex[length=1.5mm, width=1.5mm]},thick] (state)--(channel);
\draw[-{Latex[length=1.5mm, width=1.5mm]},thick] (est) --(S);
\draw[-{Latex[length=1.5mm, width=1.5mm]},thick] (t3) |- (est);
\draw (ttt) arc[start angle=0, end angle=180, radius=0.2]  (ttn); 
\draw[-{Latex[length=1.5mm, width=1.5mm]},thick] (ttn) -| (enc);
\draw[-] (point) -- (ttpoint);
\draw[-] (ttpoint) -- (ttt);
\end{tikzpicture}}
\caption{JIDAS over state-dependent channel. The sender wants to simultaneously send an ID message $i \in \mathcal{N}$ to the receiver and sense the channel state sequence $S^n$ based on the output of the noiseless strictly causal feedback link.}
\label{Fig:capacityDistortion}
\end{figure}
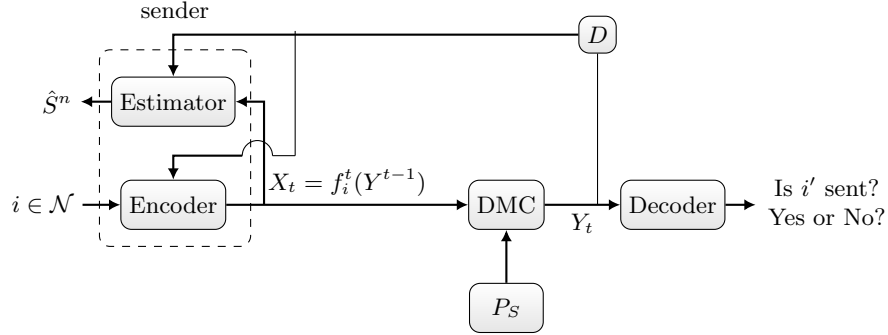

Let a DMC with random state $(\setx\times \mathcal{S}, W_S(y|x,s), \sety)$ consisting of a finite input alphabet $\setx$, a finite output alphabet $\sety$, a finite state set $\mathcal{S}$ and a conditional distribution $W_S(y|x,s)$ on $\sety$, be given. The channel is memoryless, i.e., the probability for a sequence $y^n \in \sety^n$ to be received if the input sequence $x^n \in \setx^n$ was sent and the sequence state is $s^n \in \mathcal{S}^n$ is given by 
		\begin{equation*}
		W_S^n(y^n|x^n,s^n)=\prod_{i=1}^n W_S(y_i|x_i,s_i).
		\end{equation*}
		 The state sequence $(S_1,S_2,\ldots,S_n)$ is independent and identically distributed (i.i.d.) according to the distribution $P_S$. We assume that the input $X_i$ and the state $S_i$ are statistically independent for all $i\in\{1,2,\ldots,n\}$. We assume that the channel state is  known to neither the sender nor the receiver. 
The setting is depicted in Fig.~\ref{Fig:capacityDistortion}, where the sender wants to jointly identify and sense the channel state. The sender comprises an encoder that sends a symbol $x_t=f_i^t(y^{t-1})$ for each identity $i \in \{1,\ldots,N\}$ and delayed feedback output $y^{t-1} \in \sety^{t-1}$ and a state estimator that outputs an estimation sequence $\hat{s}^n \in {\mathcal{S}}^n$ based on the feedback output and the input sequence. 
We define the per-symbol distortion as the following:
\begin{equation}
 d_t=\mathbb{E}\left[d(S_t,\hat{S}_t)\right], \quad t \in \{1,\ldots,n\}.\label{eq:persymbolDistortion}
\end{equation}
where $d\colon \mathcal{S} \times {\mathcal{S}} \to [0, +\infty)$ is a distortion function and the expectation is over the joint distribution of $(S_{t},\hat{S}_{t})$ conditioned by the ID message $i \in \{1,\ldots,N\}$.

We consider the ID capacity-distortion tradeoff as a performance metric. This metric is analogous to the one studied in \cite{performanceMetric} and is defined as the supremum of all ID rates we can achieve such that some distortion constraint on state sensing is fulfilled.
 
 The following theorem characterizes the \emph{deterministic} and \emph{randomized} ID capacity-distortion function of the state-dependent DMC $W_S$ under the per-symbol distortion constraint $d_t \leq D, \quad \forall t \in \{1,\ldots,n\}$.
\begin{theorem}[\cite{labidi2025joint}]
The \emph{deterministic} ID capacity-distortion function of the state-dependent channel $W_S$ under the per-symbol distortion constraint $d_t \leq D, \quad \forall t \in \{1,\ldots,n\}$ and w.r.t. the \emph{rate function} $\zeta_3(n,R)=2^{2^{nR}}$ is given by
\begin{equation*}
    C_{ID}^{(d)} (D,W_S,\zeta_3) = \max_{x \in \setx_D}  H\big(\mathbb{E}[W_S(\cdot|x,S)] \big),
\end{equation*}
where the set $\setx_D$ is characterized by
\begin{equation*}
    \setx_D=\left\{x \in \setx,\quad d_t \leq D, \quad \forall t \in \{1,\ldots,n\} \right\}.
\end{equation*}
The \emph{randomized} capacity-distortion function of the state-dependent channel $W_S$ under the per-symbol distortion described in \eqref{eq:persymbolDistortion} and w.r.t. the \emph{rate function} $\zeta_3(n,R)=2^{2^{nR}}$ is given by:
\begin{equation*}
    C_{ID} (D,W_S,\zeta_3) = \max_{P \in \mathcal{P}_D}  H\big(\sum_{x\in\setx}P(x)\mathbb{E}[W_S(\cdot|x,S)] \big),
\end{equation*}
where the set $\mathcal{P}_D$ is characterized by
\begin{equation*}
    \mathcal{P}_D=\left \{P \in \mathcal{P}(\setx),\quad  d_t \leq D, \quad \forall t \in \{1,\ldots,n\}\right\}.
\end{equation*}
\label{theorem:capDist2}
\end{theorem}
The \emph{deterministic} capacity-distortion formula is given by the entropy of the average output distribution induced by a single input symbol, optimized over all inputs that satisfy the distortion constraint. This reflects selecting one input that balances high output uncertainty with reliable state estimation. In the \emph{randomized} ID case, the encoder can use an input distribution instead of a single symbol, leading to a mixture output distribution. This generally enlarges the capacity by improving the trade-off between output entropy and sensing accuracy under the same distortion constraint. This makes sense because the encoder introduces additional local randomness on top of the randomness already induced through "sensing", which can further increase the capacity. Overall, the result shows that ID performance is governed by the entropy of the induced output distribution, subject to constraints imposed by state estimation accuracy.

Now, we define the average distortion as the following:
\begin{equation}
    \bar{d}^n=\frac{1}{n} \sum_{t=1}^{n}  \mathbb{E} \big[d(S_t,\hat{S}_t)\big],\label{eq:averageDistortion}
\end{equation}
where $d\colon \mathcal{S} \times {\mathcal{S}} \to [0, +\infty)$ is a distortion function and the expectation is over the joint distribution of $(S_{t},\hat{S}_{t})$ conditioned by the ID message $i \in \{1,\ldots,N\}$.
The average distortion metric is valuable because it relaxes the per-symbol fidelity requirement, allowing for minor variations in individual symbol quality as long as the overall average distortion remains below a specified threshold.
The following theorem characterizes achievable ID rates for both \emph{deterministic} and \emph{randomized} encoders under the average distortion constraint in \eqref{eq:averageDistortion} for the state-dependent DMC $W_S$.
\begin{theorem}[\cite{isit23}]
The \emph{deterministic} ID capacity-distortion function of the state-dependent
channel $W_S$ under the average distortion constraint $\limsup_{n \to \infty} \bar d^n
\le D$  and w.r.t. the \emph{rate function}
$\zeta_3(n,R)=2^{2^{nR}}$ is lower-bounded by
\begin{equation*}
    \overline{C}_{ID}^{(d)} (D,W_S,\zeta_3) {\color{black}{\geq}} \max_{x \in \setx_D}  H\big(\mathbb{E}[W_S(\cdot|x,S)] \big), 
\end{equation*}
where the set $\setx_D$ is given by 
$$\mathcal X_D = \left\{x \in \setx,\quad \limsup_{n \to \infty} \bar d^n \leq D\right \}.$$
The \emph{randomized} ID capacity-distortion function of the state-dependent channel $W_S$ under the under the average distortion constraint $\bar d^n
\le D$ (see \eqref{eq:averageDistortion}) and w.r.t. the \emph{rate function} $\zeta_3(n,R)=2^{2^{nR}}$ is lower-bounded by
\begin{equation*}
    \overline{C}_{ID} (D,W_S,\zeta_3) {\color{black}{\geq}} \max_{P \in \mathcal{P}_D}  H\big(\sum_{x\in\setx}P(x)\mathbb{E}[W_S(\cdot|x,S)] \big), 
\end{equation*}
    where the set $\mathcal{P}_D$ is given by
    $$\mathcal{P}_D=\left\{P \in \mathcal{P}(\setx),\quad \limsup_{n \to \infty} \bar d^n \leq D\right\}.$$
\end{theorem}

 In the considered setting, the communication system consisting of a sender and one or multiple receivers pursues two objectives simultaneously. First, the receiver(s) aim to identify whether a particular message or user was transmitted. Second, the sender seeks to infer information about the hidden state of the communication channel, which in practical scenarios may depend, for example, on the varying distance between sender and receiver. The channel state can be interpreted as an unknown condition of the transmission environment, such as noise, fading, or interference. Since this state is not directly observable, the sender exploits the available channel feedback to estimate it. The quality of this estimation process is quantified by a distortion measure, where smaller distortion corresponds to more accurate sensing performance. The main result reveals a fundamental trade-off between ID performance and sensing accuracy. Higher ID rates can be achieved when the induced channel outputs are more distinguishable or diverse. However, only input symbols or input distributions satisfying the required distortion constraint are admissible. In this context, both \emph{randomized} and \emph{deterministic} encoding strategies can benefit from this trade-off, since they are able to employ mixtures of several input symbols or distributions rather than relying on a single fixed channel input.

\subsubsection{Applications}

Many modern communication systems combine sensing and communication. In ID-based architectures, sensed information can be mapped directly to ID signals. Rather than first transporting sensor observations as payload and only later converting them into actions, the system immediately turns the observed state of the environment into a distributed decision problem: which devices, actuators, or agents should react, and which should remain inactive? This creates a much tighter coupling between sensing, communication, and control than in conventional architectures. One representative scenario is environment-dependent activation of autonomous micro-devices or micro-flying machines. A sensing layer observes variables such as position, temperature, humidity, chemical concentration, time window, or atmospheric conditions. Based on these observations, the system generates an ID signal corresponding to a region, operating mode, or action class. Each individual device then performs only a local ID test to determine whether it belongs to the subset that should act. In this way, sensing results are translated directly into selective activation over a shared communication medium.

\paragraph{Application towards smart agriculture:}
This principle is particularly attractive in smart agriculture and related environmental applications \cite{EP25201196.0}. Large populations of small autonomous flying agents can be activated only when local sensor data indicate that intervention is necessary. This can be weather conditions or temperature. Depending on the sensed conditions, such agents may deposit seeds, fertilizer, or protective substances in a highly targeted way. More generally, the same architecture supports environmental intervention tasks in which physical action depends directly on sensed context. The basic idea is that ID does not merely report an observation, but immediately selects the set of agents that should respond to it.

\begin{figure}[h]
\centering
\includegraphics[width=0.88\textwidth]{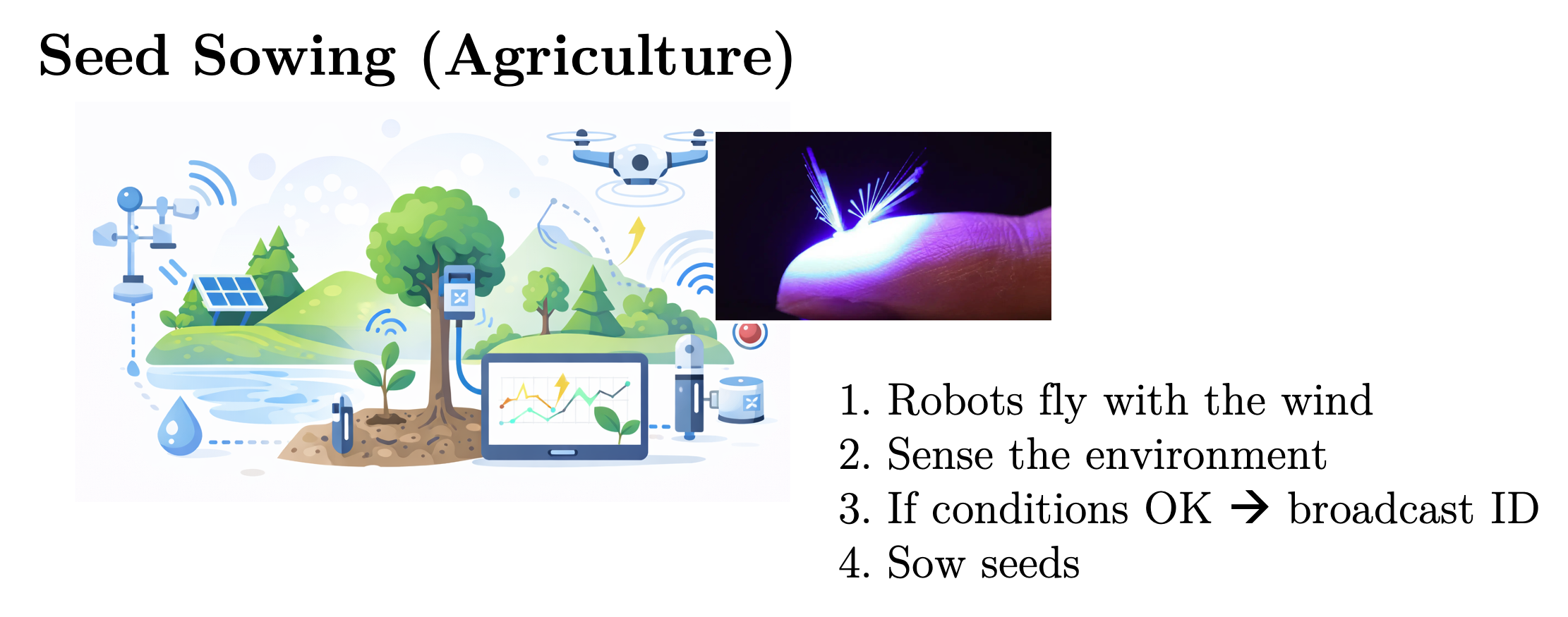}
\caption{Sensed environmental conditions are mapped directly to ID signals that selectively activate autonomous agents, such as drones in smart agriculture or smart-dust-type devices.}
\label{fig:agri}
\end{figure}

\paragraph{Application towards smart dust:}
A closely related concept appears in smart-dust-type systems \cite{EP25201196.0}. Here, the agents are extremely small devices, potentially at micrometer scale, equipped with sensing, limited processing, and communication capability. They are called smart dust, because of their minimum size and weight, capable to fly with the wind. In such scenarios, explicit point-to-point coordination becomes unrealistic because of the sheer number and simplicity of the nodes. ID-based signaling is attractive because it allows a common broadcast signal to activate only those smart-dust agents for which the sensed condition is relevant, while the remaining devices stay silent or inactive.

\paragraph{Sensing and identification of agent position:}
Another important class of examples arises when the position of an agent itself is sensed and converted into an ID decision. This is relevant whenever a system must address only those agents that are currently located in, or moving toward, a particular physical region. Possible use cases include targeted medical intervention, such as observing and activating agents in the vicinity of a tumor in cancer treatment, targeted environmental cleanup operations such as collecting garbage in the ocean, or highly specialized repair scenarios in hazardous environments, for example near damaged infrastructure in a nuclear power plant. In all these cases, the key benefit is that a large number of agents can share a common signaling layer while only the relevant subset reacts \cite{EP25201197.8}.

\begin{figure}[h]
\centering
\includegraphics[width=0.88\textwidth]{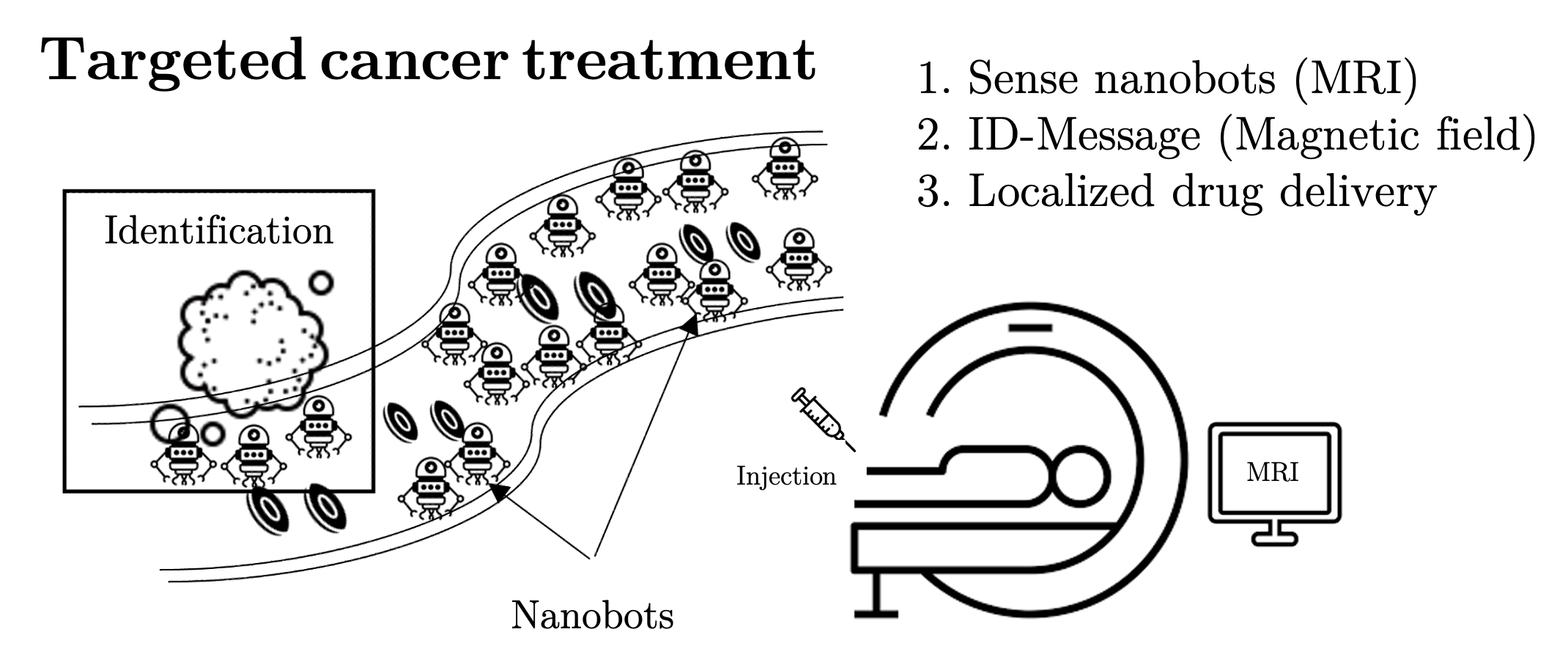}
\caption{JIDAS for targeted medical intervention with nano robots. After injection, a large population of nano robots moves through the body, while sensing and ID are used to detect a relevant pathological region, such as a tumor, and selectively activate only those agents in its vicinity. The location of the robot is detected by the MRI while the ID signal is encoded into slow magnetic field variations.}
\label{fig:cancer}
\end{figure}

\paragraph{Massive agent populations:}
More generally, the ability to address massive populations of agents over a broadcast channel also points to very large-scale intervention scenarios. 
\begin{figure}[h]
\centering
b\includegraphics[width=0.88\textwidth]{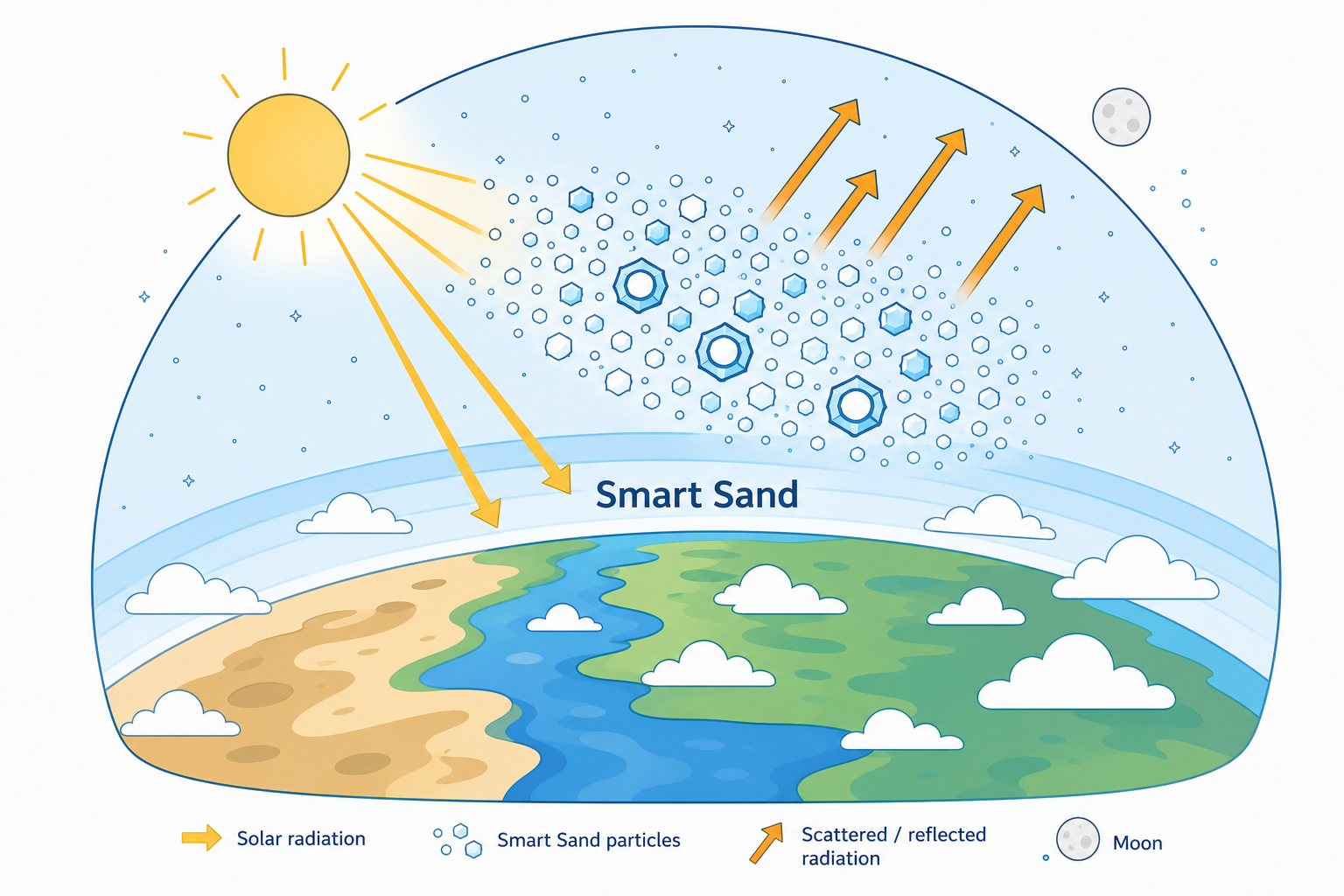}
\caption{Visionary large-scale application of JIDAS with massive populations of micro-agents. The figure illustrates how smart-dust or ``smart sand'' particles could be selectively activated to influence atmospheric conditions, for example by scattering or reflecting sunlight or ejecting chemicals. }
\label{fig:terra}
\end{figure}
In principle, ID-based activation could support coordinated chemical or material deposition by vast numbers of agents, for example in speculative terraforming settings such as atmospheric modification, ozone-layer restoration, or sunlight-reflection measures. Even if such applications remain visionary, they illustrate the scalability of the underlying communication concept: the system is designed to select relevant subsets out of enormous agent populations without individually commanding each node \cite{EP25201196.0}.

\paragraph{Communication Robustness:} An additional advantage of this application class is its robustness against occasional decision errors. In many JIDAS scenarios, the overall effect is generated collectively by a very large number of agents rather than by a single critical action. As a result, isolated false activations or missed activations are often tolerable at the system level. If, for example, a single seed is deposited unnecessarily, one smart-dust agent remains inactive, or one cleanup micro-agent reacts at the wrong time, the global task is typically not endangered. This makes JIDAS particularly attractive for applications with statistical or aggregate objectives, where reliability is achieved through population effects rather than perfect control of every individual agent. At the same time, the tolerable error level depends strongly on the application, and safety-critical scenarios still require additional safeguards to ensure that unintended actions remain bounded and harmless.

\subsection{Semantic communication} \label{subsec:semanticComm}

Semantic communication focuses on transmitting meaning rather than raw data. In the context of ID coding, this means that the receiver is not interested in reconstructing a full message, but only in deciding whether a certain meaning or action is intended. The communication objective is therefore reduced to selecting one element from a pre-agreed set of semantic actions. This aligns naturally with the ID paradigm, because the receiver performs a hypothesis test on whether a particular meaning was sent.

A representative example is high-frequency trading \cite{DE102025143864.1}. In such a setting, the communicating parties first define a semantic codebook in which specific ID messages correspond to predefined trading actions, such as buying or selling a particular stock, fund, or other financial instrument. The size of this semantic codebook is determined by the number of relevant tradable assets and the corresponding action classes, such as whether to buy or sell and in what quantity. Instead of transmitting a longer conventional instruction message, typically an XML or JSON, the sender only communicates which semantic action from the codebook is intended.
\begin{figure}[h]
\centering
\includegraphics[width=0.88\textwidth]{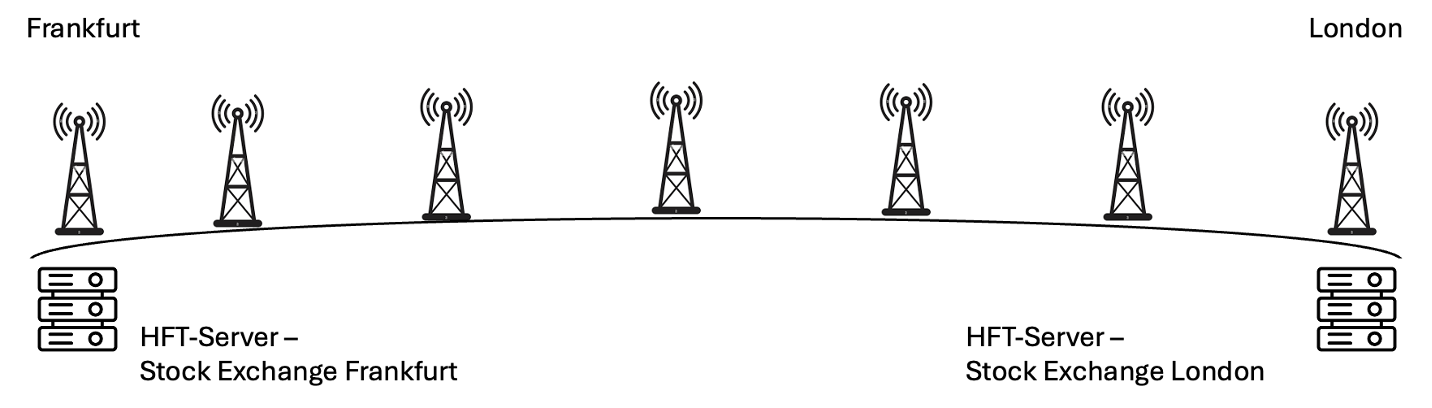}
\caption{Illustration of semantic communication for low-latency trading between the financial venues of Frankfurt and London. A trading action is encoded at the source exchange, transmitted over a chain of ultra-low-latency wireless relay links, and decoded at the destination exchange, where it is executed.}
\label{fig:comlink}
\end{figure}
Operationally, a trading action may be generated at one stock exchange, encoded as an ID message, transmitted over a low-latency channel, and decoded at another stock exchange or trading venue. A particularly instructive example is the Frankfurt--London corridor, where ultra-low-latency wireless RF links based on microwave or millimeter-wave technology have been deployed for financial-market connectivity. Publicly reported services on this route achieved latencies of about 4.192\,ms between Frankfurt and Basildon and about 4.64\,ms round trip between Slough and Frankfurt \cite{McKayFrankfurtBasildon2015,McKaySloughFrankfurt2015}. An example RF radio communication link is shown in Fig.~\ref{fig:comlink}. The exact number of active proprietary links is not publicly transparent. Once the intended action has been identified, it can be executed immediately at the target exchange. The key advantage is that ID-based signaling may require fewer transmitted bits than a classical instruction message, even if sent multiple times to counteract a type II error, thereby reducing communication overhead and potentially lowering end-to-end latency in highly time-critical settings.

\paragraph{K-identification} may be used to address and identify groups of trading targets at different levels of granularity. In such a case, a positive K-ID result does not necessarily refer to a single asset, but may identify a class or subclass of assets on which the trader should act. For example, the semantic codebook may define hierarchies such as
\begin{flushleft}
$\text{equities} \rightarrow \text{technology stocks} \rightarrow \text{information-technology stocks}, \mathrm{or}$
\end{flushleft}

\begin{flushleft}
$\text{equities} \rightarrow \text{energy stocks} \rightarrow \text{solar-power companies}.$
\end{flushleft}
Similarly, financial instruments may be grouped as
\begin{flushleft}
$\text{funds} \rightarrow \text{exchange-traded funds} \rightarrow \text{sector ETFs} \rightarrow \text{semiconductor ETFs}, \mathrm{or}$
\end{flushleft}
\begin{flushleft}
$\text{bonds} \rightarrow \text{corporate bonds} \rightarrow \text{technology-sector bonds}.$
\end{flushleft}
After identifying the intended group, the trading system may execute a predefined strategy for all assets in that group. This hierarchical use of K-ID allows semantic communication not only to encode individual trading actions, but also to efficiently trigger actions over structured asset classes and nested market segments.

The economic relevance of such a scenario is underlined by the scale of the market itself. Public industry estimates place the global high-frequency-trading market at about USD~10.36\,billion in 2024 and project continued growth to USD~16.03\,billion by 2030 \cite{GrandViewHFT2024}. These figures indicate that even comparatively small improvements in latency or execution reliability can be commercially significant.

\begin{figure}[h]
\centering
\includegraphics[width=0.88\textwidth]{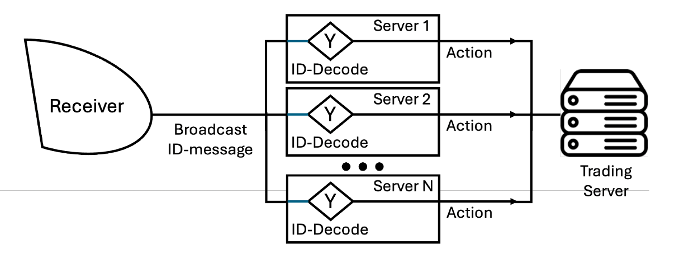}
\caption{Parallel decoding architecture for ID-based semantic communication. The received ID message is broadcast to many decoding instances running on separate servers, each testing a candidate semantic action. If a server accepts its hypothesis, the corresponding action is forwarded to the trading server for execution. This illustrates how \emph{randomized} ID can be implemented through large-scale parallel hypothesis testing in low-latency trading systems.}
\label{fig:decoder}
\end{figure}
For \emph{randomized} ID encoding, the receiver performs a hypothesis test on the received signal. This can be implemented efficiently through parallel decoding in a server or cloud infrastructure, as shown in Fig.~\ref{fig:decoder}. The received ID message is broadcast to a large number of decoding instances, each associated with a candidate semantic action, and all hypothesis tests are carried out in parallel. If one decoder accepts its hypothesis, the corresponding trade action is executed. In this way, the ID framework matches naturally with modern distributed compute architectures. For \emph{deterministic} ID encoding, the channel advantage may be smaller, but the decoding process can be simpler and faster and non-parallel, since it approaches a quasi-linear decoding procedure \cite{determinsticGaussianPau}. Both \emph{randomized} and \emph{deterministic} ID codes are therefore relevant candidates, and the preferable choice depends on the latency, complexity, and reliability requirements of the application.

A further important issue is robustness against decision errors. In a semantic trading scenario, an incorrect ID may lead to an unintended purchase or sale. Practical systems should therefore include suitable safeguards. One option is to split larger trades into smaller transaction chunks, thereby limiting the financial impact of any single erroneous decision. Another option is to transmit the same semantic instruction multiple times, for example three times with independent encodings, and determine the final action by majority voting. A third possibility is to increase the number of bits assigned to the ID message itself. In general, a larger ID description can reduce the probability of erroneous decoding, but this comes at the price of a smaller latency advantage on the communication link. The resulting system design therefore involves a trade-off between communication efficiency and decision reliability. It comes down to a risk versus competitive advantage evaluation of the high frequency trader. 

From a broader perspective, this example illustrates why semantic communication is a natural application area for post-Shannon architectures. The system is not designed to reproduce a detailed symbolic message, but to identify which action or meaning should be realized at the receiver. ID coding therefore provides a particularly attractive foundation whenever communication is valuable primarily because of the action it triggers, rather than because of the full bit string it conveys. An important direction for future work is to derive quantitative expressions for the probability of erroneous semantic decodings in such low-latency trading scenarios and to characterize the optimal trade-off between message length, latency gain, and execution risk.

\subsection{Mobile network control plane communication}

\begin{figure}[h]
\centering
\includegraphics[width=0.88\textwidth]{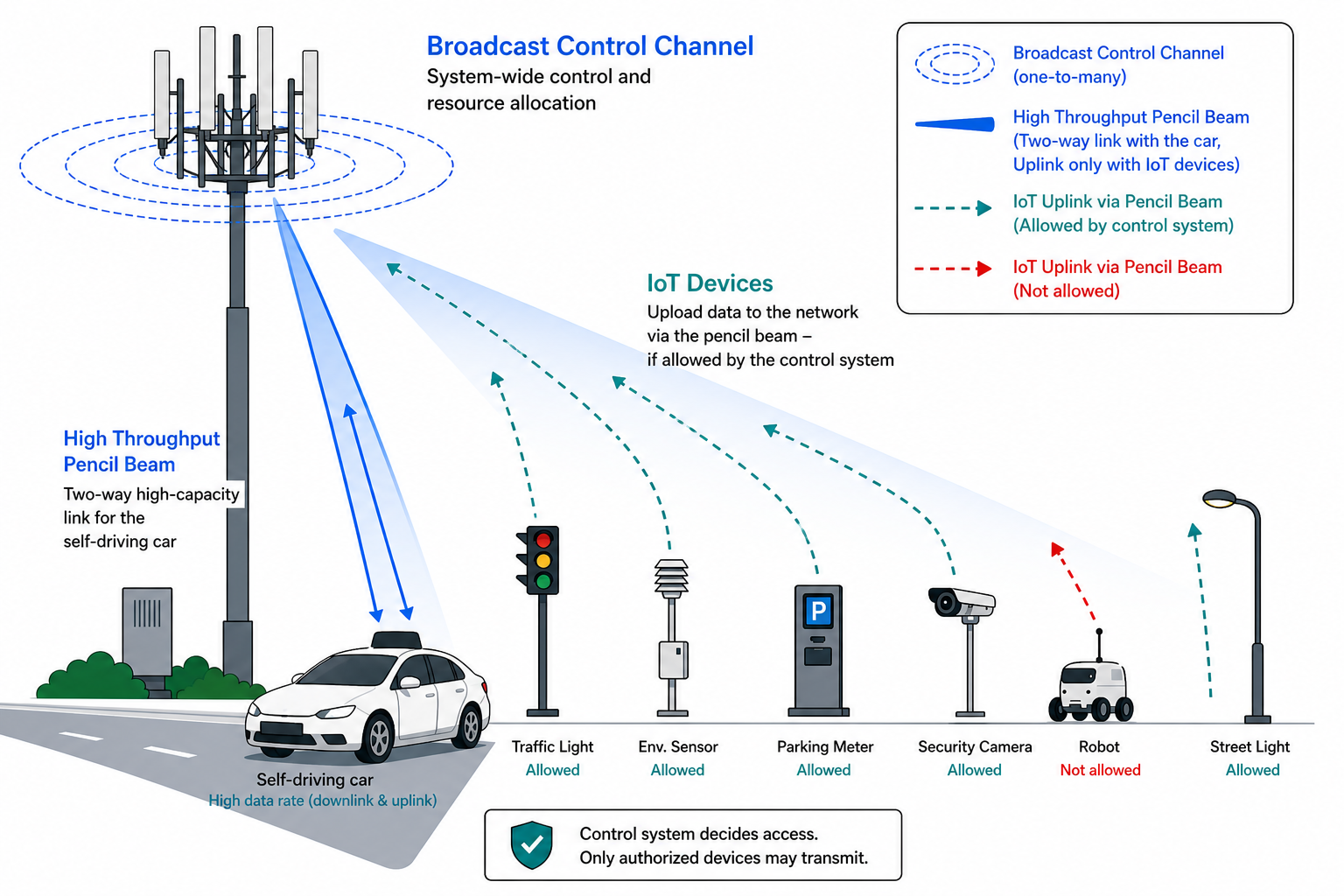}
\caption{ID-based broadcast control plane enhancement in a mobile network. A base station uses a broadcast ID signal to address a large population of IoT sensors and devices waiting to offload data, while a high-bandwidth pencil beam serves a selected user or application with detailed payload transmission. IoT sensors are triggered by ID encoded messages, when and where to use spectrum of offload their data.}
\label{fig:controlplane}
\end{figure}

A further promising application of ID-based communication arises in dense IoT deployments in mobile networks. In such scenarios, the number of sensors within a single cell may become so large that classical registration, scheduling, and control procedures no longer scale efficiently. If every sensor attempts to register, request resources, and transmit independently, the result is excessive control overhead, increased contention, and growing interference. ID-based control offers a different operating principle: unregistered sensors remain silent by default and are only activated when the mobile control system explicitly assigns them an opportunity to transmit \cite{EP24219710.1}.

The key idea is to encode these transmission opportunities through broadcast ID messages that are structured by location and time. Sensors know their own position and local time reference and therefore can perform a local hypothesis test on whether a given broadcast message is intended for them. In this way, the network does not have to address each sensor individually through explicit control signaling. Instead, it broadcasts an ID pattern describing which subset of sensors, in which region and at which time, is allowed to offload data. Only sensors that decode a positive ID activate their communication module and attempt transmission; all others remain silent.

A particularly illustrative use case is a high-capacity pencil beam following a moving car. Such a beam may be established primarily to serve the vehicle with high-bandwidth communication, but at certain times it still has residual capacity that would otherwise remain unused. This spare capacity can be exploited for opportunistic IoT data offloading. If the control system knows, or can predict using AI-based trajectory estimation, the path of the vehicle and the associated movement of the pencil beam, it can broadcast ID messages into the surrounding cell. Sensors located along the predicted track that wish to offload data listen for these ID broadcasts. If a sensor decodes a positive result, it  transmits its data in the time window in which the passing pencil beam can collect it. In this way, the beam is used not only for its primary communication task, but also as a moving high-capacity collector for local sensor traffic.

This architecture requires careful timing and coordination. The ID broadcast must be sent early enough that eligible sensors can decode, and prepare transmission, but closely enough aligned with the beam trajectory that the actual data offloading occurs when radio resources are available. The control system therefore combines mobility prediction, beam scheduling, and ID-based broadcast control into one integrated procedure. From the perspective of the sensors, the process remains simple: they only need to know their own location and time, monitor the broadcast control channel, and decide whether the current transmission opportunity applies to them.

An important property of this scheme is that it is naturally robust against occasional decision errors. If an eligible sensor does not receive or correctly decode its offloading opportunity, the consequence is usually limited, since the sensor can simply wait for the next opportunity. Conversely, if a sensor incorrectly decides that it has been selected, the result is only a bounded amount of additional interference or noise in the cell. Because the large majority of sensors remain silent unless explicitly activated, the overall disturbance remains much smaller than in a conventional random-access scenario with many simultaneously contending devices.

The main advantage of this approach is that it creates a largely silent cell in which cross-talk and uncontrolled contention among sensors are strongly reduced. Radio resources are used more efficiently because the network activates only those sensors for which transmission capacity is actually available. At the same time, the method improves spectrum usage and control-plane scalability by replacing massive device-specific coordination with compact broadcast ID. In a broader network perspective, this may also allow larger effective cell sizes or higher sensor densities to be supported than in current architectures, thereby reducing the pressure for costly network expansion. For mobile operators, ID-based IoT offloading is therefore attractive not only as a communication technique, but also as a possible architectural tool for building quieter, more scalable, and more resource-efficient radio access networks.

\subsection{Networked consensus testing}

ID codes and the paradigm of message identification can also be extended from
point-to-point hypothesis testing to networked consensus testing. In this
setting, multiple distributed nodes hold messages or local states, and the
communication goal is, as in the ID problem, not to reconstruct all messages,
but only to decide whether they are equal. This is a naturally arising goal in
scenarios involving distributed storage systems, where replicas must verify
integrity without exchanging full contents, and in networked control or
multi-robot systems such as drone swarms, where agents may first need to
determine whether they have reached a consensus on the next action they will
take \cite{rosenberger2024consensus}. Relay-based consensus testing provides a
particularly relevant architecture: edge nodes use ID codes to transmit encoded
representations of their messages to a relay, the relay performs a statistical
equality test, and only a binary decision is broadcast back to the participants
\cite{rosenberger2024consensus}. Thus, the relay acts not as a conventional
message decoder but as a goal-oriented decision node.

Recent results show that such networked consensus-testing schemes can benefit
from the use of identification codes. For deterministic encoders,
consensus-testing codes based on deterministic ID achieve capacity for several
memoryless uplink channel models, including binary adder channels and pairs of
symmetric or erasure channels, while outperforming transmission-based and
additive linear network-coding approaches \cite{rosenberger2024consensus}. In
this architecture, the relay can perform the consensus test,
and the downlink becomes essentially negligible since it only
needs to broadcast a one-bit test result (whether there is consensus or not).
Furthermore, common randomness can substantially enlarge the message set, while
purely local randomness is less useful because the relay is either no longer able
to perform the consensus test directly or there are significant performance
impairments (i.e., code cannot scale with doubly
expoenentially)~\cite{rosenberger2025stochastic}.
Networked consensus testing, therefore, illustrates how ID principles can move
beyond single-receiver message identification toward goal-oriented network
functions, where the relevant output of communication is a decision rather than
a reconstructed payload.

To illustrate the ideas behind ID-based consensus testing, consider two nodes,
Alice and Bob, who want to determine whether they have reached a consensus,
i.e., whether their messages of interest are the same. They cannot communicate
with each other directly, so they have to do it through a relay. Without ID
codes, both nodes must transmit the full content of their message. A
goal-agnostic relay would forward messages to the other node, so they would know
whether there is consensus. A goal-aware relay, on the other hand, could verify
the messages (e.g., by comparing them) and, if there is consensus, broadcast a
single bit, e.g., indicating whether there is consensus. This would require a
large amount of traffic if the messages are large, e.g., in distributed storage
applications where nodes must verify the similarity of large files stored. This
is where ID codes can help. Alice and Bob could transmit a representation of
their message to the relay (e.g., using a hash or other ID code), and the relay
could verify consensus using the received IDs. Given the scaling of ID codes,
this would enable consensus testing for large messages with reduced traffic
exchange. This reduces latency, energy, and traffic costs. Although this example
involves two nodes and a relay, this protocol can be extended to
multiple nodes through relays in any arbitrary network topology,
without reducing significantly the achieved performance
(by applying a union bound to the errors in
\cite{rosenberger2024consensus,rosenberger2025stochastic}).

\section{Identification and security}
\label{sec:IDsecure}

Security is one of the most natural application domains for ID-based communication. In many practical systems the objective is not to reconstruct a complete message, but to determine whether a user, device, or communication request is legitimate, authorized, or relevant. This aligns closely with the ID paradigm, in which the receiver performs a binary relevance test rather than full message decoding. Efficiency alone, however, is not sufficient for the deployment of real-world networks. As ID-based communication becomes the foundational architecture for critical infrastructure, ranging from decentralized disaster-warning systems and autonomous-vehicle coordination to intelligent industrial control, the security of these sparse but highly potent control signals becomes paramount. When a network's primary function is to trigger a specific, high-stakes action rather than simply to transfer a digital file, adversarial attacks such as eavesdropping, spoofing, and traffic analysis can lead to catastrophic real-world consequences.

In recent years there has been growing interest in secure ID over noisy channels. In particular, Ahlswede and Zhang investigated ID over discrete wiretap channels \cite{Idwiretapchannels}, thereby connecting ID theory with information-theoretic security. Their work revealed a remarkable difference between secure transmission and secure ID. Whereas secrecy constraints typically reduce the achievable transmission rate, the secure ID capacity coincides with the capacity of the legitimate channel whenever secure transmission is possible at all.

This observation demonstrates that secure ID can enable entirely new communication architectures. In particular, security can be achieved \emph{by design}: whenever the secrecy capacity of the underlying wiretap channel is strictly positive, secure ID can be performed without any additional asymptotic rate loss. This substantially increases the potential of physical-layer security techniques for future communication systems, and opens new perspectives for secure and scalable 6G architectures \cite{schaefer2017information}.

\subsection{Wiretap channel model}


Modern communication networks frequently rely on broadcast and wireless links that are susceptible to eavesdropping. In standard cryptography, security relies on computational complexity. However, physical-layer security guaranties security by exploiting the physical characteristics of the communication channel rather than computational assumptions. This makes it particularly valuable in scenarios where the traditional cryptography is limited \cite{schaefer2017information,PhysicalLayser6G}.
Physical-layer security offers something stronger: information-theoretic secrecy, which holds regardless of the adversary's computational resources, even against an adversary with infinite computing power. It achieves this by ensuring that the intercepted physical signal simply does not contain enough information to resolve the uncertainty about the transmitted message. From a system perspective, the wiretap model is the canonical information-theoretic answer to the question ``why can wireless be secure at all?'' It converts a physical asymmetry (geometry, antenna pattern, noise figure) into a cryptographically meaningful one: provable secrecy with no computational assumptions and no key distribution.

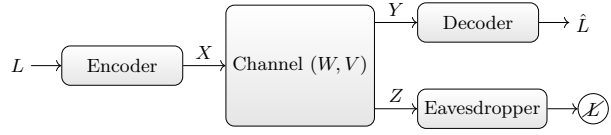
\begin{figure}[!t]
\centering
\scalebox{.8}{\tikzstyle{block} = [draw, top color=white, bottom color=white!80!gray, rectangle, rounded corners,
minimum height=2em, minimum width=2cm]
\tikzstyle{blockchannel} = [draw, top color=white, bottom color=white!80!gray, rectangle, rounded corners,
minimum height=2cm, minimum width=.1cm]
\tikzstyle{input} = [coordinate]
\usetikzlibrary{arrows}
\begin{tikzpicture}[scale= 1,font=\footnotesize]
\node[] (m) {\small $L$};
\node[block,right=.5cm of m] (enc) {\small Encoder};
\node[blockchannel, right=.7cm of enc](channel) {\small Channel
$(W,V)$};
\node[block,right= .7cm of channel.390] (bob) {\small Decoder};
\node[block,right=.7cm of channel.330] (eve) {\small Eavesdropper};
\node[right=.5cm of bob] (what) {\small $\hat{L}$};
\node[draw,circle,minimum size=.5cm,inner sep=0pt, right=.5cm of eve] (wbar) {\small $\cancel{L}$ }; 

\draw[->] (m) -- (enc);
\draw[->] (enc) -- node[above]{$X$}  (channel);
\draw[->] (channel.390) -- node[above]{$Y$} (bob);
\draw[->] (channel.330) -- node[above]{$Z$} (eve);
\draw[->] (bob) -- (what);
\draw[->] (eve) -- (wbar);
\end{tikzpicture}}
\caption{Discrete memoryless wiretap channel. Alice communicates reliably with Bob while keeping the message secret from Eve.}
\label{fig:wiretap}
\end{figure}

There are three parties: a sender Alice, a legitimate receiver Bob, and an eavesdropper Eve. We begin by recalling the classical discrete memoryless wiretap channel introduced by Wyner \cite{wyner1975wire}. Alice transmits a symbol over the primary communication medium (the \emph{main channel}) to Bob (see Fig.~\ref{fig:wiretap}), characterized by the transition probability matrix
\[
W=\{W(y|x):x\in\mathcal X,y\in\mathcal Y\},
\]
where $\mathcal X$ is the finite input alphabet and $\mathcal Y$ is Bob's output alphabet. Simultaneously, because wireless signals naturally radiate in all directions, the signal leaks to Eve through a \emph{wiretap channel}
\[
V=\{V(z|x):x\in\mathcal X,z\in\mathcal Z\},
\]
where $\mathcal Z$ is Eve's observation space. The discrete memoryless wiretap channel is therefore defined by the quintuple
\[
(\mathcal X,\mathcal Y,\mathcal Z,W,V).
\]
Each channel use is independent, so that over $n$ uses the channel kernels are the products $W^n$ and $V^n$, and the joint output distribution is $P_{YZ|X}$. In plain language: Alice chooses a symbol; the radio environment (or fiber, or optical link) sends a copy of it to Bob and a possibly different, possibly noisier copy to Eve. The question is whether Alice can encode in such a way that Bob still recovers the message reliably while Eve learns essentially nothing.



Wyner established the secrecy capacity for physically degraded wiretap channels under weak secrecy constraints \cite{wyner1975wire}.
Subsequently, Csiszár and Körner generalized the result to arbitrary discrete memoryless wiretap channels \cite{csiszar1978broadcast}. 


In Shannon's classical transmission paradigm, Alice's goal is twofold: she must ensure that Bob can reliably reconstruct her message, and that Eve learns nothing about it. To make the second requirement precise, we need to say what ``learns nothing'' means, and here the literature distinguishes two standard notions.

\paragraph{Weak, strong and semantic secrecy.}
In the following, we provide the definitions of the three secrecy notions; weak, strong, and semantic secrecy that will be used in the subsequent sections.
\emph{Weak secrecy} requires that the leakage \emph{rate} vanish, i.e., that $\tfrac{1}{n}I(L;Z^n)\to 0$ as $n\to\infty$, where $L$ is the confidential message and $Z^n$ is Eve's observation. This is a relatively mild guarantee: the total information Eve obtains may still grow without bound, as long as it grows more slowly than $n$. \emph{Strong secrecy} removes this loophole by requiring that the \emph{total} leaked information vanish, $I(L;Z^n)\to 0$, irrespective of the block length. Strong secrecy is the notion we adopt throughout, since it is the operationally meaningful one: it guarantees that, after $n$ channel uses, Eve has effectively learned nothing at all about which message was sent. We now state it formally.
A sequence of codes with confidential message $L$ taking values in $\mathcal L$ achieves \emph{strong secrecy} if
\begin{equation}
I(L;Z^n)\;\leq\;\delta_n,\qquad \delta_n\to 0\ \text{as } n\to\infty.
\end{equation}
This is stronger than per-symbol leakage: after $n$ channel uses, Eve has effectively learned nothing about which confidential message was sent.
{\emph Semantic} secrecy further strengthens this notion by requiring security against the worst-case message distribution. In semantic secrecy, the information leakage at the eavesdropper is defined as the maximization over the message distribution $P_L$ of the mutual information between the message $L$ and the channel output at the eavesdropper after $n$ uses of the channel, $Z^n$, i.e., $S_{\text{semantic}}:=\max_{P_L} I(L; Z^n)$. Semantic secrecy requires that the information leakage to Eve $S_{\text{semantic}}$ goes to zero as $n\to\infty$.







\subsection{Secure transmission codes}

We next recall the definition of \emph{randomized} transmission codes for the wiretap channel.

\begin{definition}
An $(n,M,\lambda)$ \emph{randomized} transmission code for the wiretap channel $(W,V)$ consists of a family
\[
\{(Q(\cdot|i),\mathcal D_i),\, i=1,\ldots,M\},
\]
where $Q(\cdot|i)\in \mathcal P(\mathcal X^n)$ is a probability distribution over channel input sequences and $\mathcal D_i\subseteq \mathcal Y^n$ is the decoding region associated with message $i$.

The code satisfies, for all $i\neq j$,
\begin{align}
\sum_{x^n\in\mathcal X^n} Q(x^n|i)W^n(\mathcal D_i^c|x^n)
&\leq \lambda,\\
\mathcal D_i\cap \mathcal D_j
&=\emptyset,\\
I(L;Z^n)
&\leq \lambda,
\label{eq:wiretap_strong_secrecy}
\end{align}
where $L$ is uniformly distributed over $\{1,\ldots,M\}$ and $Z^n$ denotes Eve's observation.
\end{definition}


\begin{theorem}[Wyner; Csisz\'ar--K\"orner \cite{wyner1975wire,csiszar1978broadcast}]
\label{thm:wyner_ck}
The strong-secrecy transmission capacity of a discrete memoryless wiretap channel $(W,V)$ equals
\begin{equation}
C_S(W,V)\;=\;\max_{U\to X\to YZ}\big[\,I(U;Y)-I(U;Z)\,\big],
\end{equation}
where $U$ is an auxiliary random variable forming the Markov chain $U\to X\to (Y,Z)$.
\end{theorem}

The construction behind this result is intuitive. The encoder uses two layers of randomness: an outer codebook indexed by the confidential message $m$, and an inner per-message \emph{bin} of dummy codewords. Bob, who sees less noise, can decode the bin and thereby recover $m$. Eve can decode neither and is left with essentially uniform uncertainty over the bins. This construction is sometimes called \emph{wiretap coding} or \emph{random binning}, and the inner randomization is the local randomness that confuses Eve. Put differently, instead of mapping a message to a single fixed codeword, Alice maps it to a large cloud of possible codewords and selects one at random to transmit. This deliberate injection of local randomness drowns out the information at Eve's receiver, leaving her with pure noise, while still preserving just enough structural integrity for Bob, who has the less noisy channel, to decode the message accurately.

The bracketed positive part captures the fundamental insight: the secrecy capacity is the \emph{excess} information Bob obtains over what Eve obtains, optimized over input distributions. Secrecy at the physical layer is possible exactly when Bob has an information-theoretic advantage over Eve; if Eve sees everything Bob sees, then $C_S = 0$ and confidentiality at the physical layer is hopeless. The key contrast with ID follows immediately. The gap $C(W)-C_S(W,V)$ is unrecoverable: in the classical transmission sense, security behaves like a mandatory tax on the system's efficiency. This is exactly the tax that ID, as we now show, does not pay.

A practical way to implement wiretap codes is to use a \emph{seeded modular coding scheme}, in which the security and reliability problems in the communication are addressed separately. This is achieved by designing two independent code constructions that enable us to continue using well-known error-correcting codes for reliable transmission when integrating physical-layer security into existing communication systems. That is, we can now focus on the design of novel wiretap codes to address solely the security problem, which can be regarded as pre- and post-coding operations at the transmitter and receiver, respectively. 

The first implementation of a semantic-secure wiretap code for AWGN channels \cite{torresfigueroa21a} and fading channels \cite{torresfigueroa23b} used such a scheme, and showcased how the adaptive behavior of the secrecy encoder can be fine-tuned to achieve semantic secrecy under different channel conditions and signal-to-noise ratio levels. That is, depending on the channel state information available at the transmitter, the design parameters can be dynamically chosen to maintain a desired secrecy level in terms of the distinguishing error rate \cite{frank_22a}. For an overview of existing semantic secure wiretap codes and security metrics, refer to \cite{voichtleitner24_accepted,voichtleitner25a_accepted,voichtleitner26a_accepted,voichtleitner2026b}.

At the system level, physical-layer security is regarded as complementary to cryptography for ensuring message confidentiality and they reside at different layers of the protocol stack. One additional advantage of physical-layer security is its low complexity, i.e., the latency overhead introduced by physical-layer security is negligible \cite{torresfigueroa22a}, making it a good candidate for lightweight power-constrained transceivers.

To achieve a positive secrecy rate in physical-layer security, the eavesdropper must experience a channel that is noisier than the legitimate receiver's channel. This can be realized using beam-based millimeter-wave communications, where the eavesdropper must constantly align its beam to the main or side lobes to exploit potential information leaks; failing to do so improves the secrecy performance, as experimentally demonstrated in \cite{torresfigueroa24b_accepted}.


\subsection{Secure identification codes}

Ahlswede and Zhang introduced secure ID codes for wiretap channels in \cite{Idwiretapchannels}. In contrast to secure transmission, the receiver is now not interested in reconstructing the transmitted message itself, but only in verifying whether a particular identity was sent. The eavesdropper, listening on $Z^n$, must learn nothing about which identity was active. The secure ID problem therefore asks: how many distinct identities can Alice reliably verify with Bob over a noisy channel, while keeping Eve completely ignorant of which identity was triggered?

\begin{definition}
A \emph{randomized} $(n,N,\lambda_1,\lambda_2)$ ID code for the wiretap channel $(W,V)$ is a family
\[
\{(Q(\cdot|i),\mathcal D_i),\, i=1,\ldots,N\},
\]
with
\[
Q(\cdot|i)\in\mathcal P(\mathcal X^n),
\qquad
\mathcal D_i\subseteq \mathcal Y^n,
\]
such that for all $i\neq j$ and every $\mathcal E\subseteq \mathcal Z^n$,
\begin{align}
\sum_{x^n\in\mathcal X^n}
Q(x^n|i)W^n(\mathcal D_i|x^n)
&\geq 1-\lambda_1,
\\
\sum_{x^n\in\mathcal X^n}
Q(x^n|j)W^n(\mathcal D_i|x^n)
&\leq \lambda_2,
\\
\sum_{x^n\in\mathcal X^n}
Q(x^n|j)V^n(\mathcal E|x^n)
+
\sum_{x^n\in\mathcal X^n}
Q(x^n|i)V^n(\mathcal E^c|x^n)
&\geq 1-\lambda.
\label{eq:wiretap_identification}
\end{align}
\end{definition}

The condition in \eqref{eq:wiretap_identification} ensures that the eavesdropper cannot reliably distinguish between different ID messages. It was shown in \cite{Igor} that this condition implies strong secrecy.

\begin{definition}
Let $\zeta_i$ be a \emph{rate function} as defined in \eqref{ratefct}.
\begin{itemize}
\item A secure ID rate $R$ for the wiretap channel $(W,V)$ is said to be achievable w.r.t. the \emph{rate function} $\zeta_i$ if for every $\lambda\in(0,\frac{1}{2})$ there exists an integer $n_0(\lambda)$ such that for all $n\geq n_0(\lambda)$ there exists an $(n,\zeta_i(n,R),\lambda,\lambda)$ secure ID code for $(W,V)$.

\item The secure ID capacity $C_{SID}\left( (W,V), \zeta_i\right)$ is defined as the supremum of all achievable secure ID rates.
\end{itemize}
\end{definition}

The following dichotomy theorem, proved by Ahlswede and Zhang \cite{Idwiretapchannels} (and re-derived for the compound case by Boche and Deppe \cite{CompoundChannel}), is one of the fundamental results in secure ID theory.fundamental results in secure ID theory.

\begin{theorem}[Ahlswede--Zhang Dichotomy]
Let $C(W)$ denote the Shannon capacity of the legitimate channel $W$, and let $C_S(W,V)$ denote the secrecy capacity of the wiretap channel $(W,V)$. Then the secure ID capacity is given by
\[
C_{SID}\left((W,V),\zeta_3\right) =
\begin{cases}
C(W),
& \text{if } C_S(W,V)>0,\\[1ex]
0,
& \text{if } C_S(W,V)=0.
\end{cases}
\]
\end{theorem}

This result highlights a remarkable phenomenon: whenever secure transmission is possible at a positive secrecy rate, secure ID achieves the full ID capacity of the legitimate channel with no additional asymptotic penalty. Conversely, if secure transmission is impossible, then secure ID collapses completely. The same behaviour is observed for secure storage for ID \cite{quantum2,quantum4}.

To see how such a code is built, one starts from an ordinary (non-secret) \emph{randomized} ID code on the main channel $W$, of size $N\approx 2^{2^{nC(W)}}$. Such a code exists by Ahlswede and Dueck \cite{ahlswede1989identification}, and crucially each encoder $Q(\cdot|i)$ already carries local randomness. One then uses a tiny piece of the block, say $\lceil \sqrt n\rceil$ symbols, to send, secretly via Wyner's wiretap code \cite{wyner1975wire}, an index that selects a random offset drawn from a public family of codes. The lesson is that small amounts of secrecy can have a disproportionately large impact on security.

The mechanism is worth dwelling on. When each transmitted block contains at least one bit that is unknown to the eavesdropper, that bit acts like a random switch that changes how the rest of the codeword appears from her perspective. Even though the legitimate receiver knows how to interpret the message, the eavesdropper sees many possible versions of what could have been sent. This is why the only thing that matters is whether $C_S(W,V)>0$, not how large $C_S$ actually is. The construction concatenates a secure transmission phase with a \emph{randomized} ID-encoding phase, and a channel-resolvability argument shows that Eve's output statistics become essentially independent of the identity.


For the more realistic Gaussian case, the wiretap channel can be treated in much the same manner and yields the same dichotomy result \cite{icassp,labidi2025secure}. Secure ID over the Gaussian wiretap channel is possible only when the legitimate receiver observes a better channel than the eavesdropper. In that case, the Ahlswede-Zhang dichotomy implies that secure ID achieves the full ID capacity of the legitimate Gaussian channel.

The first construction of capacity-achieving noiseless identification codes was based on Reed-Solomon codes \cite{verdu1993explicit,PerformanceAnalysis}, but its practical implementation is computationally expensive \cite{Ferrara2022Implementation}, hindering its use in production systems. A better performing code design based on Reed-Muller codes \cite{Spandri2022RMID,torresfigueroa23c} has been integrated with a semantic-secure wiretap code \cite{torresfigueroa25a_accepted}, where experimental results demonstrated that the performance of the eavesdropper for exploiting the information leakage gradually reduces when using physical-layer security, thus diminishing its ability to successfully verify the sent identity.

\subsection{Limitations of deterministic identification under secrecy constraints}

The secure ID constructions discussed above rely fundamentally on \emph{randomized} encoding. This randomness is essential for simultaneously achieving reliability, ID performance, and information-theoretic secrecy.

Such constructions are, in general, not possible for \emph{deterministic} ID codes. The reason is that information-theoretic wiretap coding inherently requires stochastic encoding in order to hide the transmitted message from the eavesdropper. In classical wiretap coding, the encoder introduces local randomness so that multiple channel input sequences may correspond to the same message. This randomization is what creates uncertainty at the eavesdropper, and it is the key mechanism for achieving strong secrecy.

\emph{Deterministic} ID codes, by contrast, associate each identity with exactly one \emph{deterministic} channel input sequence. Consequently, the encoder cannot introduce the additional local randomness needed to mask the transmitted identity, and the secrecy mechanism underlying \emph{randomized} wiretap codes cannot be transferred directly to \emph{deterministic} ID.

This distinction is particularly important because \emph{deterministic} ID already exhibits fundamentally different scaling laws from \emph{randomized} ID. While \emph{randomized} ID over discrete memoryless channels achieves double-exponential growth in the number of identifiable messages, \emph{deterministic} ID typically achieves only exponential, or intermediate super-exponential, growth, depending on the channel model.
Under secrecy constraints, the absence of encoder randomization imposes even stronger limitations.

More precisely, the secure ID constructions of Ahlswede and Zhang rely on combining:
\begin{itemize}
    \item a secure \emph{randomized} transmission code for the wiretap channel, and
    \item an ID construction based on \emph{randomized} encoding.
\end{itemize}

If the secrecy capacity of the underlying wiretap channel is zero, then no secure \emph{randomized} transmission layer can be established, and the resulting secure ID capacity collapses to zero. For \emph{deterministic} ID, the situation is even more restrictive, because the encoder itself cannot generate the local randomness required for secrecy generation.

From a system perspective, this observation highlights a fundamental trade-off between implementation simplicity and achievable security guarantees. \emph{Deterministic} ID codes are attractive because of their lower decoding complexity and explicit constructions. In addition, they also show improved performance in very noisy environments compared to classical Shannon transmission, which has been demonstrated experimentally by implementing DI using different concatenated codes for binary symmetric channels \cite{Vorobyev2024DeterministicID}, as well as using concatenated Reed-Solomon codes for the AWGN channel \cite{torresfigueroa26a_accepted}. However, \emph{randomized} ID provides substantially stronger capabilities in adversarial settings, particularly in the presence of eavesdroppers and privacy constraints.

For any application that demands robust security, whether medical implants, automated financial trading, or military swarm coordination, \emph{randomized} ID is not merely a theoretical optimization; it is an absolute mathematical necessity. The local randomness is not an artefact of the proof but the fundamental physical resource that powers the security. Designers should therefore treat encoder randomness as a first-class resource, on the same footing as bandwidth and energy.

A simple RFID tag system, for example, might use a \emph{deterministic} wake-up sequence per tag. This has the advantage of requiring no secret keys, but if an attacker learns a tag's codeword, it can spoof the system. \emph{Randomized} schemes avoid this by effectively changing the codeword each time. This is precisely where a physically unclonable function (PUF), or a quantum PUF (QPUF), earns its keep: it supplies the local entropy that \emph{randomized} ID requires, without that entropy ever having to be stored as a key \cite{JuliFarr,quantum7}.

\subsection{Identification under semantic effective secrecy}

In classical secure ID, as modelled by the wiretap channel, the primary goal is to prevent Eve from learning \emph{which} specific identity was sent. In that model, however, Eve still knows that some communication is taking place; she can easily detect the bursts of radio energy across the spectrum. In many high-stakes operational environments, such as military coordination, covert intelligence gathering, or anti-front-running in financial markets, the mere detection of a signal is enough to compromise a mission. Beyond classical secrecy constraints, then, recent work has investigated ID under stronger notions that combine secrecy and stealth requirements simultaneously. In particular, the work in \cite{rosenberger2023capacity} studies ID over discrete memoryless wiretap channels under the criterion of \emph{semantic effective secrecy}. This criterion combines semantic secrecy with stealthy communication, and therefore guarantees not only that the transmitted message remains secret, but also that the communication itself is statistically indistinguishable from innocent or expected communication behaviour.

The communication model consists of a legitimate sender Alice, a legitimate receiver Bob, and an adversarial observer Willie. Bob performs an ID test, while Willie attempts both to detect whether meaningful communication is occurring and to infer information about the transmitted message. Consequently, the communication system must simultaneously satisfy:
\begin{itemize}
    \item reliability for the legitimate receiver,
    \item semantic secrecy of the transmitted message, and
    \item stealth of the communication process itself.
\end{itemize}

The semantic secrecy and stealth requirements can be combined into a single KL-divergence requirement:
\begin{equation}
D\!\big(P_{Z^n}\,\big\|\,Q_{Z^n}\big)\;\leq\;\delta_n\to 0.
\end{equation}
Because relative entropy upper-bounds variational distance, this single requirement guarantees both semantic secrecy (Eve cannot tell which message was sent) and stealth (Eve cannot tell whether any message was sent, i.e. she cannot distinguish $P_{Z^n}$ from an innocent background reference distribution $Q_{Z^n}$, typically the channel output induced by the ``off'' input). A small number of randomly chosen inputs suffices to make the induced output distribution match a target arbitrarily closely. To fool Willie successfully, Alice cannot simply transmit encrypted data; she must carefully shape the entire statistical distribution of her physical transmission so that the channel output observed by Willie perfectly mimics a predetermined innocent reference distribution $Q_{Z^n}$.

Combining secrecy and stealth in this way leads to the notion of \emph{effective secrecy}. A central observation in \cite{rosenberger2023capacity} is that, under semantic secrecy constraints, semantic stealth and semantic effective secrecy become equivalent. Hence ensuring stealth automatically guarantees semantic secrecy as well.

\subsubsection*{The AOST construction}
This extraordinary level of stealth is achieved through a technique called \emph{approximation of output statistics and transmission} (AOST), which is deeply related to the information-theoretic concepts of channel resolvability and soft-covering. In an AOST framework, Alice uses incredibly dense layers of local randomness to synthesise a signal structure that looks, to Willie's sensors, exactly like thermal noise or ambient, expected cellular chatter. The objective is, simultaneously, to communicate reliably with the legitimate receiver and to approximate a prescribed reference output distribution at the adversary.

The paper \cite{rosenberger2023capacity} derives both lower and upper bounds on the effectively secret ID capacity. The central achievability result is summarised in the following corollary.

\begin{corollary}[\cite{rosenberger2023capacity}]
For any reference distribution $Q_Z \in \mathcal P(\mathcal Z)$, the \emph{effectively secret} ID capacity of the discrete memoryless wiretap channel $(W_{Y|X},W_{Z|X})$, with respect to the \emph{rate function} $\zeta_3(n,R)=2^{2^{nR}}$, is lower-bounded by
\begin{equation}
C_{{ESID}}\left((W_{Y|X},W_{Z|X}),Q_Z,\zeta_3\right)
\geq
\max_{P_{U,X}\in\mathcal P_{{ESID}}}
I(U;Y),
\end{equation}
where
\begin{equation}
\mathcal P_{{ESID}}
=
\left\{
P_{U,X} :
\begin{array}{l}
U-X-(Y,Z),\\
I(U;Y)\geq I(U;Z),\\
P_X \circ W_{Z|X}=Q_Z
\end{array}
\right\}.
\end{equation}
\end{corollary}

This result shows that positive effectively secret ID rates are achievable whenever the legitimate receiver has a sufficient informational advantage over the adversary, while simultaneously satisfying the stealth constraint induced by the reference distribution.

A matching converse, establishing a fundamental upper limit on the effectively achievable secret ID rates, is given below.

\begin{corollary}[\cite{rosenberger2023capacity}]
For every discrete memoryless wiretap channel,
\begin{equation}
C_{{ESID}}\left((W_{Y|X},W_{Z|X}),Q_Z,\zeta_3\right )
\leq
\max_{P_{U,X}\in\mathcal P_{{ESID}}}
I(U;Y).
\end{equation}
\end{corollary}

For general wiretap channels, a gap between the lower and upper bounds may remain. The bounds coincide, however, for an important class of channels, namely the more capable wiretap channels.

\begin{theorem}[\cite{rosenberger2023capacity}]
For more capable discrete memoryless wiretap channels,
\begin{equation}
C_{{ESID}}\left((W_{Y|X},W_{Z|X}),Q_Z,\zeta_3\right)
=
\max_{P_X\in\mathcal P_{{ESID}}}
I(X;Y),
\end{equation}
where
\begin{equation}
\mathcal P_{{ESID}}
=
\left\{
P_X :
I(X;Y)\geq I(X;Z),
\;
P_X \circ W_{Z|X}=Q_Z
\right\}.
\end{equation}
\end{theorem}

This theorem provides a complete characterization of the effectively secret ID capacity for more capable wiretap channels. In contrast to classical secure ID, the stealth requirement introduces an additional output-statistics approximation constraint, so that not every secrecy-achieving input distribution is admissible.

A key conceptual insight is that effective secrecy imposes substantially stronger constraints than classical semantic secrecy. In conventional secure ID, only a small part of the codeword structure must remain secret. Under semantic effective secrecy, by contrast, the \emph{entire communication process} must statistically resemble innocent channel behaviour, so that the complete channel output observed by the adversary must remain indistinguishable from the reference distribution.

From a system perspective, these results are highly relevant for future post-Shannon communication systems. In many emerging applications it is not enough to hide the transmitted information alone; the communication itself should also remain difficult to detect. The combination \emph{reliability $+$ semantic secrecy $+$ stealth} fits a number of real-world settings precisely:
\begin{itemize}
    \item \textit{Covert command-and-control}: a military operator broadcasts an order; only authorized field units should identify themselves as the addressee; and an adversary must learn neither the order's content (semantic secrecy) nor that an order was issued at all (stealth).
    \item \textit{Stealthy activation of distributed agents}: a controller wants to activate a subset of $10^6$ smart-dust sensors. Each sensor needs to identify whether it is in the activation set, while an adversary must neither learn the activation list nor detect that an activation campaign is in progress.
    \item \textit{Secure IoT coordination}: plain encryption hides payloads but not timing; effective-secrecy ID shapes the output statistics so that even traffic analysis cannot extract the membership of an activation set.
    \item \textit{Privacy-preserving wireless systems}: in military and homeland-security contexts, \emph{low probability of intercept} (LPI) and \emph{low probability of detection} (LPD) are mission-critical, and effective-secrecy ID provides both.
\end{itemize}

Overall, this work establishes a strong connection between ID theory, physical-layer security, stealth communication, and channel resolvability, and it highlights the fundamental role of stochastic encoding and output-statistics approximation in secure task-oriented communication systems.

\subsection{Secure randomized identification with feedback and sensing}


Recent work has extended secure ID to joint sensing and communication scenarios over state-dependent wiretap channels with noiseless feedback \cite{Gholamian2026SecureFeedback}. In many emerging 6G applications, communication systems are required not only to transmit or identify information, but also to estimate environmental states such as interference levels, channel conditions, or physical parameters. This naturally connects ID theory with integrated sensing and communication (ISAC).

The model under consideration is a state-dependent wiretap channel
\[
\big(\mathcal X\times\mathcal S,\; W_S(y|x,s),\; V_S(z|x,s),\; \mathcal Y,\; \mathcal Z\big),
\]
with i.i.d.\ random states $S^n\sim P_S$ that are unknown a priori to both parties. Bob's channel output $Y_t$ is fed back noiselessly to Alice with strictly causal delay. Alice plays a double role: she encodes an identity $i\in\mathcal N$ as a stochastic input $X^n$, and she runs an estimator $\hat S^n$ of the channel state from $(X^n,Y^n)$. Eve observes $Z^n$ through her own state-dependent channel $V_S$, and must learn neither which identity is active nor the state $S$ (which may itself reveal physical information about the environment).

The operational sequence is as follows. Alice transmits an ID signal. The physical channel state alters the signal as it travels through the air. Bob receives the output and performs his local ID hypothesis test. Eve receives the leaked output and attempts to crack the identity. Bob simultaneously sends a noisy feedback signal back to Alice. Alice uses this feedback loop both to estimate the physical channel state (the sensing component) and to generate, dynamically, the common randomness for her subsequent ID transmission.

The sensing objective is quantified through an average distortion constraint. 
Let
\[
d(S_t,\hat S_t)
\]
be a distortion measure. The average sensing distortion over a block length $n$ is then defined as
\begin{equation}
\bar d^{(n)}
=
\frac{1}{n}
\sum_{t=1}^n
\mathbb E[d(S_t,\hat S_t)], \label{eq:AverageDistSecure}
\end{equation}
where the expectation is over the joint distribution of $(S_{t},\hat{S}_{t})$ conditioned by the ID message $i \in \{1,\ldots,N\}$.
An ID rate-distortion pair $(R,D)$ is achievable if reliable and secure \emph{randomized} ID is possible while satisfying the sensing constraint
\[
\limsup_{n \to \infty } \bar d^{(n)} \leq D.
\]

The main contribution is a characterization of the \emph{randomized} secure ID capacity under sensing constraints.

\begin{theorem}[\cite{Gholamian2026SecureFeedback}]
If
\[
C_S(W_S,V_S)>0,
\]
then the \emph{randomized} secure ID capacity-distortion function under the average distortion constraint $\limsup_{n \to \infty } \bar d^{(n)} \leq D$ and w.r.t. the \emph{rate function} $\zeta_3(n,R)=2^{2^{nR}}$ is given by
\begin{equation}
C_{SID}(D,(W_S,V_S),\zeta_3)
=
\max_{P\in\mathcal P_D}
H\!\left(
\sum_{x\in\mathcal X}
P(x)\,
\mathbb E[W_S(\cdot|x,S)]
\right),
\end{equation}
where
\[
\mathcal P_D
=
\left\{
P\in\mathcal P(\mathcal X) \colon
\limsup_{n \to \infty } \bar d^{(n)} \leq D
\right\}
\]
denotes the set of admissible input distributions satisfying the sensing-distortion constraint.
\end{theorem}

The intuition is that the capacity is entropy-driven: it equals the entropy of Bob's marginal output distribution under the most uncertainty-producing admissible input, subject to the sensing-distortion constraint. Sensing and ID are coupled through the input distribution $P$, since the same waveform must produce both enough entropy at Bob (for ID) and enough informativeness for the state estimator. Highly structured, radar-like pulses sense well but constrain randomness, so the trade-off is genuine. The dichotomy tells us that secrecy imposes no additional asymptotic penalty once the underlying state-dependent wiretap channel admits some positive secrecy capacity. Feedback plays a \emph{dual role}: it generates common randomness between Alice and Bob (the engine of the doubly-exponential ID code size), and it carries the sensing returns from which Alice estimates the state.

This result reveals several remarkable properties at once. First, even in the presence of secrecy and sensing constraints, the ID problem remains fundamentally entropy-driven: the achievable ID rate is determined by the entropy of the induced output distribution at the legitimate receiver. Second, the sensing requirement restricts the admissible input distributions through the distortion constraint, so that sensing and ID become fundamentally coupled, and improving sensing accuracy may reduce the achievable ID rate. Third, secrecy introduces no additional asymptotic penalty whenever the secrecy capacity of the underlying wiretap channel is strictly positive: secure ID with sensing achieves the same capacity--distortion trade-off as the corresponding non-secure problem.

The achievability proof combines several important information-theoretic techniques: feedback-generated common randomness, \emph{randomized} ID coding, channel-resolvability and soft-covering arguments, state estimation under distortion constraints, and short secure wiretap transmission phases. In particular, noiseless feedback enables the transmitter and receiver to generate common randomness from the observed channel outputs. This common randomness is then used to construct \emph{randomized} ID codes with double-exponential message growth, while a short wiretap transmission phase ensures strong secrecy with respect to the eavesdropper.

This creates a strong connection between ID, sensing, and feedback communication. The resulting framework therefore provides a natural information-theoretic model for future task-oriented and ISAC-based communication systems. The architecture admits a clean flow-chart description: Alice senses an event and encodes an ID code; Bob performs a simple yes/no test, triggering an action when matched; and Eve hears the channel output but, with a proper code, cannot identify the target index.

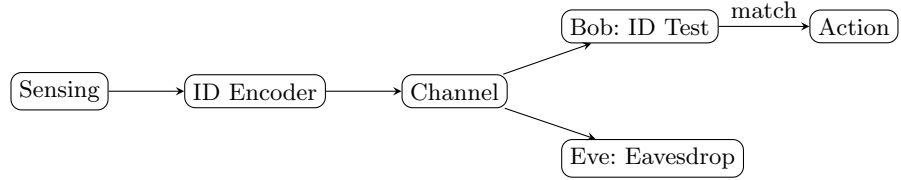
\begin{figure}[h]
\centering
\begin{tikzpicture}[->, >=stealth, node distance=1cm, every node/.style={font=\small}]
\node[draw, rounded corners] (S) {Sensing};
\node[draw, rounded corners, right=of S] (E) {ID Encoder};
\node[draw, rounded corners, right=of E] (C) {Channel};
\node[draw, rounded corners, above right=of C, yshift=-0.3cm] (B) {Bob: ID Test};
\node[draw, rounded corners, below right=of C, yshift=0.3cm] (V) {Eve: Eavesdrop};
\node[draw, rounded corners, right=1.2cm of B] (A) {Action};
\draw (S) -- (E);
\draw (E) -- (C);
\draw (C) -- (B);
\draw (C) -- (V);
\draw (B) -- node[above]{match} (A);
\end{tikzpicture}
\caption{Flow of a secure architecture: sensed information feeds the ID encoder; Bob performs a yes/no test and triggers the action only on a positive ID; Eve cannot recover the identity.}
\label{fig:jidas_flow}
\end{figure}

This setting is particularly relevant for integrated sensing and communication (ISAC), adaptive wireless networks, secure autonomous systems, distributed control architectures, task-oriented communication, and low-latency 6G coordination systems. Overall, the work demonstrates that secure ID, sensing, and feedback can be unified within a single information-theoretic framework, while preserving the characteristic double-exponential ID scaling laws.

As telecommunications transitions into the era of 6G, classical--quantum networks, and integrated sensing ecosystems, the mathematical realities of secure ID will increasingly dictate the foundational architecture of network control planes \cite{amiriquantum,amiriquantum2}. The appeal of simple \emph{deterministic} systems must be weighed against their inherent, unfixable vulnerability to wiretapping. The generation and distribution of common randomness, harvested dynamically through environmental sensing and feedback loops, will become a primary currency of secure physical-layer design.

\section{Code construction and implementation}
\label{sec:code}


A central question for ID-based communication is whether ID codes can be realized in practical communication systems with finite blocklength, moderate complexity, and implementable encoding and decoding procedures. 
This is no longer only asymptotic theory.
\emph{Randomized} ID achieves the characteristic double-exponential scaling over discrete memoryless channels, but practical implementations may require substantial resources for randomness generation, synchronization, and large-scale hypothesis testing at the receiver. \emph{Deterministic} ID does not provide the same asymptotic scaling behavior over discrete memoryless channels, but it can still achieve significant gains over classical transmission while offering substantially simpler implementations.

Therefore, we begin by considering \emph{deterministic} ID codes, since \emph{deterministic} encoders avoid the need for local encoder randomization, thereby providing a more direct path to practical implementations.

\subsection{Realization of deterministic identification codes}

While early research on \emph{deterministic} ID primarily established infor\-mation-theoretic existence results, recent advances have shown that \emph{deterministic} ID codes can also be constructed explicitly with polynomial complexity and implemented in practical communication systems. In particular, constructive \emph{deterministic} ID schemes based on concatenated coding architectures have been developed for discrete memoryless channels. For example, compared to the classical Shannon transmission, improved performance has been demonstrated in very noisy environments. This has been validated experimentally through implementations of \emph{deterministic} ID using different concatenated codes for memoryless channels \cite{Vorobyev2024DeterministicID,torresfigueroa26a_accepted}.

The key idea is to replace purely existential random-coding arguments with structured and efficiently realizable code constructions. A typical approach combines an inner code with prescribed type and minimum Hamming distance properties with an outer Reed--Solomon code. The resulting concatenated \emph{deterministic} ID code exhibits linear minimum-distance growth while maintaining polynomial complexity for both code construction and encoding.

These developments are particularly important because they provide explicit finite blocklength \emph{deterministic} ID codes. Consequently, \emph{deterministic} ID is no longer limited to asymptotic achievability arguments, but can instead be implemented using concrete codebooks and practically realizable decoding procedures. For instance, decoding regions may be represented by Hamming balls centered on \emph{deterministic} codewords, leading to explicit tradeoffs between the missed-ID and false-ID probabilities.

A particularly remarkable observation is that successful ID may still be achievable in communication regimes where reliable classical message transmission already becomes impossible. In other words, the receiver may still reliably answer the binary question of whether a specific identity was transmitted, even when full message reconstruction suffers from an error probability close to one. This phenomenon highlights one of the fundamental operational differences between classical transmission and ID.

\begin{figure}
    \includegraphics[width=\linewidth]{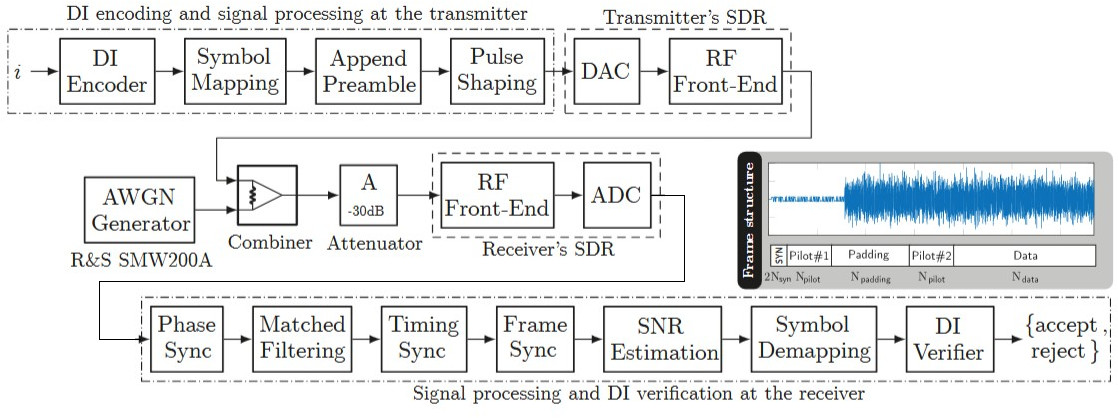}
    \caption{
    Hardware setup from \cite{Vorobyev2024DeterministicID}, including the bit- and signal-processing blocks at the transmitter and receiver. The correspondence between the synchronized time-domain received signal and the employed frame structure is also illustrated (not to scale). The padding region compensates for small timing jitters caused by the activation of the AWGN noise source.}
    \label{fig:DISetup}
\end{figure}

The practical feasibility of \emph{deterministic} ID has furthermore been demonstrated experimentally using software-defined-radio platforms \cite{Vorobyev2024DeterministicID}. In these implementations, \emph{deterministic} ID codewords are mapped onto conventional modulation schemes, such as BPSK, and transmitted over noisy wireless channels (see Fig.~\ref{fig:DISetup}). At the receiver, synchronization, symbol recovery, and the \emph{deterministic} ID test are performed successively. The experiments demonstrate that \emph{deterministic} ID can be integrated into realistic wireless communication architectures and implemented with moderate system complexity.

Overall, \cite{Vorobyev2024DeterministicID} proposed a capacity-achieving construction of \emph{deterministic} ID codes for discrete memoryless channels. The resulting codes admit efficient construction and practical encoding algorithms. Moreover, the ID error probability was analyzed for the binary symmetric channel at finite blocklengths, demonstrating that reliable ID remains possible even when the corresponding transmission error probability is close to one. Finally, the theoretical findings were validated experimentally through software-defined-radio implementations of \emph{deterministic} ID systems.

\subsection{Realization of randomized identification codes}
\label{sec:realizationRanIDCodes}


Although most of ID theory is asymptotic, several practical constructions have
been proposed, including structured codes such as Reed--Solomon and Reed--Muller
codes, as well as combinatorial designs and pseudo-random implementations
\cite{Ferrara2022Implementation,vonlengerke2023digital,vonlengerke2025tutorial,perotti2024wakeup,verdu1993explicit,PerformanceAnalysis}.
These approaches demonstrate that message ID paradigms can be realized under
finite blocklength, hardware, and latency constraints.

To the best of our knowledge, most explicit constructions so far are based on the separation construction
that originated in \cite{Feedbackidentification}, where a transmission code
is used to correct errors due to channel noise, and on top of that, one uses a
compression that is specific to the ID task. The compression works by
hashing a random number, called a seed $S$, and the message $m$
with an $\epsilon$-almost universal hash function, i.e., a function
that maps a random seed and a message to a hash such that for
any two distinct messages, the probability of a hash collision is below $\epsilon$
(see Definition~\ref{def:auhf}).

The construction of an ID code using hash functions is depicted in Fig.~\ref{fig:CodeConcatenation}.
A hash function is employed in an ID code as follows.
The sender, Alice, hashes her message $i \in \mathcal M = \set{ 1,\dots, N}$
with a random seed $S$ from the set $\mathcal S$, using a hash
function $h : \mathcal M \times \mathcal S \to \mathcal H$.
She transmits the pair $(S, h(i, S))$ to the receiver, Bob.
He hashes his message $j$ with the same seed and compares his hash value,
$h(j, S)$, against Alice's hash $h(i, S)$.
If the hashes are equal, he declares that the messages are as well, i.e., $i = j$;
otherwise he declares that $i \ne j$.
The error probability of this test is bounded by the error of the
underlying transmission code plus the collision probability of the hashes,
\begin{equation}
  \max_{i \ne j} P_S \mleft[ h(i, S) = h(j, S) \mright] \le \epsilon .
\end{equation}
In general, such hash functions can be constructed from a linear codes
~\cite{bierbrauerJohanssonKabatianskiiSmeets1993hash_geom_codes}
of dimension $\log N$ and length $\mathcal S$ over an alphabet $\mathcal H$.
Every message corresponds to one codeword $c_i$, and the random seed $S$
selects the position, i.e., the $S$-th symbol $h(i, S) = c_{i,S}$.
Then, the collision probability is determined by the minimal Hamming distance
$d_{\min}$ of
the code, i.e., the minimum for any pair of codewords of the number
of positions where the codewords have distinct symbols,
\begin{equation}
  \max_{i \ne j} P_S \mleft[ h(i, S) = h(j, S) \mright]
  = 1 - \frac {d_{\min}} {|\mathcal S|} .
\end{equation}

The hash function in \cite{Feedbackidentification} was constructed via random coding.
The first explicit construction of capacity-achieving noiseless ID codes was
given in \cite{verdu1993explicit}, in which the ID code is constructed by
concatenating Reed-Solomon codes. While they did not describe their codes in
terms of hash functions,
the performance analysis of this concatenation \cite{PerformanceAnalysis}
revealed this connection \cite[Observation 2]{PerformanceAnalysis} and showed
that this construction demands significant computation complexity at the sender
and receiver. Later on, the authors in \cite{SpandriReedMuller} extended
the previous results to Reed-Muller codes and showed that it is possible to
implement ID with latency comparable to current transmission speeds and with
arbitrarily small error. A review of ID code constructions was presented in
\cite{TopicalReviewIDcodes}. The authors compared different ID codes in the
finite-parameter regime using three main criteria: cue size (tag), ID size, and
error probability.
\begin{itemize}
  \item Some alternative methods (like special hash functions or optical
    codes) don’t perform as well as the standard PPM-based approach—they either
    have higher error rates or require more data.
  \item  It is not the CWC initialization but rather the employed linear block
    code that primarily determines the performance of an ID code in terms of the
    three investigated metrics.
  \item Reed-Solomon codes perform very well but are computationally expensive.
  \item Reed-Muller codes are cheaper to compute but have worse error performance.
\end{itemize}
 
\subsection{Reed--Muller-based identification codes}

Beyond asymptotic capacity results, practical implementation complexity
plays a central role in the realization of identification-based communication
systems. Early capacity-achieving constructions based on concatenated
Reed--Solomon codes demonstrated the theoretical feasibility of
identification coding, but also revealed significant computational
complexity and latency challenges for practical implementations.

To address these limitations, Reed--Muller-based identification
codes were proposed in \cite{Spandri2022RMID}. The construction
extends the classical polynomial-based identification framework by
using multivariate Reed--Muller polynomials instead of univariate
Reed--Solomon polynomials. This enables substantially larger
identity spaces while maintaining moderate finite-field sizes and
manageable computational complexity.

The basic idea is to interpret each identity as a multivariate
polynomial
$
p_w : \mathbb{F}_q^m \rightarrow \mathbb{F}_q
$
of bounded degree. During identification, the encoder selects a
random challenge vector and transmits both the challenge and the
corresponding polynomial evaluation (tag). The receiver recomputes
the tag locally and performs a binary verification test.
This realization naturally fits the general identification framework
based on randomized tag evaluation and verifier sets.

An important advantage of the Reed--Muller approach is that the
construction can asymptotically achieve identification capacity
while enabling substantially more efficient implementations than
previous Reed--Solomon-based schemes. In particular, recursive
evaluation strategies and optimized finite-field arithmetic allow
a significant reduction of computational latency.

Moreover, the construction provides a flexible trade-off between
identification rate, false-accept probability, and computational
complexity. While larger field sizes improve the achievable
identification rates and reduce the error probability, practical
implementations often operate with moderate field sizes and employ
multiple challenges to reduce the verification error exponentially.

Experimental implementation results reported in \cite{Spandri2022RMID}
demonstrate that Reed--Muller-based identification can achieve
very large identification spaces while maintaining practical
encoding and verification times. The obtained latency can become
comparable to classical transmission approaches despite the
double-exponential identification scaling.

\begin{figure}[ht]
    \centering
    \includegraphics[width=0.9\linewidth]{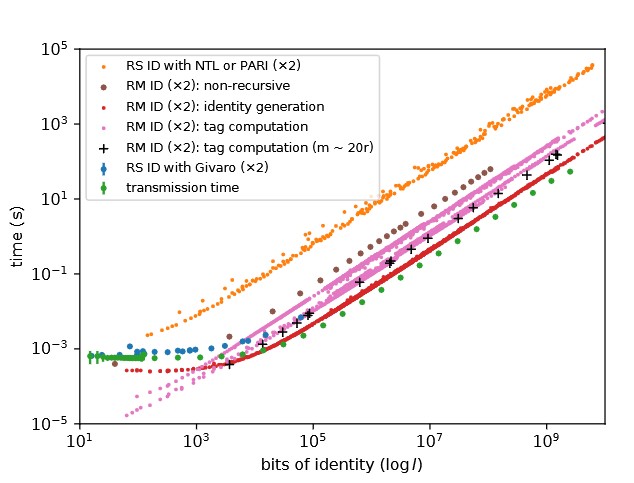}
    \caption{
    Computation time comparison between Reed--Muller-based
    identification and Reed--Solomon-based identification constructions.
    The figure illustrates the achievable reduction in latency for
    large identification spaces. Adapted from \cite{Spandri2022RMID}.}
    \label{fig:rm_latency}
\end{figure}

Fig.~\ref{fig:rm_latency} illustrates the computational latency
of Reed--Muller-based identification schemes compared to both
Reed--Solomon-based identification constructions and classical
transmission approaches. The results demonstrate that optimized
Reed--Muller implementations significantly reduce the encoding
and verification complexity while still supporting extremely
large identification spaces. In particular, the figure highlights
that recursive polynomial evaluation and optimized finite-field
operations substantially improve the practical feasibility of
identification-based communication systems.

\begin{figure}[ht]
    \centering
    \includegraphics[width=0.9\linewidth]{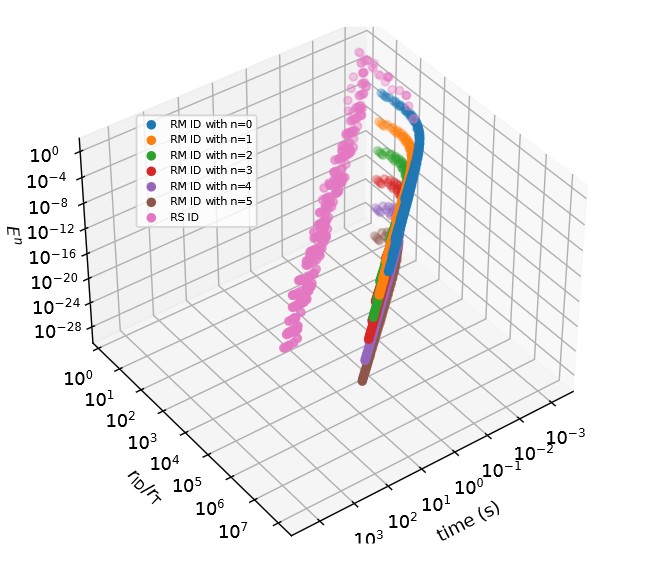}
    \caption{
    Trade-off between identification rate, latency, and
    false-accept probability for Reed--Muller-based identification
    schemes with multiple challenges. Adapted from
    \cite{Spandri2022RMID}.}
    \label{fig:rm_tradeoff}
\end{figure}

Fig.~\ref{fig:rm_tradeoff} further illustrates the trade-off
between identification rate, computational latency, and
false-accept probability. By employing multiple independent
challenges, the false-accept probability can be reduced
exponentially, while only moderately increasing the computational
complexity. This demonstrates that Reed--Muller-based
identification schemes provide a flexible design framework
that allows balancing reliability, latency, and scalability
depending on the application requirements.

These implementation results are particularly relevant for
embedded systems, monitoring architectures, and low-latency
control applications, where both scalability and computational
efficiency are essential. More generally, they demonstrate that
post-Shannon identification concepts can be realized in practice
without prohibitive implementation overhead.

These results indicate that algebraic identification-code
constructions can provide practically relevant implementations
for post-Shannon communication architectures, particularly in
low-latency control, monitoring, and embedded communication
systems.

\subsection{Efficient tag evaluation and embedded implementations}

While the information-theoretic properties of ID codes are well understood,
their practical realization requires efficient methods for tag generation,
verification, and low-complexity decoding. In particular, practical
post-Shannon communication systems must support efficient processing on
resource-constrained embedded hardware while maintaining low latency and
low energy consumption.

The work \cite{Ali2606:Quantitative} investigated practical implementations of
deterministic and randomized ID constructions on embedded platforms,
including Reed-Solomon ID (RSID), Reed-Muller ID (RMID),
concatenated Reed-Solomon ID constructions (RS2ID),
and hash-based ID schemes.
The considered implementations target verification-oriented communication
tasks in which the receiver only evaluates whether a message or event is
relevant, rather than reconstructing a complete payload.

A central practical component of many ID schemes is the efficient
evaluation of compact tags. Instead of reconstructing the full codeword,
the receiver only evaluates a small subset of symbols or algebraic
relations that are sufficient for the corresponding ID test.
This principle significantly reduces both computational complexity
and communication overhead.

\begin{figure}[ht]
    \centering
    \includegraphics[width=0.72\linewidth]{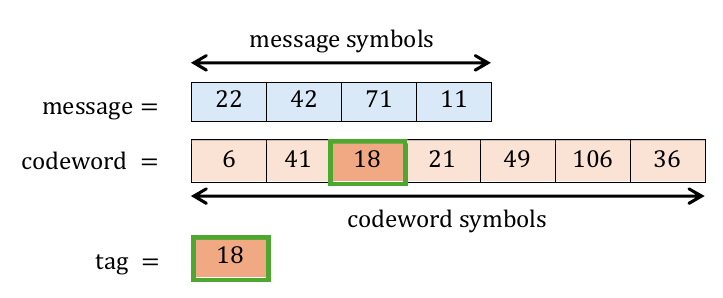}
    \caption{Illustration of tag generation in practical ID systems.
    A message vector is encoded into a codeword over a finite field,
    while only selected tag symbols are evaluated during the ID process.
    Adapted from \cite{Ali2606:Quantitative}.}
    \label{fig:tag_evaluation}
\end{figure}

Fig.~\ref{fig:tag_evaluation} illustrates the general concept of
tag generation in practical ID systems.
A message vector is mapped to a codeword over a finite field,
while only a small subset of symbols is extracted and evaluated
during the ID process.

The practical realization of such ID constructions is closely connected
to finite-field arithmetic and efficient polynomial evaluation.
For binary extension fields, tag computation can be implemented using
carryless multiplication and modular reduction operations over
Galois fields. Modern embedded processors additionally provide
hardware acceleration for such operations through SIMD instructions,
for example, ARM NEON intrinsics or dedicated polynomial
multiplication instructions~\cite{gopal2009fast}.

To evaluate the practical feasibility of these constructions,
embedded-system, the authors in~\cite{Ali2606:Quantitative}
deployed them on Raspberry Pi platforms
using optimized C++ implementations and GNU Radio integration. The considered implementations
support practical Reed--Solomon ID codes, Reed--Muller ID codes,
concatenated constructions, and hash-based schemes for different
finite-field sizes.

\begin{figure}[ht]
    \centering
    \includegraphics[width=0.82\linewidth]{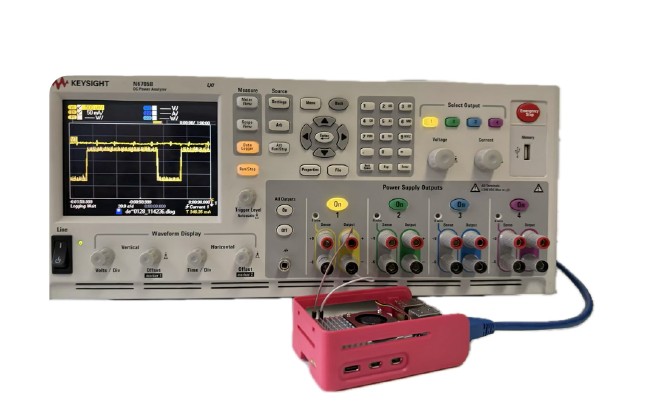}
    \caption{Experimental setup for practical latency and energy measurements
    of ID-code implementations on embedded hardware platforms.
    Adapted from \cite{Ali2606:Quantitative}.}
    \label{fig:measurement_setup}
\end{figure}

Fig.~\ref{fig:measurement_setup} illustrates the corresponding
experimental setup used for practical latency and energy measurements.
The setup combines embedded ARM-based hardware with high-resolution
power measurements to evaluate the practical efficiency of ID processing.
The power supply powers the Raspberry Pi and reports the voltage and current it delivers at a high sampling rate. The authors use this information to estimate the energy consumption of running the different ID implementations for a large number of iterations. Based on the number of iterations and total energy consumption, the authors can provide metrics such as energy per information bit for encoding and decoding different ID codes.

The obtained results demonstrate that practical ID processing can
be realized with very low latency and low energy consumption,
particularly for short verification-oriented messages.
The measurements further indicate that compact tag-based processing
can significantly reduce computational effort compared to
classical payload-oriented communication schemes.

\begin{figure}[ht]
    \centering
    \includegraphics[width=0.82\linewidth]{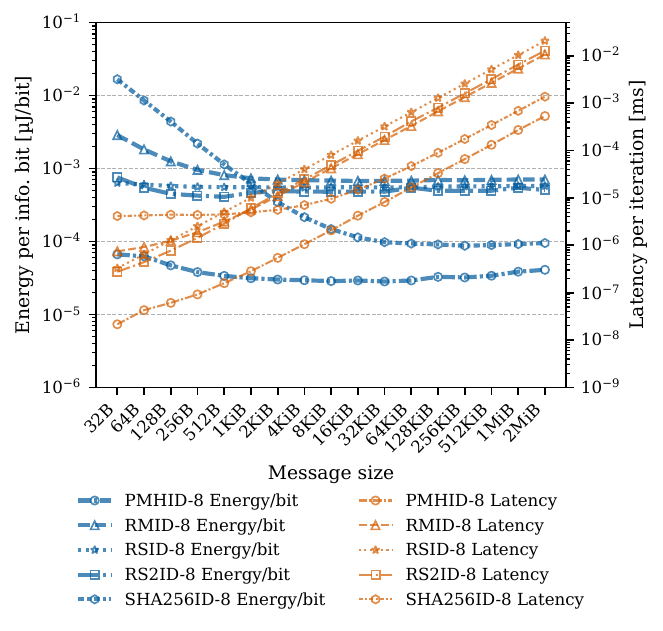}
    \caption{Representative energy-latency tradeoff for encoding and 8-bit tag of practical
    ID-code implementations on embedded hardware.
    Adapted from \cite{Ali2606:Quantitative}.}
    \label{fig:energy_latency_tradeoff}
\end{figure}

Fig.~\ref{fig:energy_latency_tradeoff} shows a representative example
of the measured energy-latency tradeoff for several practical ID
constructions. In particular, the results indicate that
an efficient tag evaluation can achieve very small energy per information bit
values even on low-power embedded systems.

The presented implementation results, therefore, demonstrate that
ID-based communication is not only theoretically attractive,
but can also be realized efficiently on practical hardware
platforms with stringent energy and latency constraints.

\section{Open challenges and future research directions}
\label{sec:open}
ID coding has evolved from a fundamental information-theoretic concept into a promising design principle for post-Shannon communication architectures. The preceding sections have shown that ID-based communication is particularly attractive when the communication task can be formulated as a sparse relevance test rather than as full message reconstruction. This applies to monitoring systems, special-purpose storage, JIDAS, and mobile-network control.

At the same time, several theoretical and practical challenges remain open before ID-based communication can become a mature systems technology. Most existing results are asymptotic, while practical deployments are constrained by finite blocklength, latency, computational complexity, synchronization, security, and integration with existing communication protocols. This section summarizes the most important open challenges and outlines promising directions for future research.
\subsection{Finite blocklength identification}

One of the most important open problems is the characterization of finite blocklength ID performance. Classical ID theory is dominated by asymptotic scaling laws, where the number of identifiable messages may grow doubly exponentially with the blocklength. In practical systems, however, the relevant regime is often short-packet communication under strict latency, energy, and bandwidth constraints.

A finite blocklength theory of ID should quantify how many identities can be reliably distinguished for a given blocklength $n$, error probabilities $(\lambda_1,\lambda_2)$, and channel model. In analogy to modern finite blocklength Shannon theory, one may ask whether ID admits meaningful second-order or dispersion-type characterizations. Such results would be highly relevant for ultra-reliable low-latency communication, massive IoT, and event-driven control systems, where decisions must be taken from short observations.

A central challenge is that ID involves two different error events. Errors of the first kind correspond to missed detections, while errors of the second kind correspond to false positives. Practical systems may weight these errors very differently. For instance, a missed alarm in a disaster-warning system and a false trade in a financial application have very different operational consequences. Future work should therefore study finite blocklength tradeoffs of the form
\[
    \log \log N^\star(n,\lambda_1,\lambda_2)
\]
for different channel classes, coding constraints, and asymmetric error requirements.

Another important direction is the development of explicit finite-length ID
codes. While there are several known constructions
based on Reed--Solomon codes, Reed--Muller codes, combinatorial designs, and hashing methods,
several aspects remain insufficiently understood,
such as their finite blocklength performance, implementation complexity, 
possibilities for hardware acceleration to reduce practical computational cost,
and robustness under practical channel impairments.

\subsection{Complexity and scalable decoding}

A second major challenge concerns computational complexity. \emph{Randomized} ID codes can achieve the characteristic double-exponential scaling, but their implementation may require large-scale hypothesis testing at the receiver. 

Future work should therefore focus on structured decoding architectures. Possible directions include hierarchical ID, sparse hypothesis testing, tree-based identity classes, compressed representations of decoding regions. Such methods could reduce receiver-side complexity while preserving the main scaling advantage of ID.

The distinction between \emph{randomized} and \emph{deterministic} ID is also important from an implementation perspective. \emph{Randomized} ID can provide superior asymptotic scaling, whereas \emph{deterministic} ID may offer simpler encoding and decoding procedures. A systematic comparison between both approaches under realistic complexity, latency, memory, and energy constraints is still missing.

In particular, practical systems require design criteria beyond ID capacity alone. Relevant metrics include encoding complexity, decoding complexity, memory size, synchronization overhead, latency, robustness to channel mismatch, and the cost of generating or distributing randomness.

\subsection{Hybrid Shannon and post-Shannon architectures}

ID-based communication should not be viewed as a replacement for classical Shannon transmission. Instead, one of the most promising research directions is the design of hybrid architectures in which ID is used for relevance filtering, activation, and control, while conventional transmission is used for payload delivery.

A typical architecture may consist of two stages. In the first stage, a short ID signal determines whether a receiver, device, record, or action is relevant. In the second stage, classical communication is activated only if more detailed information is required. Such a design naturally separates control-plane and data-plane functions. ID provides scalable, low-overhead triggering, while Shannon transmission provides reliable payload reconstruction. An example of this approach is~\cite{lengerke2022stopping}, where ID codes are used to understand if a digital twin application is out of sync with its real-life representation, and if so, message transmission is used to synchronize them again. Another example is~\cite{kutsevol2025distributed}, where vehicles in a truck platoon with limited knowledge of the system's state use ID codes to synchronize their distributed control decisions with a centralized edge server that has a global view of the system state.

This hybrid view is particularly relevant for mobile networks. ID can be used to activate subsets of devices, assign transmission opportunities, or trigger local control actions. Once a device has been selected, conventional communication can be used for high-rate data transfer. Similar two-stage mechanisms appear in monitoring systems, where ID triggers an alarm state and classical communication subsequently provides detailed warning information.

Future research should develop protocol architectures that integrate ID codes with existing random-access, scheduling, broadcast, and multi-cast mechanisms. Important open questions include how ID messages should be embedded into current control channels, how fallback transmission should be triggered after an ID decision, and how reliability guarantees should be formulated for the combined ID-transmission system.

\subsection{Identification and semantic communication}

Message ID, as introduced by Ahlswede and Dueck
\cite{ahlswede1989identification}, clearly belongs to the broader class
of goal-oriented communication systems.
Instead of reconstructing the complete transmitted message,
the receiver only verifies whether a particular message of
interest has been sent.
The communication process is therefore directly connected
to a specific task or decision problem.

The question naturally arises as to what extent ID
can also be interpreted as a form of semantic communication.
To answer this question, it is first necessary to clarify what
is meant by semantic communication.

As already discussed above, Shannon's classical communication
theory focuses on the reliable transmission of symbols over noisy
channels while intentionally ignoring the semantic content of
messages \cite{ShannonWeaver1949}.
Inspired by Weaver's extension of Shannon's framework,
many attempts have been made to include semantic and effectiveness
aspects into communication theory.

One important direction defines semantic information through
entropy-like or logic-based measures.
Bar-Hillel and Carnap \cite{CarnapBarHillel1952}
introduced a semantic information measure based on logical
probabilities.
However, their approach is limited to propositional logic
and leads to contradictions in certain scenarios.
Floridi later extended the concept by explicitly incorporating
truthfulness into the definition of semantic information
\cite{Floridi2004}.
Nevertheless, this framework also remains restricted and
depends on external truth references.
Overall, many existing approaches to semantic information
are tailored to specific applications and do not yet provide
a universally accepted mathematical framework
\cite{Gunduz2023}.

A promising direction toward a rigorous theory of semantic
communication was proposed in \cite{Gholipour2026}.
The authors argue that semantic communication should be
decomposed into simpler subproblems that can be analyzed
systematically.
For one such subproblem, where semantic degradation only
originates from the physical communication channel,
an achievable rate for reliable semantic communication
was derived.
This approach currently appears to be one of the most
promising ways to mathematically formalize semantic
communication within an information-theoretic framework.

However, alternative interpretations of semantic communication
also exist that are not necessarily compatible with this view.
For example, one may interpret joint communication and
ID systems as semantic communication systems,
where the ID stage acts as an additional semantic
query associated with the transmitted message.

An interesting idea is given in \cite{Mariani2025},
where sequential hypothesis testing is combined with
classical communication.
After message transmission, a nearest-neighbor test is used
to determine the semantic category or feature associated
with the transmitted message.
The authors explicitly relate their framework to ID
theory and suggest that concepts from ID coding
may provide useful tools for implementing semantic communication.
However, realizing such a connection rigorously would likely
require a generalized or modified notion of ID,
since the classical Ahlswede--Dueck ID framework
does not directly model semantic feature extraction. An example use case for semantic communication is provided in Section \ref{subsec:semanticComm}.

\subsection{Joint sensing, feedback, and common randomness}

JIDAS is a particularly promising direction because sensing can serve two different roles. First, sensed information can determine which identity, action, or device subset should be activated. Second, sensing and feedback can provide correlated observations that may be used to generate common randomness. For further details, see Section \ref{sec:JIDAS}.

This connection is important because common randomness can significantly enhance ID performance. In practical systems, however, common randomness is not free. It must be generated, synchronized, protected, and refreshed. Sensing-based randomness generation, therefore, raises several open questions: How much usable common randomness can be extracted from a physical environment? How robust is this randomness under noise, mobility, and adversarial observation? How should randomness be allocated between ID coding, security, and other system functions?

Another open problem is the joint optimization of sensing accuracy and ID reliability. In many applications, the system must simultaneously estimate an environmental state and decide which receivers should react. This naturally leads to capacity-distortion-reliability tradeoffs, where sensing quality, ID rate, and error probabilities must be optimized together.

\subsection{Security, authentication, and privacy}

Security is a critical issue for ID-based communication. Since an ID signal may directly trigger an action, forged or replayed signals could cause unintended activations, denial-of-service effects, or safety-critical failures. This is particularly relevant in industrial control, medical applications, financial systems, and mobile-network control.

Future systems therefore require authentication mechanisms that are compatible
with the short-message and low-latency nature of ID. Classical cryptographic
authentication may introduce overhead that reduces the advantage of ID-based
signaling. While authentication codes have been studied in coding theory
\cite{simmons1988authentication_survey} and are very similar to ID codes,
the relation of the codes and reciprocal use of code constructions for the other
problem has not been studied comprehensively.
This motivates further study of lightweight authentication,
physical-layer security, secret ID codes, and secure common randomness
generation.

Privacy is equally important. ID-based storage systems may answer only binary membership queries and therefore appear privacy-preserving at first sight. However, repeated adaptive queries may gradually reveal sensitive information. Future work should therefore investigate privacy leakage under repeated ID queries, rate-limited access mechanisms, differentially private ID protocols, and secure query auditing.

The interaction between ID, secrecy, authentication, and privacy remains a largely open research area. For further details, see Section \ref{sec:IDsecure}.

\subsection{Open information-theoretic problems}

Beyond implementation challenges, many fundamental information-theoretic questions remain open. These include \emph{deterministic} ID over continuous-alphabet channels, finite blocklength ID, multi-user ID networks, secure ID, ID with feedback, and ID under sensing constraints.

For Gaussian channels, \emph{deterministic} ID exhibits scaling behavior that differs from both classical transmission and \emph{randomized} ID. A complete understanding of such intermediate scaling regimes remains an important theoretical challenge. Similarly, ID over networks, relay channels, broadcast channels, interference channels, and massive multiple-access systems is still far less developed than classical Shannon network information theory.


Quantum ID codes also represent an important long-term direction. Since quantum
communication systems naturally involve measurement, hypothesis testing, and
state discrimination, ID may provide a useful perspective for future quantum
networks and quantum-secure communication architectures.
There are two types of ID over quantum channels \cite{winter2013qid_book}:
Once can consider ID of
classical messages over quantum channels, or the messages themselves can be
quantum states. The latter problem
can be understood as a problem of relaying or delegating a class of measurements
to a third party: Alice holds a quantum state and wants to enable Bob to perform
any pure-state (rank one) measurement on it, but she does not know which
measurement Bob will perform.
This behaves again differently from either
classical ID or transmission over quantum channels
\cite{haydenWinter2012qid,winter2013qid_book,rosenbergerBocheDeppePereg2025qid_ea}.
For instance, the
kind of dependence on shared entanglement deviates from the dependence
of classical ID on shared randomness as well as the dependence of classical or
quantum transmission on entanglement~\cite{rosenbergerBocheDeppePereg2025qid_ea}.

\bigskip

Overall, ID coding opens a broad research landscape at the intersection of information theory, communication systems, sensing, distributed control, and network architecture. Its main promise lies not in replacing Shannon communication, but in complementing it with a new communication primitive based on relevance testing and sparse activation.

Future communication systems may therefore combine ID-based signaling for scalable control, semantic triggering, and distributed decision making with classical transmission for reliable payload delivery. Bridging the gap between asymptotic ID theory and practical low-complexity system implementations remains one of the central challenges for future research.

\subsection{Migration path toward identification-based mobile networks}
\label{subsec:migration-path-mobile-networks}

A key open challenge is how to integrate ID-based communication into existing mobile-network architectures. ID should not replace classical Shannon transmission, but complement it as a control-plane function or relevance filter. Its role is to decide which devices, events, or actions are relevant before full payload communication is activated. A realistic migration path can be organized into four stages.

\paragraph{Stage 1: Application-layer overlay.}
ID-coded messages are transported over existing mobile connectivity as ordinary application-layer packets. Encoders and decoders are implemented in application servers, IoT gateways, cloud platforms, or edge applications. This stage requires no changes to the radio-access network and is suitable for early use cases such as alarm aggregation, geofencing verification, semantic action triggering, and special-purpose storage queries.

\paragraph{Stage 2: Edge-assisted identification.}
ID functions are moved closer to the network edge. Edge nodes collect ID-coded indications from many devices, perform local decoding or aggregation, and trigger application or network actions. This reduces backhaul traffic, cloud processing, and decision latency, while still relying on conventional radio-access procedures.

\paragraph{Stage 3: Radio Access Network (RAN)-assisted broadcast activation.}
The g-Node B broadcasts ID-coded activation opportunities to large device populations. Each device locally tests whether the signal applies to its identity, device class, location, time window, service type, or sensed state. Only devices with a positive ID result activate and proceed to classical transmission. For example, instead of each IoT device independently attempting random access and requesting uplink resources, the gNB may broadcast an ID-coded opportunity such as: devices of class $c$ in region $r$ during time window $t$ may transmit. Non-matching devices remain silent. This can reduce Random Access Channel load, scheduling requests, paging overhead, uplink contention, Physical Resource Block consumption, and device energy.

\paragraph{Stage 4: Native 6G identification primitive.}
In the long term, ID could become a native 6G control-plane primitive. Future networks may support dedicated ID channels, standardized identity namespaces, ID-aware scheduling, and secure activation mechanisms. Classical transmission would remain responsible for payload delivery, while ID would decide whether to transmit the payload.


\section{Conclusion}
\label{sec:conclusion}
ID theory extends the classical Shannon communication paradigm by replacing full message reconstruction with relevance testing. Instead of recovering complete payload information, receivers only decide whether a particular identity, event, action, or control instruction is relevant to them. This seemingly small conceptual shift fundamentally changes both the scaling behavior and the architectural possibilities of communication systems.

A central insight of ID theory is the double-exponential growth of the number of identifiable messages with blocklength. This property enables communication architectures operating in regimes where classical transmission-based systems face increasing scalability limitations. In particular, ID-based communication becomes attractive whenever communication tasks are sparse, event-driven, control-oriented, or semantic in nature, and where only a small subset of receivers, devices, or actions is relevant at a given time.

In this survey, we reviewed the theoretical foundations of \emph{deterministic} and \emph{randomized} ID, including their connections to common randomness generation, feedback, sensing, and distributed decision making. We further discussed how these concepts naturally lead to post-Shannon communication architectures in which communication signals are interpreted less as payload streams and more as distributed triggers, queries, activation signals, or semantic actions.

The considered application domains illustrate that this perspective is relevant far beyond classical communication scenarios. Monitoring systems, sensing-driven control, ID-oriented storage architectures, and scalable mobile-network coordination all share a common structural property: the communication objective is not the reconstruction of complete information, but the efficient selection of relevant actions, devices, events, or hypotheses from extremely large logical spaces.

At the same time, substantial challenges remain open before such architectures can become practically deployable at large scale. These challenges include finite blocklength ID theory, low-complexity decoding architectures,
privacy-preserving query mechanisms, secure activation protocols, and the integration of ID-based signaling into existing communication infrastructures. Future systems will therefore likely combine classical payload transmission with lightweight ID-based control and relevance-filtering layers.

More broadly, ID suggests a possible long-term evolution of communication systems away from purely data-centric architectures toward relevance-centric and action-oriented communication paradigms. As future networks increasingly integrate sensing, AI-driven control, semantic processing, autonomous agents, and massive machine populations, communication may no longer primarily consist of transporting complete messages, but rather of efficiently determining which information, action, or decision is relevant to which part of the system.

For this reason, ID should not only be viewed as a specialized coding-theoretic concept, but also as a promising architectural principle for future large-scale distributed communication systems.

\section*{Acknowledgement}

The authors gratefully acknowledge the financial support of the Federal Ministry of Research, Technology and Space of Germany (BMFTR) within the programme “Souverän. Digital. Vernetzt.”, joint project 6G-life (grant numbers 16KIS2414, 16KIS2413K and 16KIS2415).
This work was further supported by the BMFTR Quantum Programme, including the projects QUIET (grants 16KISQ093 and 16KISQ0170), QD-CamNetz (grants 16KISQ077 and 16KISQ169), and QSTARS (grants 16KIS2611 and 16KIS2602).
Additional support was provided by the German Research Foundation (DFG) within the project “Post-Shannon Theory and Implementation” (grants DE1915/2-1 and BO 1734/38-1).
The authors also acknowledge financial support from the Federal Ministry of Research, Technology and Space of Germany (BMFTR) within the project Internet of Bio-Nano-Things (IoBNT) under grant number 5310223. The authors also acknowledge
funding by the German Research Foundation as part of Germany’s
Excellence Strategy – EXC 2050/2 – Project ID 390696704 – Cluster of
Excellence “Centre
for Tactile Internet with Human-in-the-Loop” (CeTI) of Technische
Universität Dresden.

\bibliographystyle{splncs04}

\bibliography{IEEEfull,long,references}

\end{document}